\documentclass[apj]{emulateapj}

\newcommand {\Lya}    {Ly$\alpha$}   
\newcommand {\Lyb}    {Ly$\beta$}    
\newcommand {\Lyg}    {Ly$\gamma$}

\newcommand {\HI}     {\ion{H}{1}}   
\newcommand {\OIV}    {\ion{O}{4}}
\newcommand {\OVI}    {\ion{O}{6}}   
\newcommand {\CIII}   {\ion{C}{3}}   
\newcommand {\CIV}    {\ion{C}{4}}
\newcommand {\NIII}   {\ion{N}{3}}
\newcommand {\NV}     {\ion{N}{5}}
\newcommand {\SiIV}   {\ion{Si}{4}}
\newcommand {\SiIII}  {\ion{Si}{3}}
\newcommand {\SiII}   {\ion{Si}{2}}

\newcommand {\FeII}   {\ion{Fe}{2}}
\newcommand {\CII}    {\ion{C}{2}}
\newcommand {\CI}     {\ion{C}{1}}
\newcommand {\NiII}   {\ion{Ni}{2}}
\newcommand {\NeVIII} {\ion{Ne}{8}}

\newcommand {\kms}    {km~s$^{-1}$}
\newcommand {\NHI}    {$N_{\rm HI}$}
\newcommand {\tnma}{\tablenotemark{a}}
\newcommand {\tnmb}{\tablenotemark{b}}
\newcommand {\tnmc}{\tablenotemark{c}}
\newcommand {\nd}  {\nodata}
\newcommand {\lam}    {$\lambda$}
\newcommand {\FUSE}  {{\it FUSE}} 
\newcommand {\HST}  {{\it HST}}
\newcommand {\dndz}  {$d{\cal N}/dz$}
\newcommand {\flux}  {$\rm erg~cm^{-2}~s^{-1}~\AA^{-1}$}
\newcommand {\etal}  {et~al.} 
\newcommand {\cd}    {cm$^{-2}$}  

\shorttitle{HST/COS Survey of the Low-$z$ IGM}
\shortauthors{Danforth \etal}

\begin{document}

\title{An HST/COS Survey of the Low-Redshift Intergalactic Medium.\\ I. Survey, Methodology, and Overall Results
\footnote{Based on observations made with the NASA/ESA {\it Hubble Space Telescope}, obtained from the data archive at the Space Telescope Science Institute. STScI is operated by the Association of Universities for Research in Astronomy, Inc. under NASA contract NAS5-26555.}}

\author{
 Charles W. Danforth, 
 Brian A. Keeney, 
 Evan M. Tilton, 
 J. Michael Shull, 
 John T. Stocke, 
 Matthew Stevans\altaffilmark{1}, 
 Matthew M. Pieri\altaffilmark{2}, 
 Blair D. Savage\altaffilmark{3}, 
 Kevin France, 
 David Syphers\altaffilmark{4}, 
 Britton D. Smith\altaffilmark{5},
 James C. Green, 
 Cynthia Froning\altaffilmark{1}, 
 Steven V. Penton\altaffilmark{6}, \&
 Steven N. Osterman\altaffilmark{7}
}

\affil{CASA, Department of Astrophysical \& Planetary Sciences, University of Colorado, 389-UCB, Boulder, CO, USA 80309; danforth@colorado.edu}
\altaffiltext{1}{Department of Astronomy, University of Texas at Austin, Austin, TX 78712}
\altaffiltext{2}{A*MIDEX, Aix Marseille Universit{\'e}, CNRS, LAM, UMR7326, Marseille, FR}
\altaffiltext{3}{Department of Astronomy, University of Wisconsin, Madison, WI, USA, 53706}
\altaffiltext{4}{East Washington University, Cheney, WA, USA, 99004}
\altaffiltext{5}{Institute for Astronomy, University of Edinburgh, Royal Observatory, Edinburgh EH9 3HJ, UK}
\altaffiltext{6}{Space Telescope Science Institute, Baltimore, MD, USA, 21218}
\altaffiltext{7}{The Johns Hopkins University Applied Physics Lab, Laurel, MD, USA, 20723}

\begin{abstract}

We use high-quality, medium-resolution {\it Hubble Space Telescope}/Cosmic Origins Spectrograph (\HST/COS) observations of 82 UV-bright AGN at redshifts $z_{\rm AGN}<0.85$ to construct the largest survey of the low-redshift intergalactic medium (IGM) to date: 5138 individual extragalactic absorption lines in \HI\ and 25 different metal-ion species grouped  into 2611 distinct redshift systems at $z_{\rm abs}<0.75$ covering total redshift pathlengths $\Delta z_{\rm HI}=21.7$ and $\Delta z_{\rm OVI}=14.5$.  Our semi-automated line-finding and measurement technique renders the catalog as objectively-defined as possible.  The cumulative column-density distribution of \HI\ systems can be parametrized $d{\cal N}(>N)/dz=C_{14}(N/10^{14}$~\cd$)^{-(\beta-1)}$, with $C_{14}=25\pm1$ and $\beta=1.65\pm0.02$.  This distribution is seen to evolve both in amplitude, $C_{14}\propto(1+z)^{2.3\pm0.1}$, and slope $\beta(z)=1.75-0.31\,z$ for $z\le0.47$.  We observe metal lines in 418 systems, and find that the fraction of IGM absorbers detected in metals is strongly dependent on \NHI.  The distribution of \OVI\ absorbers appear to evolve in the same sense as the \Lya\ forest.  We calculate contributions to $\Omega_b$ from different components of the low-$z$ IGM and determine the \Lya\ decrement as a function of redshift.  IGM absorbers are analyzed via a two-point correlation function in velocity space.  We find substantial clustering of \HI\ absorbers on scales of $\Delta v=50-300$ \kms\ with no significant clustering at $\Delta v\ga1000$~\kms.  Splitting the sample into strong and weak absorbers, we see that most of the clustering occurs in strong, $N_{\rm HI}\ga10^{13.5}$ \cd, metal-bearing IGM systems.  
The full catalog of absorption lines and fully-reduced spectra is available via the Mikulski Archive for Space Telescopes (MAST) as a high-level science product at {\tt http://archive.stsci.edu/prepds/igm/}. 
\end{abstract}



\section{Introduction}

The low-redshift intergalactic medium (IGM) holds many important clues to complete our understanding of cosmology.  Even after nearly 14 Gyr of evolution, only a small fraction of baryonic matter has collapsed into luminous objects (galaxies, groups, clusters) while $\sim80\%$ or more still exists as a diffuse, often unvirialized IGM, the distribution and characteristics of which are only beginning to be measured \citep[e.g.,][]{Shull12a}.  Furthermore, there is a complicated and poorly understood interplay between the IGM, circumgalactic medium (CGM), and stars and gas in galaxies.  These gaseous reservoirs provide raw material which is subsequently formed into stars and galaxies.  These, in turn, enrich the IGM via outflows driven by supernovae, radiation pressure, and active galactic nuclei \citep[AGN][]{OppenheimerDave08,Smith11}.

Diffuse intergalactic gas is currently quite difficult to observe in emission \citep[Frank \etal\ 2012; but see also][]{Steidel11,Martin14a,Martin14b}.  The most sensitive method for detecting most of the gas is through absorption-line spectroscopy using bright background objects (typically AGN) to provide an ultraviolet continuum.  The highest concentration of strong gas-diagnostic lines is in the rest-frame far-ultraviolet (FUV) band from $\sim2000$~\AA\ shortward to the Lyman edge at 912~\AA.  Investigating the FUV at low redshift requires optimized spectrographs above Earth's UV-blocking atmosphere.  Thus, there have been a series of space-based UV spectrographs, both as primary-science instruments on space-borne observatories ({\it Copernicus, International Ultraviolet Explorer, Hopkins Ultraviolet Telescope, Far-Ultraviolet Spectroscopic Explorer}) and instruments installed aboard the {\it Hubble Space Telescope} or \HST: Faint Object Spectrograph (FOS), Goddard High-Resolution Spectrograph (GHRS), Space Telescope Imaging Spectrograph (STIS), and now the Cosmic Origins Spectrograph (COS).  

COS is the fourth-generation UV spectrograph onboard \HST\ and is optimized for medium-resolution ($\lambda/\Delta\lambda\approx18,000$, $\Delta v\approx17$ \kms) spectroscopy of point sources in the 1135--1800~\AA\ band \citep{Green12,Osterman11}.  COS has an effective area that is an order of magnitude larger in the FUV ($\lambda<1800$~\AA) than previous spectrographs.  Furthermore, the excellent scattered light control and low background detectors of COS mean that fainter objects ($F_\lambda\la10^{-14}$ \flux) can be observed than with previous instruments, and higher quality spectra of bright targets can be obtained in much less time.  In its first five years of science operations \citep{Green12}, COS accumulated an unprecedented archive of hundreds of AGN spectra collected under a broad range of scientific programs from AGN physics to studies of the interstellar medium (ISM) in our own Galaxy.  Regardless of their original scientific intent, many of these spectra are suitable for quasar absorption-line studies of the IGM. 

In this paper, we build on the heritage of many previous low-$z$ IGM absorber catalogs from \HST/FOS \citep{Bahcall93,Bahcall96,Jannuzi98,Weymann98}, \HST/GHRS \citep{Penton1}, \FUSE\ \citep{Danforth05,Danforth06}, and \HST/STIS \citep{Penton04,Lehner07,DS08,Tripp08,Danforth10a,Tilton12}.  This current survey represents the largest sample of low-$z$ absorbers to date and is more sensitive than previous studies in most cases.  It was designed to be a general-purpose survey with applications to a wide variety of extragalactic astrophysics: 82 extragalactic sight lines covering a combined redshift pathlength in \HI\ of $\Delta z=21.7$, 5138 intergalactic absorption lines comprising 2611 distinct redshift systems, and detections of 25 different metal-ion species.  We describe the sample selection criteria, data reduction methods, and semi-automated measurement techniques in Section~2.  We present the catalog and substantial electronic resources available to users of the survey in Section~3.  Overall results from the survey are given in Section~4.  In Section~5, we present some of the more detailed findings from the survey, including the evolution of \HI\ and metals in the IGM at $z\le0.47$ and the radial-velocity clustering properties of IGM absorbers (the two-point correlation function).  Important survey parameters, fitted quantities, and initial findings are summarized in Section~6.

\section{Data Analysis}

\subsection{Sight Line Selection}

In the past five years, over 400 extragalactic targets have been observed with COS.  The main objective of this project is to develop a comprehensive, statistical catalog of intervening absorbers in the low-redshift universe, in particular weak \HI\ and metal-line systems.  Obtaining high signal-to-noise data was our first priority in choosing which targets to include, which precludes many of the archival datasets.  With a few exceptions, we include only spectra with a typical signal-to-noise of $S/N\ga15$ per ($\sim17$ \kms) COS resolution element in the \Lya\ forest \citep[see][for full discussion of the acheivable S/N in COS data]{Keeney12}.

Secondly, we require reasonably distant targets to maximize the redshift pathlength probed by each sight line.  Nearby AGN, particularly Seyfert galaxies, are typically bright and often have exquisite spectra, but their available IGM pathlength is short and often contaminated by absorption intrinsic to the AGN.  We require a sight line to have $\Delta z_{\rm Ly\alpha}\ge 0.05$ of unobstructed pathlength for inclusion in our survey.  Conversely, high-$z$ targets sample long IGM pathlengths, but they often suffer from a high line density that makes line identification difficult.  The two moderate-resolution FUV channels of COS (G130M and G160M) are only sensitive to \HI\ absorption through \Lya\ at $z\la0.47$ and through \Lyb\ and higher-order Lyman transitions at $0.1\la z\la 0.9$.  Thus we set an upper limit on source redshift of $z_{\rm AGN}<0.9$.  Similarly, we concentrate on AGN observed with both the G130M and G160M gratings for the longest possible wavelength coverage.  We augment the target list with a set of G130M-only observations of targets at $z\la0.2$ where the entire \Lya\ forest falls within the G130M grating.

The COS FUV archive also contains $\ga130$ extragalactic sight lines observed with the low-resolution ($\Delta v\ga100$ \kms) G140L grating.  While these data are often of higher quality than those from IUE or \HST/FOS or low-resolution modes of GHRS or STIS, they are sensitive only to the strongest intervening absorbers, and we have not utilized them in this survey.  However, these low-resolution and/or low-S/N data are useful for studies of AGN continua \citep[e.g.][]{Shull12b,Stevans14}, for surveys of strong absorption lines, and as flux-qualification observations for future medium-resolution observations.

AGN with strong absorption lines (BALs, mini-BALs) were not explicitly excluded from the survey.  There are a few AGN with strong intrinsic absorption (e.g., RBS\,542), but these systems are not on the level of a BAL or mini-BAL.  In any case, absorption obviously intrinsic to the absorber was excluded from analysis (as discussed in S\ref{intrinsic_abs}).

To constitute our survey, we selected 82 AGN sight lines from the archive which met these criteria; 68 were observed with both FUV medium-resolution gratings ($1135-1800$~\AA\ at $\sim17$ \kms\ resolution) in the redshift range $0.058 \le z_{\rm AGN} \le 0.852$ and an additional 14 AGN at $0.07 \le z_{\rm AGN}\le 0.2$ had coverage in G130M only ($1135-1450$~\AA).  Astronomical target information is presented in Table~1.  Most of the AGN observed in Cycles~18--20 under the Guaranteed Time Observation programs (GTO; PI-Green) are included, along with numerous archival datasets collected under various Guest Observer programs.  Observational and programatic details are presented in Table~2.

Individual exposures for each target were obtained from the Mikulski Archive for Space Telescopes (MAST).  We start with {\tt x1d.fits} files (individual exposures reduced to one-dimensional spectra) processed uniformly by the pipeline software circa mid-2014\footnote{\tt http://www.stsci.edu/hst/cos/pipeline/CALCOSReleaseNotes/notes/}.  All exposures of each sight line were combined to maximize the $S/N$ of each spectrum.  The calibrated, one-dimensional spectra for each target were coadded into a continuous spectrum, usually over the useful range $\sim1140-1790$~\AA\ employing a custom IDL procedure developed and extensively tested at University of Colorado.  

\clearpage
\LongTables
\begin{deluxetable*}{lllrrc}
\tabletypesize{\scriptsize}
\tablecolumns{6} 
\tablewidth{0pt} 
\tablecaption{{\it HST}/COS Sight Lines}
\tablehead{\colhead{Sight Line}   &
         \colhead{R.A. (J2000)}  &
         \colhead{Dec. (J2000)}  &
	   \colhead{$z_{\rm AGN}$}&
         \colhead{Flux\tnma} &
         \colhead{AGN type}
         }
\startdata 
PHL\,2525                 & 00 00 24.42   & -12 45 47.8 & 0.1990 &  17.7 & QSO          \\
PG\,0003$+$158            & 00 05 59.24   & +16 09 49.0 & 0.4509 &   7.7 & Sy1.2        \\
PG\,0026$+$129            & 00 29 13.71   & +13 16 04.0 & 0.1420 &  19.1 & Sy1          \\
QSO\,0045$+$3926          & 00 48 18.98   & +39 41 11.6 & 0.1340 &   9.3 & Sy1          \\
HE\,0056$-$3622           & 00 58 37.39   & -36 06 05.0 & 0.1641 &  17.3 & Sy1          \\
RBS\,144                  & 01 00 27.13   & -51 13 54.5 & 0.0628 &  22.1 & Sy1          \\
B0117$-$2837              & 01 19 35.70   & -28 21 31.4 & 0.3489 &  11.6 & Sy1        \\
Ton\,S210                 & 01 21 51.51   & -28 20 57.8 & 0.1160 &  35.9 & Sy1        \\
HE\,0153$-$4520           & 01 55 13.20   & -45 06 12.0 & 0.4510 &  15.8 & QSO          \\
PG\,0157$+$001            & 01 59 50.25   & +00 23 41.3 & 0.1631 &  20.4 & Sy1.5        \\
3C\,57                    & 02 01 57.16   & -11 32 33.1 & 0.6705 &   6.6 & Sy1.2        \\
3C\,66A                   & 02 22 39.61   & +43 02 07.8 & $>$0.3347 &   6.8 & BL\,Lac        \\
HE\,0226$-$4110           & 02 28 15.19   & -40 57 14.6 & 0.4934 &  21.4 & Sy1        \\
HE\,0238$-$1904           & 02 40 32.50   & -18 51 51.0 & 0.6310 &  14.0 & QSO          \\
UKS0242$-$724             & 02 43 09.60   & -72 16 48.4 & 0.1018 &  10.3 & Sy1.2        \\
PKS\,0405$-$123           & 04 07 48.43   & -12 11 36.7 & 0.5740 &  32.1 & Sy1.2        \\
RBS\,542                  & 04 26 00.70   & -57 12 01.8 & 0.1040 &  33.1 & Sy1.5        \\
HE\,0435$-$5304           & 04 36 50.80   & -52 58 49.0 & 0.4300 &   2.5 & QSO          \\
RX\,J0439.6$-$5311        & 04 39 38.64   & -53 11 31.6 & 0.2430 &   3.5 & Sy1          \\
RX\,J0503.1$-$6634        & 05 03 03.93   & -66 33 45.9 & 0.0640 &   6.4 & Sy1          \\
PKS\,0552$-$640           & 05 52 24.50   & -64 02 10.8 & 0.6800 &  11.1 & AGN          \\
PKS\,0558$-$504           & 05 59 47.39   & -50 26 51.9 & 0.1372 &  28.1 & NLSy1/FSRQ   \\
IRAS\,L06229$-$6434       & 06 23 07.68   & -64 36 20.7 & 0.1290 &   8.0 & FSRQ         \\
PKS\,0637$-$752           & 06 35 46.50   & -75 16 16.8 & 0.6500 &   8.0 & FSRQ        \\
S5\,0716$+$714            & 07 21 53.45   & +71 20 36.4 & $>$0.2315 &  24.3 & BL\,Lac      \\
SDSS\,J080908.13$+$461925 & 08 09 08.14   & +46 19 25.7 & 0.6563 &   7.2 & QSO          \\
PG\,0804$+$761            & 08 10 58.61   & +76 02 41.6 & 0.1000 &  79.6 & Sy1        \\
PG\,0832$+$251            & 08 35 35.80   & +24 59 41.0 & 0.3298 &   3.5 & QSO          \\
PG\,0838$+$770            & 08 44 45.26   & +76 53 09.5 & 0.1310 &   8.5 & Sy1        \\
PG\,0844$+$349            & 08 47 42.45   & +34 45 04.4 & 0.0640 &  25.5 & Sy1        \\
Mrk\,106                  & 09 19 55.36   & +55 21 37.4 & 0.1230 &  11.4 & Sy1        \\
SDSS\,J092554.43$+$453544 & 09 25 54.44   & +45 35 44.5 & 0.3295 &   5.1 & QSO          \\
SDSS\,J092909.79$+$464424 & 09 29 09.78   & +46 44 24.0 & 0.2400 &  11.5 & QSO          \\
SDSS\,J094952.91$+$390203 & 09 49 52.93   & +39 02 03.8 & 0.3656 &   7.8 & QSO          \\
RXS\,J09565$-$0452        & 09 56 30.18   & -04 53 17.0 & 0.1550 &   4.6 & Sy1          \\
PG\,0953$+$414            & 09 56 52.41   & +41 15 22.1 & 0.2341 &  38.1 & QSO          \\
PG\,1001$+$291            & 10 04 02.59   & +28 55 35.2 & 0.3297 &  10.6 & Sy1          \\
FBQS\,J1010$+$3003        & 10 10 00.70   & +30 03 22.0 & 0.2558 &   2.5 & QSO          \\
Ton\,1187                 & 10 13 03.20   & +35 51 23.0 & 0.0789 &  17.3 & Sy1.2        \\
PG\,1011$-$040            & 10 14 20.68   & -04 18 40.5 & 0.0583 &  16.2 & Sy1.2        \\
1ES\,1028$+$511           & 10 31 18.50   & +50 53 36.0 & 0.3604 &   2.6 & BL\,Lac        \\
1SAX\,J1032.3$+$5051      & 10 32 16.10   & +50 51 20.0 & 0.1731 &   1.1 & AGN          \\
PG\,1048$+$342            & 10 51 43.90   & +33 59 26.7 & 0.1671 &   5.9 & Sy1        \\
PG\,1049$-$005            & 10 51 51.48   & -00 51 17.6 & 0.3599 &   9.2 & Sy1.5        \\
PMN\,J1103$-$2329         & 11 03 37.60   & -23 29 30.0 & 0.1860 &   2.1 & BL\,Lac/FSRQ   \\
HS\,1102$+$3441           & 11 05 39.80   & +34 25 34.4 & 0.5088 &   3.6 & QSO          \\
SBS\,1108$+$560           & 11 11 32.20   & +55 47 26.0 & 0.7666 &   4.9 & QSO          \\
PG\,1115$+$407            & 11 18 30.30   & +40 25 54.0 & 0.1546 &  10.5 & Sy1        \\
PG\,1116$+$215            & 11 19 08.60   & +21 19 18.0 & 0.1763 &  45.7 & Sy1        \\
PG\,1121$+$422            & 11 24 39.18   & +42 01 45.0 & 0.2250 &   7.4 & Sy1        \\
SBS\,1122$+$594           & 11 25 53.79   & +59 10 21.6 & 0.8520 &   2.7 & QSO          \\
Ton\,580                  & 11 31 09.50   & +31 14 05.0 & 0.2895 &   9.5 & Sy1/FSRQ   \\
3C\,263                   & 11 39 56.99   & +65 47 49.2 & 0.6460 &  10.4 & FR2/Sy1.2    \\
PG\,1216$+$069            & 12 19 20.93   & +06 38 38.5 & 0.3313 &  12.3 & NLSy1        \\
PG\,1222$+$216            & 12 24 54.45   & +21 22 46.3 & 0.4320 &  17.0 & Blazar       \\
3C\,273                   & 12 29 06.70   & +02 03 08.7 & 0.1583 & 461.5 & FSRQ/Sy1.0   \\
Q\,1230$+$0115            & 12 30 50.00   & +01 15 21.5 & 0.1170 &  36.3 & NLSy1        \\
PG\,1229$+$204            & 12 32 03.61   & +20 09 29.4 & 0.0630 &  19.2 & Sy1        \\
PG\,1259$+$593            & 13 01 12.90   & +59 02 07.0 & 0.4778 &  15.3 & Sy1        \\
PKS\,1302$-$102           & 13 05 33.00   & -10 33 19.0 & 0.2784 &  14.9 & FSRQ/Sy1.2   \\
PG\,1307$+$085            & 13 09 47.01   & +08 19 48.3 & 0.1550 &  25.0 & Sy1.2        \\
PG\,1309$+$355            & 13 12 17.75   & +35 15 21.1 & 0.1829 &  10.1 & QSO          \\
SDSS\,J135712.61$+$170444 & 13 57 12.60   & +17 04 44.0 & 0.1500 &   4.7 & QSO          \\
PG\,1424$+$240            & 14 27 00.39   & +23 48 00.0 & $>$0.6035& 15.2 & BL\,Lac        \\
PG\,1435$-$067            & 14 38 16.15   & -06 58 20.7 & 0.1260 &  16.2 & QSO          \\
Mrk\,478                  & 14 42 07.47   & +35 26 23.0 & 0.0791 &  26.4 & NLSy1        \\
Ton\,236                  & 15 28 40.60   & +28 25 29.7 & 0.4500 &   6.1 & Sy1.2        \\
1ES\,1553$+$113           & 15 55 43.04   & +11 11 24.4 & $>$0.4140 &  15.9 & BL\,Lac/FSRQ   \\
Mrk\,876                  & 16 13 57.20   & +65 43 010.0 & 0.1290 &  39.5 & Sy1          \\
PG\,1626$+$554            & 16 27 56.12   & +55 22 31.5 & 0.1330 &  22.8 & Sy1        \\
H\,1821$+$643             & 18 21 57.30   & +64 20 36.0 & 0.2968 &  53.0 & Sy1.2        \\
PKS\,2005$-$489           & 20 09 25.39   & -48 49 53.7 & 0.0710 &  22.3 & BL\,Lac        \\
Mrk\,1513                 & 21 32 27.92   & +10 08 18.7 & 0.0630 &  17.5 & Sy1.5        \\
RX\,J2154.1$-$4414        & 21 54 51.06   & -44 14 06.0 & 0.3440 &   9.8 & Sy1          \\
PHL\,1811                 & 21 55 01.50   & -09 22 25.0 & 0.1920 &  56.2 & NLSy1        \\
PKS\,2155$-$304           & 21 58 52.07   & -30 13 32.1 & 0.1165 &  73.1 & BL\,Lac      \\
RBS\,1892                 & 22 45 20.31   & -46 52 11.8 & 0.2000 &  20.3 & Sy1          \\
IRAS\,F22456$-$5125       & 22 48 41.20   & -51 09 53.2 & 0.1000 &  19.3 & Sy1.5        \\
MR\,2251$-$178            & 22 54 05.88   & -17 34 55.3 & 0.0640 &  34.3 & Sy1.5        \\
PMN\,J2345$-$1555         & 23 45 12.46   & -15 55 07.8 & 0.6210 &   7.0 & FSRQ         \\
PG\,2349$-$014            & 23 51 56.12   & -01 09 13.1 & 0.1737 &  29.8 & Sy1.2      \\
H\,2356$-$309             & 23 59 07.93   & -30 37 40.9 & 0.1651 &   1.8 & BL\,Lac        
\enddata      
 \tablenotetext{a}{Median observed continuum flux in the COS/FUV band in units of $10^{-15}~\rm erg~cm^{-2}~s^{-1}~\AA^{-1}$.}
\end{deluxetable*}

\LongTables
\newpage\begin{deluxetable*}{lccccccc}
\tabletypesize{\scriptsize}
\tablecolumns{8} 
\tablewidth{0pt} 
\tablecaption{COS Observation Details}
\tablehead{\colhead{Sight Line}   &
	   \colhead{Obs. Date}    &
	   \colhead{Exp. (ksec)}& 
	   \colhead{$S/N$}& 
	   \colhead{Exp. (ksec)}& 
	   \colhead{$S/N$}& 
	   \colhead{PI} &
	   \colhead{Program}  \\        
       \colhead{Name}   &
	   \colhead{(year-mo)}    &
	   \colhead{(G130M)}& 
	   \colhead{(G130M)\tnma}& 
	   \colhead{(G160M)}& 
	   \colhead{(G160M)\tnma}& 
	   \colhead{Name}   &
	   \colhead{Number} }
\startdata 
PHL\,2525                 & 2012-10 &   2.1 & 22 &  2.8 & 15  & Fox      & 12604\\
PG\,0003$+$158            & 2011-10 &  10.4 & 27 & 10.9 & 22  & Green    & 12038\\
PG\,0026$+$129            & 2011-10 &   1.9 & 21 & \nd  & \nd & Veilleux & 12569\\
QSO\,0045$+$3926          & 2009-10 &  13.5 & 40 & 17.2 & 31  & Rich     & 11632\\
                          & 2010-09 &       &    &      &     & Arav     & 11686\\
HE\,0056$-$3622           & 2012-07 &   5.0 & 32 &  5.7 & 20  & Fox      & 12604\\
RBS\,144                  & 2012-04 &   2.4 & 26 &  3.0 & 19  & Fox      & 12604\\
B0117$-$2837              & 2011-06 &   5.2 & 28 &  8.5 & 23  & Thom     & 12204\\
Ton\,S210                 & 2011-06 &   5.0 & 49 &  5.5 & 32  & Thom     & 12204\\
HE\,0153$-$4520           & 2009-12 &   5.2 & 30 &  5.9 & 23  & Green    & 11541\\
PG\,0157$+$001            & 2012-01 &   1.8 & 21 & \nd  &\nd  & Veilleux & 12569\\
3C\,57                    & 2011-08 &  11.0 & 29 &  8.7 & 17  & Green    & 12038\\
3C\,66A                   & 2012-11 &  12.6 & 27 &  7.2 & 17  & Stocke   & 12863\\
                          & 2012-11 &       &    &      &     & Furniss  & 12863\\
HE\,0226$-$4110           & 2010-02 &   6.8 & 41 &  7.8 & 26  & Green    & 11541\\
HE\,0238$-$1904           & 2009-12 &   6.5 & 30 &  7.5 & 26  & Green    & 11541\\
UKS0242$-$724             & 2011-06 &   2.1 & 20 &  3.2 & 14  & Misawa   & 12263\\
PKS\,0405$-$123           & 2009-08 &  24.2 & 76 & 11.1 & 35  & Noll     & 11508\\
                          & 2009-12 &       &    &      &     & Green    & 11541\\
RBS\,542                  & 2010-06 &  20.4 & 67 & 15.9 & 41  & Howk     & 11692\\
                          & 2010-06 &       &    &      &     & Arav     & 11686\\
HE\,0435$-$5304           & 2010-04 &   8.4 & 16 &  8.9 & 10  & Green    & 11520\\
RX\,J0439.6$-$5311        & 2010-02 &   8.2 & 21 &  8.9 & 12  & Green    & 11520\\
RX\,J0503.1$-$6634        & 2010-09 &   4.7 & 21 &  3.9 & 12  & Howk     & 11692\\
PKS\,0552$-$640           & 2009-12 &   9.3 & 26 &  8.3 & 22  & Howk     & 11692\\
PKS\,0558$-$504           & 2010-05 &   1.1 & 20 &  0.7 &  9  & Howk     & 11692\\
IRAS\,L06229$-$6434       & 2009-12 &   8.7 & 34 &  8.0 & 22  & Howk     & 11692\\
PKS\,0637$-$752           & 2009-12 &   9.6 & 26 &  8.7 & 18  & Howk     & 11692\\
S5\,0716$+$714            & 2011-12 &   6.0 & 41 &  8.3 & 32  & Green    & 12025\\
SDSS\,J080908.13$+$461925 & 2010-10 &   3.1 & 17 &  5.0 & 14  & Tumlinson& 12248\\
PG\,0804$+$761            & 2010-06 &   5.5 & 53 &  6.3 & 47  & Arav     & 11686\\
PG\,0832$+$251            & 2011-04 &   6.1 & 15 &  6.8 & 13  & Green    & 12025\\
PG\,0838$+$770            & 2009-09 &   8.9 & 34 &  6.3 & 18  & Green    & 11520\\
PG\,0844$+$349            & 2012-03 &   1.9 & 23 & \nd  &\nd  & Veilleux & 12569\\
Mrk\,106                  & 2011-05 &   6.5 & 32 &  7.6 & 21  & Green    & 12029\\
SDSS\,J092554.43$+$453544 & 2010-10 &   4.4 & 17 &  7.1 & 16  & Tumlinson& 12248\\
SDSS\,J092909.79$+$464424 & 2010-10 &   2.4 & 20 &  2.9 & 14  & Tumlinson& 12248\\
SDSS\,J094952.91$+$390203 & 2010-09 &   2.3 & 16 &  2.8 & 13  & Tumlinson& 12248\\
RXS\,J09565$-$0452        & 2010-10 &   7.7 & 20 & \nd  &\nd  & Wakker   & 12275\\
PG\,0953$+$414            & 2011-10 &   4.8 & 42 &  5.6 & 30  & Green    & 12038\\
PG\,1001$+$291            & 2012-03 &   6.2 & 26 &  6.8 & 21  & Green    & 12038\\
FBQS\,J1010$+$3003        & 2011-05 &  10.8 & 20 & 10.8 & 11  & Green    & 12025\\
Ton\,1187                 & 2011-01 &   2.0 & 19 & \nd  &\nd  & Wakker   & 12275\\
PG\,1011$-$040            & 2010-03 &   5.4 & 34 &  4.7 & 22  & Green    & 11524\\
1ES\,1028$+$511           & 2011-05 &  14.7 & 23 & 14.6 & 13  & Green    & 12025\\
1SAX\,J1032.3$+$5051      & 2011-10 &  11.4 & 12 & 11.3 &  7  & Green    & 12025\\
PG\,1048$+$342            & 2011-04 &   7.8 & 25 & 11.0 & 18  & Green    & 12024\\
PG\,1049$-$005            & 2011-06 &   2.3 & 15 &  2.8 & 14  & Tumlinson& 12248\\
PMN\,J1103$-$2329         & 2011-07 &  13.3 & 19 & 13.3 & 11  & Green    & 12025\\
HS\,1102$+$3441           & 2010-01 &  11.4 & 21 & 11.3 & 15  & Green    & 11541\\
SBS\,1108$+$560           & 2011-05 &   8.4 &  5 &  8.9 & 15  & Green    & 12025\\
PG\,1115$+$407            & 2010-06 &   5.1 & 26 &  5.7 & 17  & Green    & 11519\\
PG\,1116$+$215            & 2011-10 &   4.7 & 43 &  5.5 & 32  & Green    & 12038\\
PG\,1121$+$422            & 2011-04 &   5.0 & 23 &  5.8 & 14  & Green    & 12024\\
SBS\,1122$+$594           & 2009-11 &   9.9 & 17 & 10.5 & 13  & Green    & 11520\\
Ton\,580                  & 2010-01 &   4.9 & 24 &  5.6 & 19  & Green    & 11519\\
3C\,263                   & 2010-01 &  15.4 & 40 & 18.0 & 28  & Green    & 11541\\
PG\,1216$+$069            & 2012-02 &   5.1 & 27 &  5.6 & 21  & Green    & 12025\\
PG\,1222$+$216            & 2013-12 &   3.4 & 21 &  7.7 & 26  & Stocke   & 13008\\
3C\,273                   & 2012-04 &   4.0 & 78 & \nd  &\nd  & Green    & 12038\\
Q\,1230$+$0115            & 2010-07 &  11.0 & 50 & 11.0 & 40  & Arav     & 11686\\
PG\,1229$+$204            & 2012-04 &   1.9 & 18 & \nd  &\nd  & Veilleux & 12569\\
PG\,1259$+$593            & 2010-04 &   9.2 & 39 & 11.2 & 29  & Green    & 11541\\
PKS\,1302$-$102           & 2011-08 &   6.0 & 31 &  6.9 & 23  & Green    & 12038\\
PG\,1307$+$085            & 2012-06 &   1.8 & 23 & \nd  &\nd  & Veilleux & 12569\\
PG\,1309$+$355            & 2011-12 &   1.9 & 15 & \nd  &\nd  & Veilleux & 12569\\
SDSS\,J135712.61$+$170444 & 2011-06 &   4.2 & 18 &  6.8 & 12  & Tumlinson& 12248\\
PG\,1424$+$240            & 2012-04 &   3.8 & 24 &  7.9 & 25  & Stocke   & 12612\\
PG\,1435$-$067            & 2012-02 &   1.9 & 20 & \nd  &\nd  & Veilleux & 12569\\
Mrk\,478                  & 2012-01 &   1.9 & 21 & \nd  &\nd  & Veilleux & 12569\\
Ton\,236                  & 2011-09 &   6.6 & 20 &  9.4 & 17  & Green    & 12038\\
1ES\,1553$+$113           & 2009-09 &  10.8 & 38 & 11.9 & 30  & Green    & 11520\\
                          & 2011-07 &       &    &      &     & Green    & 12025\\
Mrk\,876                  & 2010-04 &  12.6 & 59 & 11.8 & 42  & Arav     & 11686\\
                          & 2010-04 &       &    &      &     & Green    & 11524\\
PG\,1626$+$554            & 2011-06 &   3.3 & 33 &  4.3 & 22  & Green    & 12029\\
H\,1821$+$643             & 2009-07 &  12.0 & 59 &  0.5 & 14  & Hartig   & 11484\\
                          & 2012-07 &       &    &      &     & Green    & 12038\\
PKS\,2005$-$489           & 2009-09 &   2.5 & 28 &  1.9 & 17  & Green    & 11520\\
Mrk\,1513                 & 2010-10 &   5.5 & 33 &  4.8 & 23  & Green    & 11524\\
RX\,J2154.1$-$4414        & 2010-06 &   8.2 & 32 &  8.5 & 25  & Green    & 11541\\
PHL\,1811                 & 2010-10 &   3.5 & 42 &  3.1 & 27  & Green    & 12038\\
PKS\,2155$-$304           & 2012-07 &   4.6 & 49 & \nd  &\nd  & Green    & 12038\\
RBS\,1892                 & 2012-07 &   2.2 & 23 &  2.9 & 16  & Fox      & 12604\\
IRAS\,F22456$-$5125       & 2010-06 &  15.1 & 52 & 11.9 & 31  & Arav     & 11686\\
MR\,2251$-$178            & 2011-09 &   4.6 & 38 &  5.4 & 31  & Green    & 12029\\
PMN\,J2345$-$1555         & 2013-08 &   4.1 & 17 &  7.6 & 16  & Stocke   & 13008\\
PG\,2349$-$014            & 2011-10 &   1.8 & 25 & \nd  &\nd  & Veilleux & 12569\\
H\,2356$-$309             & 2013-06 &  17.0 & 19 & \nd  &\nd  & Fang     & 12864
\enddata      
 \tablenotetext{a}{Median $S/N$ per resolution element in the G130M and G160M channels.}
\end{deluxetable*}

\subsection{Data Reduction and Processing}

The default pipeline software that produces the {\tt x1d.fits} files is capable of combining \HST/COS FUV data taken at multiple FP-POS positions for a single CENWAVE, but it does not combine data across multiple CENWAVE positions nor does it combine data taken with the G130M and G160M gratings.  Our IDL procedure {\tt coadd\_x1d} was written to serve this purpose, as were several other routines used by other groups, including the {\tt counts\_coadd} code used by the COS Halos and COS Dwarfs teams \citep[e.g., ][]{Tumlinson11,Werk14,Bordoloi14} and the {\tt coscombine} code used by the Wisconsin group \citep[e.g., ][]{Savage14,Fox15,Wakker15}.  While all of these routines share the common goal of optimally combining data taken across multiple FP-POS and CENWAVE positions and gratings, they do have some differences in implementation, including the procedure by which individual exposures are cross-correlated and interpolated onto a common wavelength scale.  However, the primary philosophical difference between these codes is in the units of their output.  Our procedure, {\tt coadd\_x1d}, preserves the flux units of the input {\tt x1d.fits} files, while the other two routines create spectra whose fluxes are in units of counts/sec, which simplifies the Poissonian error treatment.  Since the {\tt x1d.fits} files contain the flux, net counts, and gross counts as a function of wavelength for each exposure, our procedure uses the gross counts at a given wavelength to determine a Poissonian uncertainty that is valid in the low-count regime \citep[e.g., ][]{Gehrels86}.  This uncertainty array is then multiplied by an empirical sensitivity function, which is given by the ratio of the flux and net counts arrays, to convert the uncertainty values into the units of the flux array before proceeding with coaddition.

\citet{Danforth10b} describe the basic algorithm of {\tt coadd\_x1d}, but many refinements have been made over the intervening years. The revised coaddition algorithm proceeds sequentially as follows: 
\begin{enumerate} 
\item The {\tt x1d.fits} files are opened, loaded into input arrays, and examined individually. 
\item The data quality (DQ) flags from the {\tt x1d.fits} files are parsed to remove bad pixels.  Pixels with some DQ values\footnote{A particular DQ flag is set by flipping one of 15 bits in an unsigned integer.  Of these 15 bits, nine are handled identically by {\tt coadd\_x1d} and the default {\tt calcos} pipeline, three are handled more leniently by {\tt coadd\_x1d} (pixels with DQ values of 8, 16, 32 are rejected by {\tt calcos} but only de-weighted by {\tt coadd\_x1d}), and three are handled more strictly by {\tt coadd\_x1d} (DQ values of 4, 1024, 4096 are de-weighted by {\tt coadd\_x1d} and ignored by {\tt calcos}).} are instead de-weighted by artificially increasing their uncertainty so that they contribute half as much to the coadded spectrum than they otherwise would.
\item The input data are binned (i.e., Nyquist sampled) if requested by the user.  At this stage the exposure-level uncertainties are optionally modified to incorporate the non-Poissonian noise characterized in \citet{Keeney12}, ensuring that the uncertainty values reflect the RMS fluctuations about the continuum level.  While neither of these steps are default behaviors of {\tt coadd\_x1d}, all data for this paper were processed with these flags. 
\item A reference exposure is chosen and all other exposures are aligned to it using cross correlation in the regions surrounding strong ISM lines (one cross-correlation feature per detector segment).  This procedure provides a first-order wavelength alignment between exposures, which are in the heliocentric frame output by the {\sc calcos} pipeline.  However, there may be residual shifts of up to a few resolution elements at different wavelengths in the coadded data \citep{Tripp08,Savage14,Wakker15}. 
\item The flux level of each exposure is scaled to match the flux in the reference exposure, and uncertainty is scaled by this same factor.  The majority (75/82) of the AGN sight lines were observed during only a single epoch, and the changes in source brightness over the observing period tended to be negligible \citep[only 8\% of exposures required a flux scaling of greater than 10\%; however, see][for a dramatic counter-example]{Danforth13}. Observations at different epochs occasionally have significantly different fluxes; these cases are scaled to match the first data epoch.  We assume no changes in spectral shape or absorption/emission properties between epochs or exposures.  Absolute velocity calibration of arbitrary \HST/COS data to better than $\sim10$ \kms\ is problematic owing to the precise location of the spectrum on the detector and the complicated wavelength solution.
\item The individual exposures are interpolated onto a common wavelength scale.  It is important to use nearest-neighbor interpolation (i.e., shift and add) to minimize the non-Poissonian noise introduced by the coaddition process \citep[e.g., correlated noise from linear interpolation;][]{Keeney12}.
\item Combine the individual exposures using one of several weighting methods (arithmetic mean, exposure-time weighting, inverse-variance weighting, or signal-to-noise weighting).  We use the default inverse-variance weighting for this analysis. 
\item Finally, the quantities of interest (wavelength, coadded flux, and coadded uncertainty at a minimum) are returned to the user.
\end{enumerate} 

Next, continua were fitted to each of the coadded data sets using a semi-automated continuum-fitting technique developed for fitting optical SDSS spectra \citep{Pieri13} and adapted for use in higher-resolution FUV datasets.  First, the spectra were split into segments of 5-15~\AA\ width (width of the bins was adjusted qualitatively based on how smooth the AGN continuum was and the density of strong absorption features).  Continuum pixels within each segment were identified as those for which the flux-to-error ratio was less than $1.5\sigma$ below the median flux/error value for all the pixels in the segment.  Thus, absorption lines (flux lower than the segment median) were excluded, as were regions of unusually strong noise (error higher than segment average).  The process was iterated until minimal change occured in the population of continuum pixels between one iteration and the next, or until only 10\% of the original pixels in the segment remained classified as continuum.  The median value of the continuum pixels was then recorded as a continuum flux node for the 5-15~\AA\ segment and a spline function was fitted between the nodes.  

The continuum fit of each spectrum was then checked manually and adjusted as needed.  The continuum identification and spline-fitting processes worked well in regions where continua varied smoothly, but it was often poor in regions of sudden change.  For instance, spline fits perform poorly at the sharp peaks of emission lines (cusps), in the Galactic \Lya\ absorption trough (1210--1220~\AA), at the absorption edge of partial Lyman limit systems \citep{Stevans14}, and at the edges of the detectors ($\lambda\la1140$~\AA, $\lambda\ga 1790$~\AA).  The spline fits in these regions were replaced by piecewise-continuous, low-order Legendre polynomial fits.

\begin{figure}
  \epsscale{1.2}\plotone{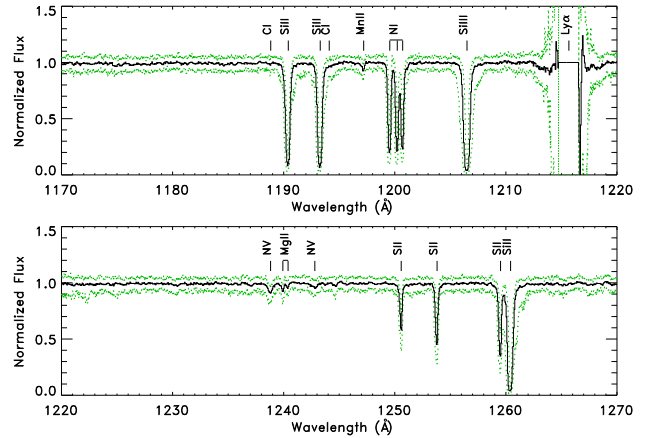}
  \caption{Median-combined ``Galactic foreground'' coaddition of 82
  normalized AGN spectra in our survey reveals the Galactic foreground
  absorption present in most extragalactic spectra.  A line-dense 100
  \AA\ spectral region near the blue end of the COS/FUV spectral range
  is shown.  Dotted curves show $\pm1\sigma$ normalized flux values in
  the sample at each wavelength.  Identified Galactic absorption lines
  are labeled.  The full spectrum is available at {\tt
  http://archive.stsci.edu/prepds/igm/}.}  \label{fig:superstack}
\end{figure}

\subsection{Absorber Identification and Measurement}

A weakness in previous quasar absorption-line surveys \citep[e.g.,][hearafter DS08]{Penton1,DS08} was the subjective nature of identifying absorption features in normalized data.  We partially corrected for this bias here by implementing an automated line-finding and measurement algorithm modeled on that used in the FOS Key Project work \citep{Schneider93}.  We believe this yields a more objectively-defined catalog of absorption lines.  
 
First, we generated a ``line-less error vector'' $\bar{\sigma}(\lambda)$ by interpolating the error $\sigma(\lambda)$ in regions defined as continuum over those identified as non-continuum during the continuum fitting process.  We calculated an equivalent width vector $W(\lambda)$ by convolving the normalized flux pixels with a representative line profile.  The line profile is the COS line spread function \citep{Kriss11} convolved with a Gaussian with Doppler parameter, $b=20$ \kms, typical of narrow absorption lines in the ISM and IGM.  The error vector $\bar{\sigma}(\lambda)$ was convolved with the same line spread function, and a crude significance-level vector was calculated as $SL(\lambda)=W(\lambda)/\bar{\sigma}(\lambda)$.  Initial absorber locations were identified by finding local maxima in significance level where $SL\ge3$.  We repeated the procedure using broader convolved-Gaussian templates ($b=50$ \kms\ or 100 \kms\ typical of broad or blended absorption lines) although the line-finding technique is not sensitive to the exact choice of kernel width.  Any additional line locations found with the broad kernels were added to the list.  

Next, Voigt profiles convolved with the COS line spread function were fitted to the normalized data at each of the identified $SL>3$ locations.  All lines were {\em initially} fitted as \Lya, and the fit parameters were allowed to vary over the range $8<b<150$ \kms, $10<\log\,N~\rm (cm^{-2})<16.5$, and $\Delta v=\pm50$ \kms.  Adjacent lines were fitted simultaneously.  The significance level of each fit \citep{Keeney12} and a goodness-of-fit parameter similar to a reduced $\chi^2$ value were determined for each component.  (Note that, since noise features are correlated between pixels, this is not a formal reduced $\chi^2$ measurement.)  Highly significant lines ($SL\ge10$) with mediocre fit qualities were refitted with two components, and the resulting one- and two-component fits were compared with an $F$-test \citep{Press92}.  The two-component fit was adopted only if it was better than the single-component fit by more than $5\sigma$, chosen empirically based on a large sample of blended lines in the data.  

The centroid wavelength, $b$-value, equivalent width, and column density (with uncertainties) of each feature were recorded in an initial line list along with the fit significance level and $\bar{\chi}^2$.  Additionally, the local S/N per resolution element in the normalized data was determined as $S/N\equiv1/\sigma_r$ where $\sigma_r$ is the RMS of continuum pixels when smoothed to a resolution element.

In the majority of cases, the fully-automated line-finding and measurement procedure produced acceptable fits.  However, some automated line fits were deemed ``pathological'', in cases when the solution represented the minimum $\chi^2$ over the parameter search space, but the fit was clearly not representative of the absorption feature.  The most common of these cases were fits to weak lines adjacent to strong features in which saturated pixels would drive the line profile to a spurious fit.  Line measurements with $\bar{\chi}^2>1$ were presented for manual refitting during the line identification process below, and components were often added or removed based on qualitative assessment. 

The automated line-finding algorithm detected $\sim$30 very broad ($b>75$ \kms), weak (3-4$\sigma$) absorbers some of which are not subjectively obvious in the data but meet the $>3\sigma$ significance criterion.  Some of these features were clearly suspect as regions of ambiguous continuum or statistical fluctuations in the noise and were rejected.  However, an objective treatment of absorption features was one of the main drivers for our automated line-finding systems and we included most of them in the final line catalog.  We caution users that they should interpret these features with a critical eye in any detailed analysis.

\begin{deluxetable*}{lllc||lllc}
\tabletypesize{\footnotesize}
\tablecolumns{8} 
\tablewidth{0pt} 
\tablecaption{Galactic Absorption Lines in Far-UV HST/COS Spectra}
\tablehead{\multicolumn{4}{c}{Strong Lines} &
\multicolumn{4}{c}{Weak Lines} \\
  \colhead{Ion}    &
  \colhead{$\lambda$ (\AA)} &
  \colhead{$f$} &
  \colhead{Ref.\tnma} &
  \colhead{Ion}    &
  \colhead{$\lambda$ (\AA)} &
  \colhead{$f$} &
  \colhead{Ref.\tnma}  }
\startdata
N\,I     &  1134.17\tnmb& 0.0146  & 1 & Fe\,II   &  1142.37     & 0.0040 & 1 \\  
N\,I     &  1134.41\tnmb& 0.0287  & 1 & Fe\,II   &  1143.23     & 0.0192 & 1 \\  
N\,I     &  1134.98\tnmb& 0.0416  & 1 & C\,I     &  1157.91     & 0.0212 & 1 \\ 
Fe\,II   &  1144.94     & 0.0830  & 1 & C\,I     &  1194.00     & 0.0124 & 1 \\ 
P\,II    &  1152.82     & 0.245   & 1 & Mn\,II   &  1197.18     & 0.148  & 2 \\ 
Si\,II   &  1190.42     & 0.292   & 1 & N\,V     &  1238.82     & 0.156  & 1 \\ 
Si\,II   &  1193.29     & 0.582   & 1 & Mg\,II   &  1239.94     & 0.000621 & 3 \\ 
N\,I     &  1199.55\tnmb& 0.132   & 1 & Mg\,II   &  1240.39     & 0.000351 & 3 \\ 
N\,I     &  1200.22\tnmb& 0.0869  & 1 & N\,V     &  1242.80     & 0.0777 & 1 \\ 
N\,I     &  1200.71\tnmb& 0.0432  & 1 & C\,I     &  1277.25     & 0.1314 & 4 \\ 
Si\,III  &  1206.50     & 1.63    & 1 & C\,I     &  1280.14     & 0.0481 & 4 \\ 
H\,I \Lya&  1215.67\tnmb& 0.4164  & 1 & Ni\,II   &  1317.22     & 0.0571 & 5 \\ 
S\,II    &  1250.58     & 0.00543 & 1 & C\,I     &  1328.83     & 0.0899 & 4 \\ 
S\,II    &  1253.81     & 0.0109  & 1 & Cl\,I    &  1347.24     & 0.153  & 1 \\ 
S\,II    &  1259.52     & 0.0166  & 1 & Ni\,II   &  1370.13     & 0.0588 & 5 \\ 
Si\,II   &  1260.42     & 1.18    & 1 & Ni\,II   &  1454.84     & 0.0323 & 6 \\ 
O\,I     &  1302.17\tnmb& 0.0480  & 1 & Ni\,II   &  1467.76     & 0.0099 & 6 \\ 
Si\,II   &  1304.37     & 0.0863  & 1 & C\,I     &  1560.31     & 0.1315 & 7 \\ 
C\,II    &  1334.53     & 0.128   & 1 & Fe\,II   &  1611.20     & 0.0014 & 1 \\ 
C\,II*   &  1335.66     & 0.128   & 1 & C\,I     &  1656.93     & 0.1488 & 7 \\ 
Si\,IV   &  1393.76     & 0.513   & 1 & Ni\,II   &  1703.41     & 0.0060 & 6 \\ 
Si\,IV   &  1402.77     & 0.254   & 1 & Ni\,II   &  1709.60     & 0.0324 & 6 \\ 
Si\,II   &  1526.71     & 0.133   & 1 & Ni\,II   &  1741.55     & 0.0427 & 6 \\ 
C\,IV    &  1548.20     & 0.1899  & 1 & Ni\,II   &  1751.91     & 0.0277 & 6 \\ 
C\,IV    &  1550.78     & 0.09475 & 1 &          &          &       &  \\ 
Fe\,II   &  1608.45     & 0.0577  & 1 &          &          &       &  \\ 
Al\,II   &  1670.79     & 1.74    & 1 &          &          &       &  
\enddata
\tablenotetext{a}{
Vacuum wavelengths and oscillator strength references: 
1- \citet{Morton03}, 
2- \citet{TonerHibbert05},
3- \citet{KelleherPodobedova08},
4- \citet{JenkinsTripp01},
5- \citet{JenkinsTripp06},
6- \citet{Fedchak00},
7- \citet{JenkinsTripp11}
}
\tablenotetext{b}{Geocoronal airglow is often associated with this transition in addition to Galactic absorption.}
\label{tab_ismlines}
\end{deluxetable*}


\clearpage

There are $\sim25$ strong absorption lines arising in the Milky Way ISM that present a foreground to every extragalactic absorption spectrum (Figure~\ref{fig:superstack}, Table~\ref{tab_ismlines}).  Another $\sim25$ weaker lines, particularly those of \NV, \CI, and \NiII, are seen in some of the sight lines and can be confused with IGM absorption.  To characterize these, we performed a simple median-combined coaddition of all 82 normalized COS AGN spectra in the observed (heliocentric) frame.  This ``Galactic foreground'' spectrum reaches a very high S/N ($\sim100$) and shows the foreground lines frequently present in the data.  It is available on-line as a MAST high-level science product as described in Section~3.  Since the relative strengths, centroids, and line widths of these Galactic lines vary from one sight line to another, the foreground spectrum cannot be used as a true Galactic flat field.  However, it is useful for identifying individual absorption features as Galactic or intergalactic in nature and in masking regions of spectra which are insensitive to IGM absorption features.  

Measured lines that are consistent with ISM features typically seen in FUV spectra are flagged as probable ISM absorption.  If these lines are not subsequently re-identified as an IGM line or a blend of IGM and ISM features, they are ignored in the remainder of this analysis.  Table~\ref{tab_ismlines} lists the commonly-seen Galactic absorption features in the $1135\rm~\AA<\lambda<1800$~\AA\ spectral range.  We note that Table~\ref{tab_ismlines} is not a comprehensive list of all FUV transitions of reasonable strength; some lines would be visible in some spectra but for blending with a stronger, omnipresent ISM line (e.g., the weak \CI\ 1260.74 line which, even when other comparable \CI\ lines are seen, will always be blended with strong \SiII\ 1260.42 absorption).  High-velocity Galactic absorption present in many datasets ($\Delta v\la500$ \kms) is flagged by using the same ISM template offset by a constant velocity.  ISM lines that have not been re-identified as IGM lines or blends are recorded in the line lists for completeness along with measured equivalent widths, but no attempt has been made to measure accurate column densities or detailed component structure for them.  

The automatic line-fitting procedure assumed that all features were \Lya\ forest lines (statistically, the most common IGM line).  However, this is not always the case.  Line identification was by far the most time-intensive part of the process, since automation in this area was not reliable.  After pathological fits were corrected and $v\approx0$ ISM lines were identified and flagged, the remaining lines were identified by a variety of means.  For sight lines covering a long pathlength ($\Delta z\ga0.15$), correlated absorption in multiple transitions of a species (e.g., \HI\ \Lya\ and \Lyb) was the most productive technique.  For instance, overplotting $z_{\rm Ly\alpha}=(\lambda/1215.67\rm\AA)-1$ versus flux and $z_{\rm Ly\beta}=(\lambda/1025.72\rm\AA)-1$ versus flux shows correlated absorption at the redshift of any moderately-strong \HI\ systems ($N\rm_{HI}\ga10^{14}~cm^{-2}$).  Comparisons of \Lyb/\Lyg\ reveal strong \HI\ absorbers at $z_{\rm abs}>0.47$ where \Lya\ has redshifted past the end of the COS/G160M detector.  Occasionally, metal-ion doublet transitions \OVI\ (1031.93, 1037.64~\AA) and \CIV\ (1548.20, 1550.77~\AA) revealed IGM systems where \HI\ absorption was weak, not present, or blended with other absorption lines.  Identifying the redshifts of the strongest absorption systems in this way allows us to identify single-transition metal ions (\CIII\ \lam977.02, \SiIII\ \lam1206.50, \NIII\ \lam989.80) and other species in some cases (e.g., \CII, \SiII, \OIV, etc.).  \FUSE\ data, where present, were used to help confirm these correlated absorber systems, but is not otherwise measured; see \citet{Tilton12} for a complete list of \FUSE+\HST/STIS IGM absorbers.  \Lya\ line fits were adjusted as needed and those lines identified as something other than \Lya\ were refitted manually with the appropriate atomic parameters.  

Regardless of how it was identified, each extragalactic system redshift was checked for corresponding absorption in the transitions most commonly seen in the IGM: \HI\ \Lya-$\delta$, \OVI, \SiIII/IV, \CIII/IV, \NV; see Table~3 of DS08 for details.  A small number of systems also showed absorption in additional ions (e.g., \CII, \FeII/III, \SiII, \NeVIII, \OIV, \NIII, and higher-order Lyman lines).  

\subsubsection{Line Identification Ambiguities}

Because \Lya\ 1215.67~\AA\ is typically the strongest transition in any diffuse IGM absorber, many low-column density systems are single-line identifications for which higher order Lyman transitions and metals are too weak to be observed.  As in our previous papers, we assume that these are weak \Lya\ lines unless that assumption is inconsistent with either the source redshift ($z_{\rm abs}>z_{\rm AGN}$ or $z_{\rm abs}<0$), a non-detection upper limit in another transition, or a more plausible identification in another identified system.

We have identified all absorption features to the best of our abilities, but some ambiguities exist.  Throughout, we assume that any system with more than one line is unambiguous.  However, single-line identifications are still in the majority and, in principle present cases where the line identification is ambiguous.  The majority ($\sim75$\%) of IGM systems consist of single \Lya\ absorbers without any confirming \Lyb\ or metal-ion detection at the same redshift.  This uncertainty is greatest for the longer sight lines (higher $z_{AGN}$) since there are more intervening systems and more possibilities for unusual metal-ion absorbers.  Short pathlengths offer a more compact \Lya\ forest with fewer realistic possibilities for line identification; in such cases, \Lya\ is usually the only plausible identification.  Nevertheless, a few of these single-line systems may be metal-line absorption from a different redshift or instrumental features.  

At $z\ga0.47$, \Lya\ shifts beyond the red end of the COS G160M detector; thus there are 13 $z_{\rm AGN}>0.47$ sight lines with incomplete coverage of the \Lya\ forest.  In these sight lines, some weak absorbers identified as \Lya\ at $0.24<z_{\rm abs}<0.47$ could instead be strong \Lyb\ systems at $z>0.47$.  For stronger lines, the $z_{\rm abs}>0.47$ \Lyb\ interpretation can be ruled out with a \Lyg\ upper limit (i.e., the predicted \Lyg\ absorber is strong enough that it should appear in the data).  However, there are $\sim40$ weak lines in our survey with this \Lya/\Lyb\ ambiguity.  From the ratio of oscillator strengths and rest wavelengths, the column density ratio $N_{\rm Ly\beta}/N_{\rm Ly\alpha}=(f\,\lambda)_{\rm Ly\alpha}/(f\,\lambda)_{\rm Ly\beta}=6.2$ for a weak line of a given observed equivalent width.  Since the frequency of \HI\ absorbers $\partial^2{\cal N}(N)/\partial N\,\partial z\propto N^{-\beta}$ where $\beta\approx1.6$, \Lyb\ absorbers should make up only $6.2^{-1.6}=5\%$ of the ambiguous sample or $\sim2/40$ weak \Lya\ absorbers at $0.24<z_{abs}<0.47$ in 13 sight lines.

The inverse case (weak \Lya\ lines at $z_{\rm abs}<0.47$ misidentified as \Lyb\ or \Lyg\ at $z_{\rm abs}>0.47$) should not be present in this survey, since we require any $z_{\rm abs}>0.47$ absorber to be confirmed with detections in at least two transitions at that redshift (typically \Lyb$+$\Lyg\ or metal-ion doublets).


\subsubsection{Instrumental Features}

As with any instrument, \HST/COS suffers from a variety of instrumental artifacts which can masquerade as real spectral features.  These include ``gain sag'', areas of decreased sensitivity near detector edges, and fixed pattern noise.  Use of quality flags in the coaddition process described above minimizes the impact of most of these effects, but some instrumental features are inevitably present in the data.  Fixed pattern noise is the most common feature and manifests itself as small undulations across the face of an otherwise smooth continuum.  Efforts to fully characterize this noise, much less correct for it, have proven elusive \citep{Wakker15}.  A small fraction of our weak absorption features are undoubtedly instrumental rather than real features.

\subsection{IGM Absorber Analysis}\label{sec_sysdef}

\subsubsection{Significance Levels}

Absorption features are initially identified using the simple approximation for significance level as described above.  During the fitting process, we use a more rigorous significance-level formula 
\begin{equation}
SL=(S/N)_1\,\frac{W_\lambda}{\Delta\lambda}\,\frac{1}{w(x_{\rm opt})},
\end{equation}
where $(S/N)_1$ is the signal-to-noise ratio per single pixel, $W_\lambda$ is the equivalent width, and $\Delta\lambda$ is the pixel width in Angstroms.  The quantity $w(x_{\rm opt})$ is an empirical function that describes the correspondence between $(S/N)_1$ and the signal-to-noise level for data binned to an optimal number of pixels $x_{\rm opt}$.  The parameter $x_{\rm opt}$ is a function of the observed wavelength and the Doppler $b$-parameter of the feature.  Full details are given in equations (4), (7), and (9-11) of \citet{Keeney12}\footnote{An IDL routine to implement the significance level calculation of \citet{Keeney12} can be found at {\tt http://casa.colorado.edu/$\sim$danforth/science/cos/costools.html}.}.

Throughout the analysis, we retain any features measured with $SL>3$.  However, even $3\sigma$ features are statistically common; a typical COS spectrum covers approximately 8000 resolution elements between 1150~\AA\ and 1800~\AA.  There could be $\sim20$ spurious $3\sigma$ features with $b$-values typical of narrow absorbers in each COS/FUV spectrum given normally-distributed (Gaussian) noise.  A more stringent $4\sigma$ detection threshold should result in less than one spurious feature per spectrum.  

The COS noise characteristics are poorly constrained and are not purely Gaussian in nature \citep{Keeney12}.  However, a $4\sigma$ detection threshold will still result in fewer spurious detections than a $3\sigma$ criterion; for this reason, we set the following criteria for inclusion in our catalog.  Single-line detections must be $\ge4\sigma$.  However, if a $4\sigma$ prior exists from another line detection, we relax the threshold to $\ge3\sigma$ for lines in other transitions at the same redshift.  For example, a single weak absorber measured at $3.5\sigma$ would be rejected from the statistics, but retained in the line catalog.  However, if the weak absorber can be interpreted as a metal line or higher-order Lyman transition of a stronger absorber, it would be accepted and used in later analysis.  Practically, this means that the \Lya\ detection threshold is set at $4\sigma$, while the higher-order Lyman line and metal-line threshold is $3\sigma$ since metal absorption is almost never seen without an \HI\ prior.  We use the same criteria for calculating the total pathlength $\Delta z(N)$ observed at column density $N$.  

\subsubsection{Intrinsic Absorbers} 
\label{intrinsic_abs}
Galactic features are easily identified as discussed above.  However, absorption from intervening (IGM) gas can sometimes be difficult to differentiate from that arising from gas associated with the AGN host galaxy itself (intrinsic absorbers with $z_{\rm abs}\approx z_{\rm AGN}$).  There is no definitive way to identify an absorber as one or the other since AGN outflow features have been observed at very large relative velocities.  We automatically flag any absorber with $\Delta v=(cz_{\rm AGN}-cz_{\rm abs})/(1+z_{\rm AGN})\le 1500$ \kms\ as a possible intrinsic system.  The exceptions to the $\Delta v<1500$ \kms\ proximity rule are the four BL\,Lacertae objects in the sample.  These objects (1ES\,1553$+$113, 3C\,66A, S5\,0716$+$714, and PG\,1424$+$240) have source redshift lower limits defined by observed narrow IGM absorbers.  Since outflows are sometimes seen at higher velocities, we also manually check any metal-line systems at $1500<\Delta v<5000$ \kms\ for obvious AGN-intrinsic properties \citep[e.g.,][]{Dunn07,Ganguly13}.  We flag as intrinsic all systems in this redshift range that show strong absorption in high ions (\OVI, \NeVIII, \NV, etc.), strong high ions but weak \HI, strongly non-Gaussian profiles, or doublet equivalent width ratios close to 1:1 (which may result from partial covering of the source).  Similarly, we mask out $cz_{\rm abs}<500$ \kms\ as possible high-velocity absorption associated with our Galaxy or absorption from the Local Group\footnote{A subset of COS observations were designed to probe M\,31, the Magellanic Clouds and Stream, or high-velocity structures in the Milky Way halo.  Thus, we exclude the lowest-velocity ($cz\le 500$ \kms) regions from our IGM survey to limit ``double counting'' of absorbing structures.}.  Absorbers in these regions are measured and reported in the tables, but are not included in IGM statistics, nor are these redshift ranges included in the total IGM pathlength calculations.  

\begin{figure}
  \epsscale{1}\plotone{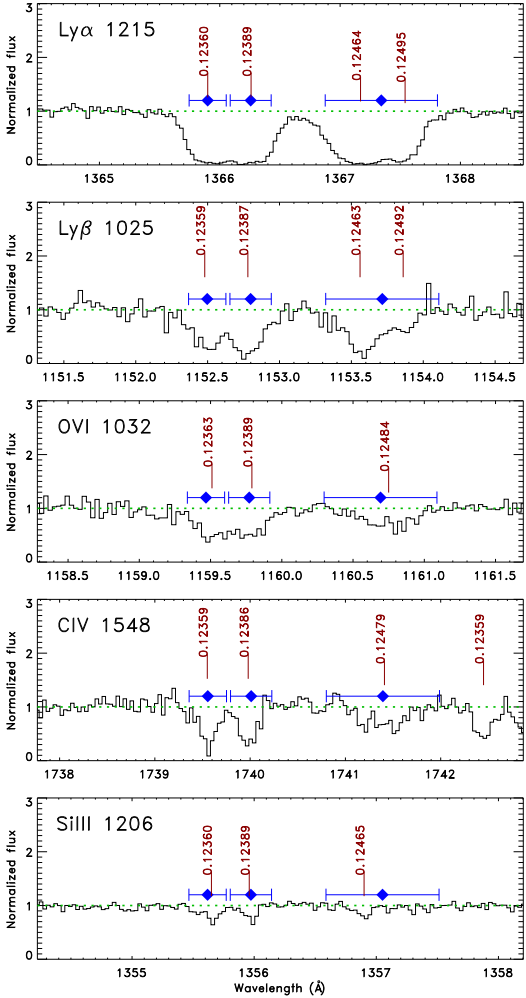}
  \caption{Absorption in the PG\,1216$+$069 sight line illustrates the
  distinction between components and systems.  Individual components
  are denoted with vertical ticks, labeled by redshift, and grouped
  together into three absorbing systems (blue horizontal bars).  Any
  components that fall within the system velocity range are included
  in their respective systems.  At first glance, the absorbers near
  $z\approx0.1237$ and $z\approx0.1247$ appear similar.  However, the
  former complex shows consistent velocity structure in all lines so
  the components can be unambiguously assigned to one or the other of
  two narrow ($\Delta v_{\rm sys}\approx40$ \kms) systems at
  $z=0.12360$ or $z=0.12390$.  The redder system is more ambiguous;
  while the H\,I absorption clearly shows two components, the O\,VI
  and C\,IV absorption shows a single, broad component which cannot be
  unambiguously assigned to either H\,I component.  For this reason,
  we maintain a single, broader ($\Delta v_{\rm sys}=115$ \kms) system
  at $z=0.12474$.}  \label{fig:sysex}
\end{figure}	

\subsubsection{Components and Systems}
In the following analyses, we classify IGM absorbers in two ways.  First is the self-explanatory {\it component} analysis where individual fitted velocity components are analyzed as measured.  However, it is often more useful to discuss {\it systems} of components \citep[e.g.,][]{Tripp08} at nearly the same redshift.  Systems are composed of components in the same sight line within a narrow velocity window, which allows lines of different species to be directly related, even if there are small velocity misalignments between them.  We adopt this method to manage the ambiguity often present in associating components of one ion with components in another (Figure~\ref{fig:sysex}) or even the ambiguity within a given transition composed of multiple, blended components.  Detailed analysis of an individual system may reveal more accurate associations \citep{Savage14,Stocke14}, but this depth of analysis is beyond the scope of our large, semi-automated survey.  Initially, we defined systems by stepping through the components in a sight line in order of line strength and grouping any other components measured within $\Delta v_{\rm sys}=30$ \kms\ into the same system.  Systems with stronger, broader lines used $\Delta v_{\rm sys}=c\,W_r/\lambda_0$ to account for ambiguously blended absorption components.

The automatic system-definition algorithm accounted for most strongly-blended absorption components and velocity calibration uncertainties between different wavelength regions in the data.  The majority of systems (1888/2611$\approx$72\%) were composed of a single line, and for these there is no distinction between component and system nomenclatures.  However, all 723 systems composed of more than a single component, whether multiple, blended components of the same transition or components in multiple transitions, were checked manually.  In most cases, the automatic system definitions were confirmed, but the redshifts $z_{\rm sys}$ and velocity widths $\Delta v_{\rm sys}$ for some systems were adjusted manually to account for obvious, unambiguous component structure.  Finally, all measured components were checked to make sure they appear in one, and only one, absorbing system.  An example of three systems composed of four components in \Lya\ is shown in Figure~\ref{fig:sysex}.  

Systems tend to be broader when more components are included: the 435 two- or three-component systems have a median width close to the nominal minimum $\Delta v_{\rm sys}\approx 30$ \kms.  The 240 IGM systems with 4-9 components tend to be broader with $\Delta v_{\rm sys}=36^{+26}_{-6}$ \kms.  The 48 richest systems (composed of 10 or more individual components) have a system width of $\Delta v_{\rm sys}=49^{+34}_{-18}$ \kms, and there are only 12 systems with 100 \kms\ $<\Delta v_{\rm sys}<300$ \kms.

\subsubsection{Consistency Checks}
The final step in the creation of a line list is to check all line measurements and identifications for consistency.  Each line list is screened for features identified as IGM at $z_{\rm abs}<0$ or $z_{\rm abs}\gg z_{\rm AGN}$.  Unidentified lines at $<3\sigma$ are rejected.  Unidentified $>4\sigma$ features are examined and identifications are attempted (though see discussion above regarding unidentified lines).  ``Orphaned'' lines, such as those identified as metal lines or higher-order Lyman lines but without any other detections at the same redshift, are flagged and examined.  

We check all multiplet transitions (\HI, \OVI, \NV, \SiIV, \CIV, \NeVIII, etc.) for consistency in column density.  For instance, if a line is identified as \OVI\ 1032~\AA\ at a particular redshift, the 1038~\AA\ \OVI\ transition must appear as a detection of consistent strength, a blend with a stronger feature, or a non-detection consistent with the $3\sigma$ minimum equivalent width at the expected location in the data.  Discrepancies are reevaluated manually.  

\HI\ column density and $b$-value measurements based on single Lyman line profiles tend to overestimate the line width and underestimate the column density \citep{Danforth10a}.  We determine column density and $b$-values for \HI\ via a curve of growth (CoG) fit to multiple Lyman lines.  CoG fits are performed for systems at $z>0.1$ (where \Lyb\ is covered in COS data) with $N_{\rm HI}>10^{14}$~\cd\ in any individual Lyman line.  Any outlying single-line measurements or poor CoG solutions are reevaluated manually.  In most cases, the CoG \HI\ solution is more accurate than a single-line \Lya\ measurement.  However, while the CoG solution to a system with multiple velocity components tends to preserve total column density in the lines, the $b$-value may be artificially broad.  We do not attempt CoG solutions to metal-ion lines because most of their absorption lines are not heavily saturated.  Neither is there enough $f\lambda$ contrast between transitions for a well-constrained CoG solution.

\subsection{Biases and Systematics}

There should be little bias in the IGM sample due to the choice of AGN in this study.  The sight lines were chosen from a large archive of \HST/COS observing programs reflecting a broad range of scientific objectives.  The one unifying characteristic of the sight lines is that they are toward UV-bright targets.  Outside a relatively small proximity zone, where the AGN may have undue effect on the local photoionization, the source luminosity has no bearing on the properties of the IGM sample.  It can be argued that using UV-bright sources selects against strong \HI\ absorption at any redshift along the line of sight; Lyman Limit systems with $\log N_{\rm HI}\ge17.2$ are opaque ($\tau_0>1$) to Lyman continuum photons at $\lambda<(1+z_{\rm abs})\times 912$~\AA, but this is only true for $z_{\rm abs}\ga 0.4$ where the Lyman continuum redshifts sufficient to block a substantial part of the FUV band.  Lyman Limit systems at $z<1$ are rare \citep[\dndz$\approx 0.33$;][]{Stevans14,Ribaudo11}, but our sample includes one such case (SBS\,1108$+$560, $z_{\rm LLS}=0.4632$).  Generally the redshifted Lyman continuum flux is only moderately absorbed \citep{Shull12b,Stevans14}, even in the rare, strong \HI\ systems at higher $z$.  Most of our AGN are at $z_{\rm em}<0.4$ where LLSs make no difference to the observed spectrum in the \Lya\ forest.

There is a small bias in that fifteen of the 82 sight lines were originally proposed because they probe specific galaxies near the AGN sight line.  Measured covering factor of \Lya\ absorbers with $N>10^{13}$~\cd\ at $R<R_{vir}$ are $\sim80\%$ for $L>L*$ and $0.1-L*$ galaxies \citep{Stocke13}.  Prochaska et al. (2011) finds even higher covering factors for $0.1-1L*$ galaxies, statistically consistent with 100\%.  \citet{Tumlinson11} find similarly high covering factors for \OVI\ around $L>L*$ star-forming galaxies.  While \citealt{Stocke13} find that dwarfs have smaller covering factors ($\sim50\%$ at $R<R_{vir}$) it turns out that that sample was heavily biased by Virgo Cluster sightlines.  Bordoloi (private communication) finds very high covering factors ($>80\%$) for \Lya\ at $R<R_{vir}$ in his COS dwarfs sample.  Given the observed high covering factor of \Lya\ absorption detected within $\pm400$ \kms\ of a galaxy velocity, this implies that $\sim15$ metal-line absorbers are present in our overall sample due to this selection bias.

\begin{figure}
  \epsscale{1.1}\plotone{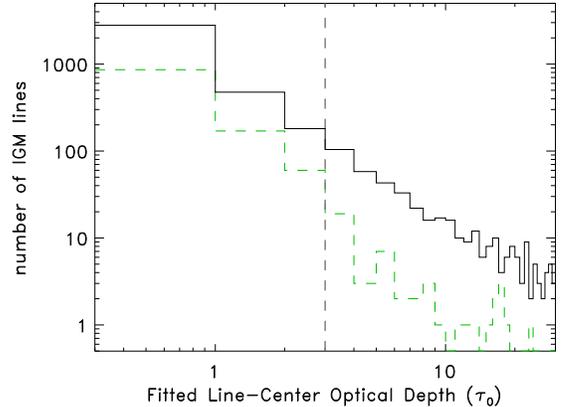}
  \caption{Histogram of fitted line-center optical depth ($\tau_0$)
  for 3966 H\,I Lyman lines (black solid) and 1155 metal-ion lines
  (green dashed) identified as IGM absorption.  Given the typical
  quality of the COS spectra ($S/N\ga20$), column density measurements
  of mildly-saturated lines ($\tau_0\la3$) should be accurate.
  However, $\sim13\%$ of Lyman lines are saturated $\tau_0>3$, so
  column density measurements of these single lines should be taken as
  lower limits.  Saturated metal lines are much less common:
  $\sim5\%$.}\label{fig_lcod}
\end{figure}

Column densities inferred from single, saturated absorption lines can underestimate the true column density as determined via a more robust curves of growth, sometimes dramatically (e.g., the $cz=1590$ \kms\ \HI\ absorber toward 3C\,273 first measured by \citet{Weymann95} based on a single \Lya\ absorber, subsequently corrected with a multi-line curve-of-growth by \citet{Sembach01}.  For the vast majority of \HI\ Lyman lines, COS absorption line profiles are well-resolved and our Voigt profile fits should give accurate single-line column density measurements even in the case of moderate saturation in the line core ($\tau_0=1-3$).  However, stronger Lyman lines should be conservatively treated as lower limits on the column density.  A line-center optical depth $\tau_0=3$ corresponds to a transmitted flux of 5\%.  Given the typical S/N of 20 in the sample, we feel this is a conservative limit on where single-line column densities can be trusted.  Of the 5121 IGM absorption lines in the catalog, 579 (11\%) show $\tau_0\ge3$ (Figure~\ref{fig_lcod}).  Subdividing by transition, 12\% of \Lya\ absorbers are saturated at $\tau_0\ge3$ as are 15\% of higher-order Lyman lines.  Saturated metal-line absorbers are rare, comprising only 5\% of the total sample.  These are also listed as lower limits in the line lists.  

Most of our analysis in the remainder of the paper is based on system-level measurements where \NHI\ is determined via a robust curve of growth for stronger absorbers.  Lower limits on column density of individual lines should not be an issue.  However, we list the individual column density measurements as lower limits in the line catalog.  There are 118 \Lya\ absorbers at low-redshift ($z<0.1$) where \Lyb\ is not available for a CoG measurement.  These systems are among those excluded from the uniform sample discussed in Section~\ref{sec:zlimit}. 

Some science programs were designed to probe individual extragalactic objects or near-field structures (the Magellanic Stream, M\,31).  This presents a bias for a few specific absorbers but the population of absorbers along the sight line as a whole is unbiased.  In a few other cases, adjacent sight lines may be sampling the same structures in absorption, creating a ``double-counting'' bias.  The closest pair of sight lines in this survey is separated by $9.3$\arcmin\ and three other pairs are separated by less than a degree on the sky.  The redshift range over which the sight lines are separated by less than 1 Mpc (where correlated absorption may be expected) is only $\Delta z\approx0.1$ or $\sim0.5$\% of the total survey pathlength.  The overall bias on the IGM absorber catalog from sight line selection is very small.

Most of the absorbers in the catalog are relatively narrow features ($b<50$ \kms).  The median doppler width and $\pm1\sigma$ range for IGM components are $b=33^{+17}_{-14}$ \kms.  Automated line-finding codes search for lines at $b=20$, 50, and 100 \kms, and profile-fitting codes fit features with a parameter range $5<b<150$ \kms.  However, COS is optimized for the detection of narrow lines, and there is an inherent bias in this catalog toward lines close to the resolution limit.  We define a sensitivity as a function of $b$-value, $S(b)\equiv[W_{\rm min}(b)/W_{\rm min}(b=10)]^{-1}$ where $W_{\rm min}(b)$ is the minimum equivalent width for a line of a Doppler parameter $b$ in data of a given $S/N$.  There is a slight dependence of $S(b)$ on observed wavelength, but it is insensitive to $S/N$.  For $b=30$ \kms, we see $S(30)=0.66$; COS data are only 2/3 as sensitive to 30 \kms\ lines as they are to those near the resolution limit of the data ($b\sim10$ \kms).  The sensitivity gets worse for broader lines: $S(50)=0.5$, $S(100)=0.35$, and $S(200)=0.23$.  Thus, there is a bias against weak, broad lines in the data, even with accurate knowledge of the AGN continuum and binning appropriate to the line width.  In addition, the continuum fitting routine uses bins of 5-15 \AA\ in width, so features broader than several Angstroms ($b\ga300$ \kms), whether they be actual absorption lines, AGN emission lines, flux calibration uncertainties, or other instrumental features, tend to be fitted as part of the continuum.

The two medium-resolution COS/FUV gratings cover the \Lya\ forest for absorbers out to $z\approx0.47$.  For weak absorbers, column density can be determined from measurements of the \Lya\ line alone with reasonable accuracy.  However, \HI\ column density measurements based on saturated \Lya\ lines alone tend to underpredict the true column density \citep{Shull00,Sembach01,Wakker15}.  CoG fits to multiple \HI\ transitions tend to give a more realistic \NHI\ value for $N_{\rm HI}\ga10^{14}~\rm cm^{-2}$.  In a typical COS/FUV dataset, the \Lyb\ transition redshifts into the G130M detector at $z\approx0.1$.  In order to limit column density uncertainty between stronger \HI\ systems at low redshift (measured from \Lya\ alone) and higher redshift (where multi-line CoG solutions are used), we define a {\em uniform}, redshift-limited subsample of IGM systems: weak \HI\ systems ($\log\,N_{\rm HI}<13.5$) are included regardless of redshift, but stronger absorbers ($\log\,N_{\rm HI}\ge13.5$) are included only at $z_{\rm abs}\ge 0.1$ where a more accurate, multi-line CoG solution is possible; either \Lya$+$\Lyb\ or \Lyb$+$\Lyg\ at minimum.  The limited sample (2256 systems) is only slightly smaller than the full COS sample (2577 systems) since only the stronger systems at low redshift are excluded.  However, this limited sample provides a more uniform analysis of the distribution of \HI\ absorbers in the low-redshift IGM.  \label{sec:zlimit}

Given the huge number of IGM absorber systems ($>$2600) and the substantial pathlength ($\Delta z>20$) used for their discovery, the sample variance should be small and adds negligible uncertainty to the Poisson errors.  However, based on a previous jack-knife style resampling exercise done for a smaller sample \citep{Penton04}, we estimate that cosmic variance can add significant uncertainty to our sample at $\log N_{\rm HI}\geq15$ (i.e., sub-sample sizes with ${\cal N}\la100$ absorbers).  For order of magnitude sized bins starting at $\log N_{\rm HI}=15$, 16, and 17, we estimate that cosmic variance adds (in quadrature) 17\%, 33\% and 86\% to the Poisson errors in those bins.  Since metal-line systems are almost always attached to \HI\ systems, the cosmic variance in those samples is small despite their small numbers.

\section{The Catalog}

\begin{figure*}
  \epsscale{.98}\plotone{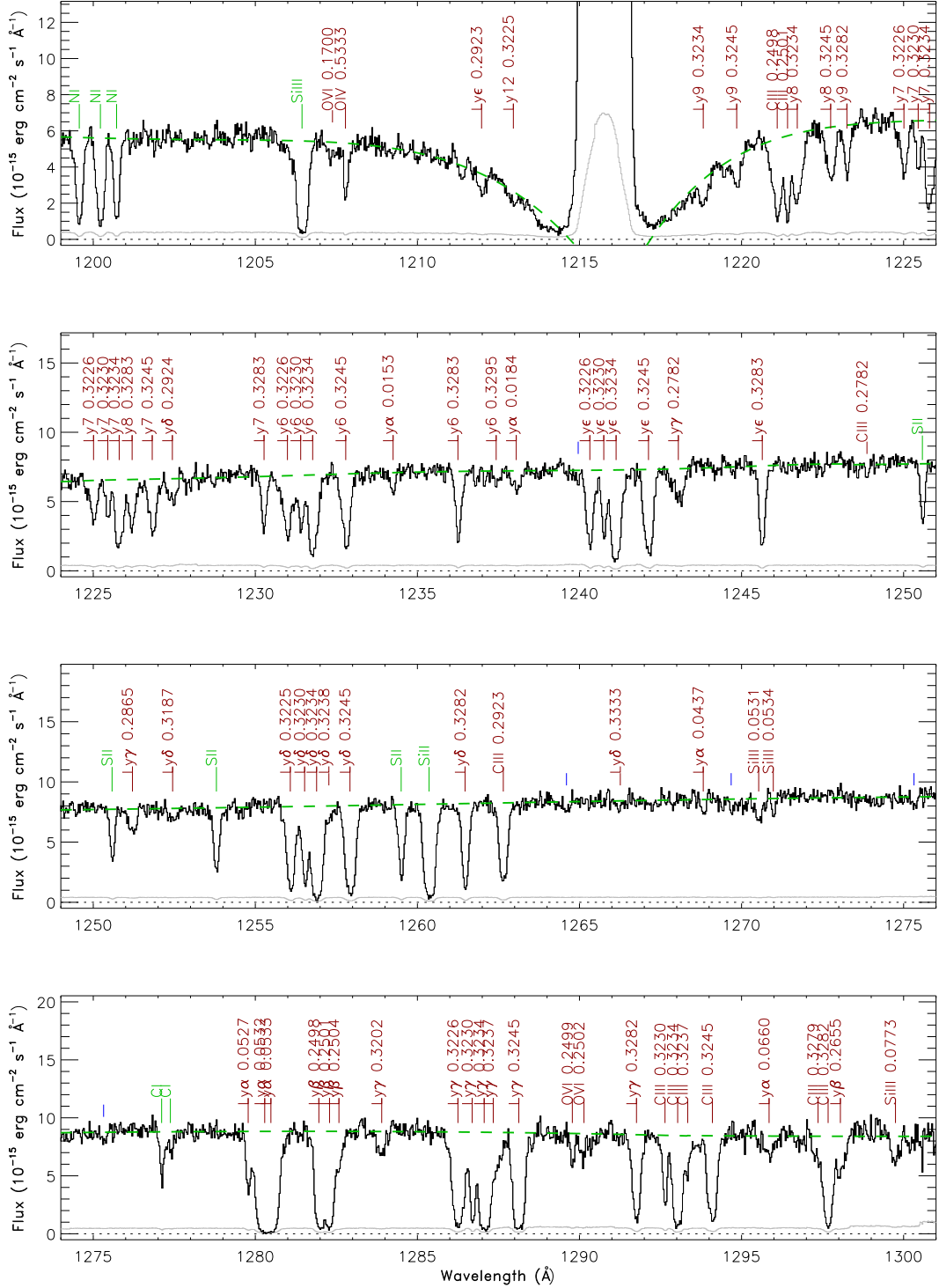}
  \caption{Sample atlas page for 3C\,57.  Coadded flux and error
  vectors are shown in black and grey, respectively.  Green dashed
  line shows the continuum.  Each page covers 100 \AA\ of the COS
  spectrum.  Red text and ticks denote IGM absorbers (identified by
  species and redshift).  Green ticks show $v\approx0$ ISM features.
  Blue ticks mark lines with no identification.}  \label{fig:atlas_sample}
\end{figure*}

To facilitate further analysis, and as a service to the community, we present data products as a High-Level Science Product (HLSP) in a partnership with the Mikulski Archive for Space Telescope (MAST) at {\tt http://archive.stsci.edu/prepds/igm/}.  The following products are included for each of the 82 AGN sight lines in this survey:
\begin{itemize}
\item A fully-reduced, coadded spectrum in heliocentric wavelength coordinates as discussed in Section~2 in {\sc fits} format including wavelength, flux, error, and local exposure time vectors.  Also included are the ``line-less'' error, continuum fit, continuum fit uncertainty, and a flag for pixels used in the continuum fit.
\item A full list of all absorption features measured at $SL>3$ sorted by observed wavelength in {\sc ASCII} format.  Each line list table includes line wavelength, significance level (as described in Section~\ref{sec_sysdef}), line identity, redshift of the feature, equivalent width and $1\sigma$ equivalent width uncertainty.  Extragalactic lines additionally have listed Doppler $b$-values and measured $\log N$ values, both with uncertainties, and the reduced $\chi^2$ value of the line fit.  Three final columns are used to flag likely Galactic (ISM) lines, likely AGN intrinsic lines, and the approximate line-center optical depth $\tau_0$.
\item A table in {\sc ASCII} format of extragalactic absorption lines grouped by system redshift as described in Section~\ref{sec_sysdef}.  Only extragalactic absorbing systems at $z>0$ are listed.  The first two columns identify each system by system redshift, $z_{\rm sys}$, and velocity half-width, $\Delta v_{\rm sys}$ followed by the same columns of measured quantities in the line lists and counts of the number of total lines which comprise the system and the number of metal lines.
\item A multi-page ``atlas'' showing the reduced data, continuum fit, and identified lines.  The first panel of the atlas shows an overview of the entire spectrum in context with some observational details from Tables~1 and~2 above.  Subsequent panels show the entire range of COS data in 25\AA\ segments with identified absorption features marked with vertical ticks and labeled where possible.  Red ticks and labels denote $z>0$ features with abbreviated line-ID and redshift.  Green ticks show $v\approx0$ ISM features.  A sample page of the seven-page atlas for 3C\,57 is shown in Figure~\ref{fig:atlas_sample}.
\end{itemize}

In addition to the individual sight lines, we provide the median-combined Galactic foreground spectrum, both in {\sc fits} format and as a multi-page atlas in the same format as the individual sight line atlases.  A portion of this spectrum is shown in Figure~1.

\clearpage
\subsection{Erratum}
We note that a previous version of the HLSP was released in early 2014.  While similar in content to this 2016 catalog, this version was based on an earlier data reduction pipeline.  A number of instrumental artifacts were included as bone-fide IGM absorption lines.  In the intervening period, our data reduction pipeline was improved tremendously.  The current version of the HLSP should be largely free of instrumental features.  Also, the signal-to-noise of the coadded data has been improved by $\sim10\%$ in most cases due to better coaddition techniques.  Data from the 2014 IGM catalog should be discarded in favor of the current version.

\section{Results}

We measure 5138 significant IGM absorption components which comprise 2611 distinct absorbing systems along 82 sight lines over total \HI\ redshift pathlength $\Delta z=21.7$.  This is the largest catalog of low-$z$ IGM absorption to date and represents three times more sight lines and absorbers than any previous low-redshift IGM absorber study \citep{Lehner07,DS08,Tripp08,Tilton12} with greatly-improved sensitivity to weaker absorbers ($\Delta z\approx18$ for $N_{\rm HI}\approx10^{13}~\rm cm^{-2}$).  The catalog is intended to be useful in many areas of astrophysics, and many follow-on studies are planned on specific, more focused areas of research \citep[e.g.,][ on the IGM metal evolution]{Shull14b}.  Here, we present some global results from the catalog.  Unless otherwise noted, all statistics refer to IGM {\em systems}, not individual components.

\subsection{\HI\ Absorber Distribution}\label{sec:h1dndz}

The normalized frequency of absorbing systems $d{\cal N}/dz$ and the bivariate distribution of \HI\ absorbers as a function of column density and redshift, $\partial^2{\cal N}/\partial N\,\partial z$, are important quantities in low-$z$ cosmology.  Line density is an important probe of cosmic structure in the linear regime and, in comparison with simulations, yields constraints on density, the spectral energy distribution and intensity of the ionizing background \citep{Kollmeier14,Shull15}, the evolution of structures, thermal structures, and the processes of galaxy feedback \citep[e.g.,][]{DaveTripp01,CenFang06,Dave10,Smith11,Shull12a,Kollmeier14,Shull15}.  

Calculating \dndz\ accurately requires both good counting statistics and knowledge of the pathlength $\Delta z(N)$ covered in the survey for absorbers of a given column density $N$.  Since sensitivity is neither a constant between sightlines nor in different spectral ranges, we must correct for incompleteness in the weak absorbers by calculating the effective redshift pathlength, $\Delta z(N_{\rm min})$, sensitive to minimum column density, $N_{\rm min}$.  The effective absorption redshift pathlength $\Delta z(N)$ is calculated in a manner similar to that used in DS08, and we refer the reader there for details.  Briefly, we calculate the $3\sigma$ minimum equivalent width, $W_{\rm min}(\lambda)$, as a function of wavelength in each spectrum ($4\sigma$ for \Lya\ detections as discussed above).  We use the Galactic foreground COS template described above to mask portions of the spectrum with strong Galactic ISM regions.  The $W_{\rm min}(\lambda)$ vector is then translated into $N_{\rm min}(z)$ for each ion.  Since many species have line multiplets, the minimum column density $N_{\rm min}(z)$ is different for different transitions of the same species.  For instance, the \OVI\ 1038~\AA\ transition is only half as sensitive to absorption as the stronger \OVI\ 1032~\AA\ line.  At $z_{\rm abs}=0.178\pm0.003$ the stronger transition lies within the strong absorption from the Galactic DLA at 1213-1218~\AA\ where we have no sensitivity, but the weaker \OVI\ line lies in clear continuum at 1223~\AA.  Thus, we can still detect \OVI\ absorption at that redshift, albeit at half the usual sensitivity.  As in the line lists, the pathlength at $(cz_{\rm AGN}-cz_{\rm abs})/(1+z_{\rm AGN})\le1500$ \kms\ and $cz_{\rm abs}<500$ \kms\ is excluded to limit AGN-intrinsic and Local Group absorbers, respectively.  In cases where probable intrinsic absorbers are seen at $cz_{\rm abs}/(1+z_{\rm AGN})>1500$ \kms\ with respect to the AGN, that limit is used instead.

\begin{deluxetable}{lrrcccc}
\tabletypesize{\footnotesize}
\tablecolumns{7} 
\tablewidth{0pt} 
\tablecaption{Absorption Pathlengths $\Delta z$ and Completeness Limits}
\tablehead{\colhead{}&
           \multicolumn{2}{c}{Maximum} &
           \colhead{$z_{\rm abs}$ }&
           \multicolumn{3}{c}{$\log\,N$ completeness} \\
           \colhead{Species}    &
           \colhead{$\Delta z$ }&
           \colhead{$\Delta X$\tnma}&
	   \colhead{range}     &
           \colhead{75\%}     &
           \colhead{50\%}     &
           \colhead{25\%}     }
 \startdata 
   H\,I      &  21.74 & 35.92 &$<0.75$\tnmb & 13.09 & 12.93 & 12.77 \\
   H\,I\tnmc &  19.23 &\nodata&$<0.75$\tnmc & 13.04 & 12.89 & 12.73 \\
   O\,VI     &  14.49 & 27.73 &$0.1-0.73$   & 13.56 & 13.42 & 13.28 \\
   N\,V      &  19.33 & 28.09 &$<0.45$      & 13.31 & 13.18 & 13.03 \\
   C\,IV     &   8.85 & 10.33 &$<0.16$      & 13.26 & 13.12 & 12.96 \\
   Si\,IV    &  14.60 & 18.96 &$<0.28$      & 12.81 & 12.64 & 12.48 \\
   Si\,III   &  20.03 & 29.83 &$<0.49$      & 12.33 & 12.20 & 12.04 \\
   C\,III    &  10.40 & 22.61 &$0.16-0.8$   & 12.84 & 12.70 & 12.56 \\
   Ne\,VIII  &   2.32 &  7.54 &$0.47-0.85$  & 14.08 & 13.88 & 13.70 \\
 \enddata
\tablenotetext{a}{Co-evolving absorber pathlength (Eq.~2).}
\tablenotetext{b}{In principle, the Lyman continuum redshifts out of the COS/G160M detector at $z=0.97$.  However, in practice, H\,I is most reliably identified and measured using \Lya\ or \Lyb\ which limits the effective range to $z<0.75$.}
\tablenotetext{c}{$z$-limited, uniform sample described in Section~\ref{sec:zlimit}.}
  \label{tab:deltaz}
\end{deluxetable}

\begin{figure*}
  \epsscale{1.2}\plotone{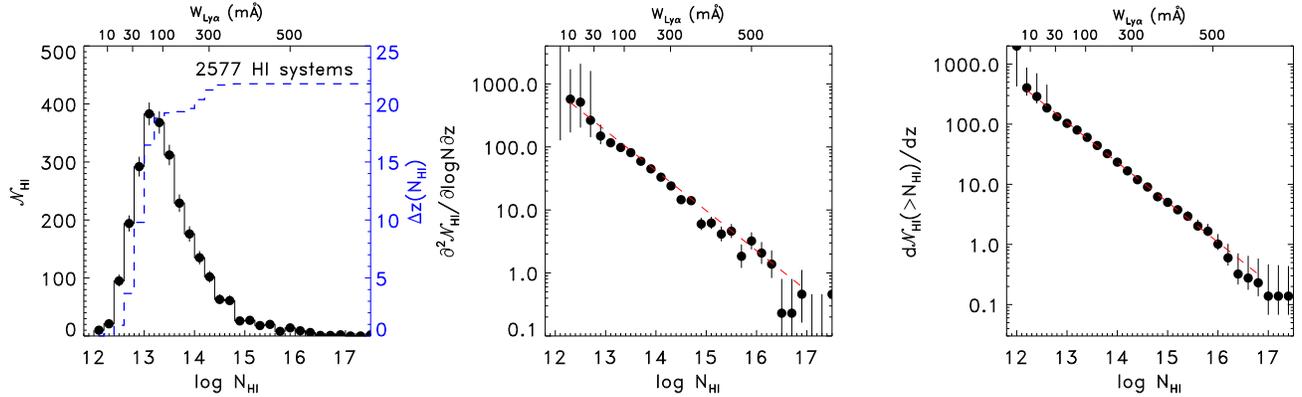}
  \caption{H\,I detection statistics for the full ($0\le z_{\rm abs} <
  0.75$) sample of 2577 IGM H\,I systems.  Left panel shows the number
  of systems $\cal N$ per 0.2 dex column density bin.  Error bars are
  one-sided Poisson uncertainty corresponding to $\pm1\sigma$.  The
  dashed curve and right-hand axis show the effective pathlength
  $\Delta z$ as a function of column density.  The differential
  absorber frequency $\partial^2{\cal N}(N)/ \partial \log N\partial
  z$ (Eq.~2) is shown in the middle panel.  The right panel shows the
  integrated system frequency $d{\cal N}(>N)/dz$.  Power law fits to
  the differential and cumulative distributions are shown as dashed
  lines.  Equivalent width values for $b=25$ \kms\ \Lya\ lines are
  shown on the top axis of each panel.}  \label{fig:dndz_h1}
\end{figure*}

\begin{figure*}
  \epsscale{1.2}\plotone{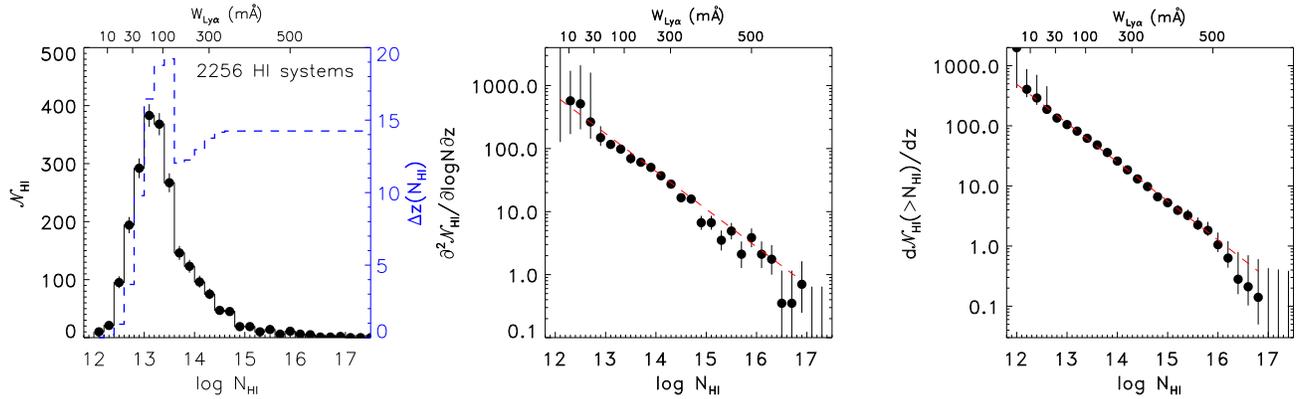}
  \caption{Same as Fig.~\ref{fig:dndz_h1} but for the ``uniform''
  sample (2256 IGM H\,I systems); weak systems at all redshifts are
  used, but stronger systems ($\log\,N\ge 13.5$) are only included at
  $z_{\rm abs} \ge 0.1$ where multi-line CoG solutions are available.}
  \label{fig:dndz_h1_zlim}
\end{figure*}

Detection limits as a function of redshift are assembled for each species (\HI, \OVI, \CIV, etc.) in each sight line, and the accumulated pathlengths at various limiting column densities and redshift ranges are totalled.  For strong \HI\ absorbers at any redshift, the total pathlength $\Delta z_{\rm HI}=21.74$, but the survey completeness has dropped to $\le50$\% at $\log\,N_{\rm HI}\le12.93$.  The detailed effective redshift path $\Delta z(N)$ is used to correct for completeness in our counting statistics to calculate \dndz.  Maximum pathlengths in redshift $\Delta z$ and co-evolving absorber pathlength $\Delta X$ \citep{BahcallPeebles69} are related by
\begin{equation}
\Delta X=\int_{z_1}^{z_2}{\frac{(1+z)^2}{[\Omega_m (1+z)^3+\Omega_\Lambda]^{1/2}}\,dz},
\end{equation}
and listed in Table~\ref{tab:deltaz} for \HI\ and seven metal species.  The fourth column shows the approximate redshift range over which an ion can be effectively identified in COS/FUV data.  The final three columns show the column densities at which the IGM survey is 75\%, 50\%, and 25\% complete.  

Figure~\ref{fig:dndz_h1} shows the distribution of total \HI\ systems as a function of column density, together with the effective path length $\Delta z(N_{\rm HI})$ and the differential distribution ($\partial^2{\cal N}/\partial \log N \partial z$) after it is corrected for completeness.  Figure~\ref{fig:dndz_h1_zlim} shows the same distribution for the more uniform $z$-limited sample as described in Section~\ref{sec:zlimit}.  Summing over column density down to a given $N_{\rm HI}$, we find the cumulative absorption system frequency $d{\cal N}(>N)/dz$, which can be fitted with a power law of the form 
\begin{equation}
\frac{d{\cal N}(>N)}{dz}=C_{14}\,\left(\frac{N}{10^{14}\rm~cm^{-2}}\right)^{-(\beta-1)}. \label{eq:dndzfit}
\end{equation}
Here, we use the traditional notation of $\beta$ as the index to the {\em differential} distribution $d^2{\cal N}/dN\,dz\propto N^{-\beta}$; the slopes of the cumulative and differential distributions differ by one.  Fitting the differential distribution of the uniform $z$-limited sample of 2256 absorbers, we find $\beta=1.65\pm0.02$ for systems in the range $12\le \log\,N \le 17$ with a normalization constant $C_{14}=25\pm1$ at the fiducial column density of $N_{\rm HI}=10^{14}\rm~cm^{-2}$ (corresponding to $W_{\rm r}\approx240$~m\AA\ for $b=25$ \kms).  The full sample of 2577 systems has fit parameters $\beta=1.67\pm0.01$ and $C_{14}=23\pm1$.  For ease of comparison with other observations and with simulation results, we list column density bins, number of absorption systems per bin, redshift pathlength, and $\partial^2{\cal N}/\partial N \partial z$ values for the uniform \HI\ subsample in Table~\ref{tab:dndz}.

In previous surveys \citep[e.g.,][]{Danforth05,DS08}, we adopted $\beta=2$ as the critical index separating ``top-heavy'' and ``bottom-heavy'' scenarios, with total IGM mass dominated by the few high-column density systems or the many lower-column density systems.  However, this assumption is simplistic as it ignores the dependence on column density of the ionization fraction (and metallicity in the case of metal ion absorbers).  These issues are discussed in detail in \citet{Shull12a} who used simulations to derive the $N$-dependence of thermal phases and metallicities in the IGM.

The sample quantities in the COS \HI\ systems are similar to those found in previous low-$z$ IGM surveys using smaller samples; \citet{Tilton12} found $\beta_{\rm HI}=1.68\pm0.03$ in a sample of 746 absorbers over pathlength $\Delta z_{\rm HI}=5.38$.  Slopes of $\beta_{\rm HI}=1.65\pm0.07$, $1.76\pm0.06$, and $1.73\pm0.04$ were found in smaller \Lya/\HI\ surveys by \citet{Penton04}, \citet{Lehner07}, and DS08, respectively.  Numerical simulations also agree, e.g., $\beta_{\rm HI}=1.70$ at $z=0$ from \citet{Dave10}.  

It has been suggested \citep{Rudie13} that a small break in the power law nature of \dndz\ may exist for IGM absorbers at $z\approx 2-3$ corresponding to the association of strong absorbers with galaxy halos and weaker absorbers with unvirialized intergalactic matter.  We searched for a fit to a broken power law to the cumulative \dndz\ in the uniform \HI\ sample with both slopes and the column density of the breakpoint as free parameters.  No significant break is found in our low-$z$ sample. 

\begin{deluxetable}{ccrccrc}[t]
\tabletypesize{\footnotesize}
\tablecolumns{7} 
\tablewidth{0pt} 
\tablecaption{H\,I and O\,VI $\partial^2{\cal N}(N)/\partial \log N\, \partial z$ Values}
\tablehead{
  \colhead{$\log\,N$}  &
  \colhead{${\cal N}_{\rm HI}$} &
  \colhead{$\Delta z_{\rm HI}$} &
  \colhead{$\frac{\partial^2{\cal N}(N_{\rm HI})}{\partial\log N\, \partial z}$} &
  \colhead{${\cal N}_{\rm OVI}$} &
  \colhead{$\Delta z_{\rm OVI}$} &
  \colhead{$\frac{\partial^2{\cal N}(N_{\rm OVI})}{\partial\log N\, \partial z}$} 
          }     
\startdata
12.2-12.4&  21&  0.18&$570^{+1000}_{-400}$&      \nd &\nd & \nd \\ 
12.4-12.6&  95&  0.93&$510^{+1000}_{-310}$&      \nd &\nd & \nd \\ 
12.6-12.8& 194&  3.67&$260^{+1000}_{-120}$&      \nd &\nd & \nd \\ 
12.8-13.0& 292&  9.79&$150^{+80}_{-40}$&        2&  0.24&$ 42^{+150}_{-37}$  \\
13.0-13.2& 383& 16.46&$120\pm10$&	        6&  0.91&$ 33^{+78}_{-25}$  \\
13.2-13.4& 368& 18.78&$ 98\pm5$&	       17&  4.13&$ 21^{+33}_{-9}$  \\
13.4-13.6& 267& 19.23&$ 69^{+17}_{-12}$&       40&  9.16&$ 22^{+7}_{-5}$  \\
13.6-13.8& 146& 12.03&$ 61\pm5$&	       62& 12.65&$ 25^{+4}_{-3}$  \\
13.8-14.0& 123& 12.25&$ 50\pm5$&	       69& 13.81&$ 25\pm3$  \\
14.0-14.2&  96& 12.96&$ 37\pm4$&	       47& 14.13&$ 17^{+3}_{-2}$   \\
14.2-14.4&  75& 13.75&$ 27^{+4}_{-3}$&	       24& 14.36&$ 8.4^{+2.1}_{-1.7}$  \\
14.4-14.6&  47& 14.17&$ 17^{+3}_{-2}$&	       10& 14.43&$ 3.5^{+1.5}_{-1.1}$  \\
14.6-14.8&  45& 14.24&$ 16^{+3}_{-2}$&	        2& 14.46&$0.69^{+0.91}_{-0.45}$  \\
14.8-15.0&  19& 14.25&$  6.7^{+1.9}_{-1.5}$&    1& 14.48&$0.35^{+0.79}_{-0.29}$  \\
15.0-15.2&  19& 14.25&$  6.7^{+1.9}_{-1.5}$&     \nd &\nd & \nd \\ 
15.2-15.4&  10& 14.25&$  3.5^{+1.5}_{-1.1}$&     \nd &\nd & \nd \\ 
15.4-15.6&  14& 14.25&$  4.9^{+1.7}_{-1.3}$&     \nd &\nd & \nd \\  
15.6-15.8&   6& 14.25&$  2.1^{+1.3}_{-0.8}$&     \nd &\nd & \nd \\  
15.8-16.0&  12& 14.25&$  4.2^{+1.6}_{-1.2}$&     \nd &\nd & \nd \\  
16.0-16.2&   6& 14.25&$  2.1^{+1.3}_{-0.8}$&     \nd &\nd & \nd \\  
16.2-16.4&   4& 14.25&$  1.4^{+1.1}_{-0.7}$&     \nd &\nd & \nd \\  
16.4-16.6&   0& 14.25&$0.0^{+0.65}_{-0.00}$&     \nd &\nd & \nd \\  
16.6-16.8&   1& 14.25&$0.35^{+0.81}_{-0.29}$&    \nd &\nd & \nd \\  
16.8-17.0&   2& 14.25&$0.70^{+0.93}_{-0.45}$&    \nd &\nd & \nd \\   
17.0-17.2&   1& 14.25&$0.35^{+0.81}_{-0.29}$&    \nd &\nd & \nd \\  
\enddata
\label{tab:dndz}
\end{deluxetable}

\begin{figure}
  \epsscale{1}\plotone{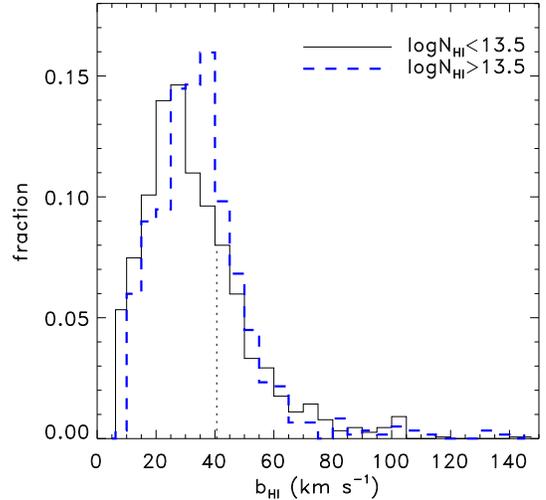}
  \caption{Distribution of line widths for H\,I systems in the uniform
  sample.  To limit broadening introduced by multiple components, we
  analyze H\,I lines fitted with only a single resolved velocity
  component.  The weak ($\log\,N\le13.5$) and strong ($\log\,N>13.5$)
  samples show median b-values of 29.6 \kms\ and 33.8 \kms,
  respectively.  The dotted line marks $b=40.6$ \kms, the line width
  corresponding to pure thermal broadening at $T=10^5$~K.  The FWHM is
  $1.67 b$.} \label{fig:bvaldist}
\end{figure}

\subsubsection{Line-Width Distribution}

Measuring the Doppler $b$ parameter of an absorption system is more prone to error than measuring the column density due to several systematic effects.  The profile-fitting routines used here take into account the instrumental line spread function \citep{Kriss11}, but the fitted $b$ is still the quadratic sum of thermal and non-thermal line widths.  Unresolved component structure and noise in the line profile can both affect the measured $b$ value of a system.  Profiles fitted to \Lya\ components are known to systematically over-estimate the $b$ value \citep{Danforth10a} so single-transition $b$ measurements should be used with caution.  Curve-of-growth fits can provide a more accurate measurement of $b_{\rm HI}$ in absorbers with a simple velocity structure, but it is likely that most strong \HI\ absorbers contain multiple, unresolved velocity components. 

We limit these effects by examining the $b$-value distribution of only those \HI\ systems fitted with a single velocity component.  Figure~\ref{fig:bvaldist} shows that the distribution of $b$ values for weak, single-component systems ($N_{\rm HI}\le10^{13.5}$~\cd) has a median b-value of $29.6\pm0.5$ \kms\ and a $\pm1\sigma$ range of $17-48$ \kms.  Stronger single-component systems ($N_{\rm HI}>10^{13.5}$~\cd) have a larger median ($b=33.8\pm0.5$ \kms) though a similar range ($20-49$ \kms).  The difference in distributions suggests that stronger absorbers may include unresolved velocity components, artificially increasing the line width.  However, it may not be significant since, in the uniform sample, strong \HI\ systems are analyzed through a multi-transition CoG while the weak systems are measured via a single \Lya\ component.

Both strong and weak \HI\ line-width distributions have a long tail toward broader widths.  Since the width of a line is a product of both thermal and non-thermal contributions, the measured $b$ parameter of an absorption line can set an upper limit on the temperature of the gas.  For \Lya\ components, $T=10^5$~K corresponds to $b=40.6$~\kms\ if all broadening is thermal in nature.  These broad \Lya\ absorbers (BLAs) can provide a tracer of WHIM gas independent of metallicity.  The weak and strong samples in Figure~\ref{fig:bvaldist} show BLA fractions of $\sim26$\% and $\sim28$\%, respectively.  However, we caution that unresolved velocity components can contribute non-thermal broadening to a line profile, so the BLA fractions quoted above are more realistically upper limits to their true values.  For more discussion, see \citet{Richter06,Lehner07,Danforth10a,Savage10,Savage14}.  In a more intensive evaluation of 85 \OVI\ components from this sample by \citet{Savage14}, 45 of which have well-aligned \HI\ \Lya\ components, only 14 components have \HI\ and \OVI\ line widths indicative of $T>10^5$ K gas.  Therefore, the fraction of BLAs which unambiguously trace WHIM gas is a small fraction of the total \HI$+$\OVI\ absorber population.


\begin{figure}
  \epsscale{1.2}\plotone{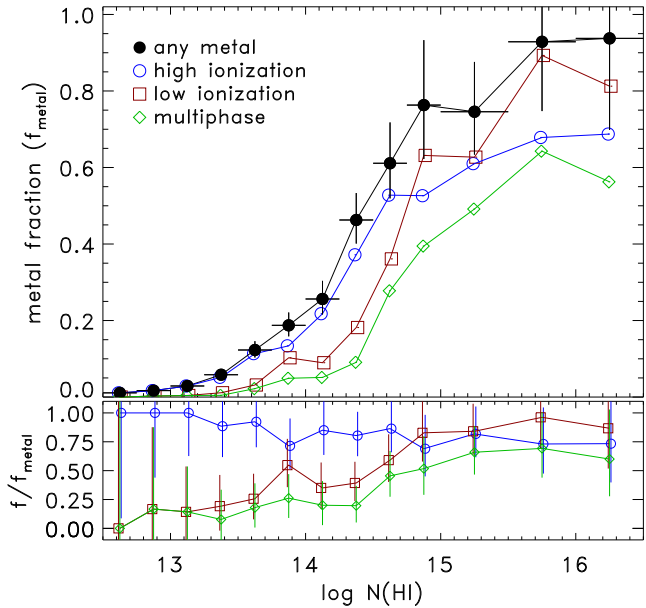}
  \caption{Fraction of IGM H\,I absorbers in which metal ions are also
  detected as a function of H\,I column density.  Black points (filled
  circles) show the fraction of IGM systems with a detection in any
  metal species, while blue (open circles) and red (open squares) show
  the fraction with detections in highly-ionized metals (O\,VI, N\,V,
  C\,IV, Ne\,VIII) and low-ionization metals (Si\,II/III/IV,
  C\,II/III), respectively.  Green diamonds show the fraction of
  multi-phase systems, i.e., systems with both high- and
  low-ionization detections.  The bottom panel shows the relative
  fraction of high- and low-ionization detections as a function of
  detections in any metal ion.}  \label{fig:metalfraction}
\end{figure}

\subsection{Metal Absorbers}

The majority of extragalactic absorption components are Lyman-series lines of \HI\ ($4234/5138\approx82$\%), and the majority of those ($2946/4234\approx70$\%) are \Lya.  Metals are detected in 418 IGM systems or $\sim 15$\% of all IGM systems.  This fraction is strongly dependent on \HI\ column density (Figure~\ref{fig:metalfraction}).  For weak \HI\ systems with $12.5\le \log N_{\rm HI}\le 13.5$, metals appear in only 48/1465 ($\sim3$\%).  For systems $13.5\le\log\,N_{\rm HI}< 14.5$, the metal fraction rises to 177/818 ($\sim22$\%).  For the strongest IGM systems ($\log N_{\rm HI}\ge 14.5$), metals are nearly ubiquitous (157/212$\approx74$\%). 
Two effects probably combine to produce the steeply rising metal fraction seen in Figure~\ref{fig:metalfraction}.  First is the finite sensitivity of COS spectra to weak absorbers; metal lines in the IGM tend to be weaker than their corresponding \Lya\ absorbers and thus may not be above the detection limit for weak \HI\ lines.  Second, weak absorbers are often associated with the diffuse IGM far from galaxies where metallicity is likely to be very low, while strong absorbers are often associated with higher-metallicity galaxy halos.  The 50\% ``cross-over'' point for metal-bearing absorbers is $\log\,N_{\rm HI}\approx14.5$, the same column density where $\sim50\%$ of absorbers are found within one virial radius of a galaxy \citep{Stocke14}. 

If we subdivide the metal systems into highly-ionized species (\OVI, \NV, \CIV) and low-ionization species (Si\,II/III/IV, C\,II/III), we see that systems with high-ionized metals are relatively more common than low-ionization metals at $\log N_{\rm HI}<14$, while they become comparable in stronger \HI\ systems (Figure~\ref{fig:metalfraction}, bottom panel).  This is easily explained by recognizing that \HI\ is, itself, a low-ionization species.  Systems in which \OVI\ and other highly-ionized metals can exist will have a lower neutral fraction and thus lower $N_{\rm HI}$ for a given total column density.  

The most common metal detection is in one or both lines of the \OVI\ doublet (1031.93, 1037.64~\AA) which appear in 280 IGM systems.  Lithium-like doublet transitions of \CIV\ (70 systems) and \NV\ (59 systems) are also common.  Absorption in the strong, singlet lines of \SiIII\ (\lam1206, 123 systems) and \CIII\ (\lam977, 115 systems) and doublet \SiIV\ (\lam\lam1393, 1402; 45 systems) trace moderately-ionized IGM.  In addition, transitions of twenty other metal ions are observed in very small numbers; the total census of metal-ion species seen in IGM absorbers is O\,I/II/III/IV/VI, N\,II/III/IV/V, C\,II/III/IV, Si\,II/III/IV, S\,IV/V/VI, Fe\,II, Ne\,VIII, P\,II, and Al\,II.  We discuss the statistics of the more common metal-ions below.  A comparison of these with higher-redshift surveys in the literature is presented in \citet{Shull14b}.  

\begin{figure*}
  \epsscale{1.2}\plotone{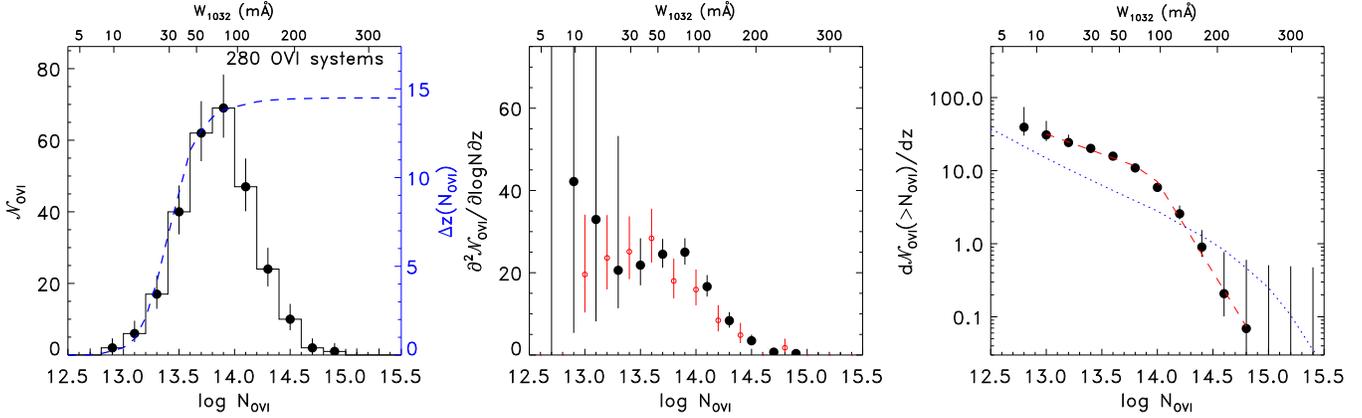}
  \caption{Statistics of 280 O\,VI systems in the COS IGM survey.  The
  left panel shows the number ${\cal N}_{\rm OVI}$ of absorbers per
  0.2 dex column density bin.  Error bars are one-sided Poisson
  uncertainties corresponding to $\pm1\sigma$.  Completeness is shown in
  the form of total pathlength $\Delta z(N_{\rm OVI})$ (dashed blue
  line, right axis).  Approximate equivalent width in the stronger
  line of the doublet is given on the top axis.  The middle panel
  shows the differential frequency of absorbing systems
  $\partial^2{\cal N}/\partial \log N\partial z$.  Open, red points
  show the corresponding distribution in the FUSE$+$STIS sample
  presented in \citet{Tilton12} but post-processed to use the same
  convention of systems we use here.  Right panel shows the cumulative
  frequency $d{\cal N}(>N)/dz$.  A single power law clearly does not
  represent the data, but a broken power law (red dashed line) with
  parameters $\beta_{\rm strong}=2.5\pm0.2$, $\beta_{\rm
  weak}=0.56\pm0.16$, $\log\,N_{\rm break}=14.0\pm0.1$, $C_{\rm
  break}=9.9\pm1.4$ provides a statistically better fit.  A
  distribution from a cosmological simulation \citet{Smith11}
  is shown as a blue dotted line in the right panel.  }
  \label{fig:o6_dndz}
\end{figure*}

\begin{figure*}
  \epsscale{1.0}\plottwo{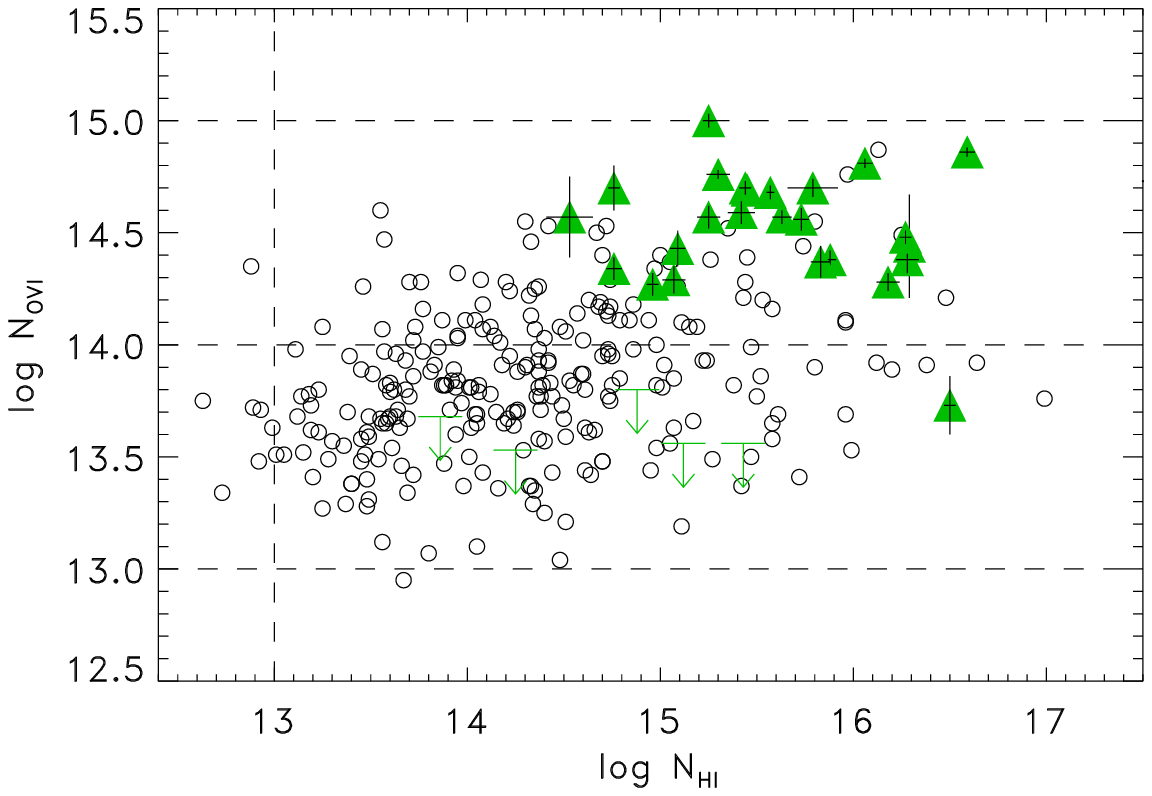}{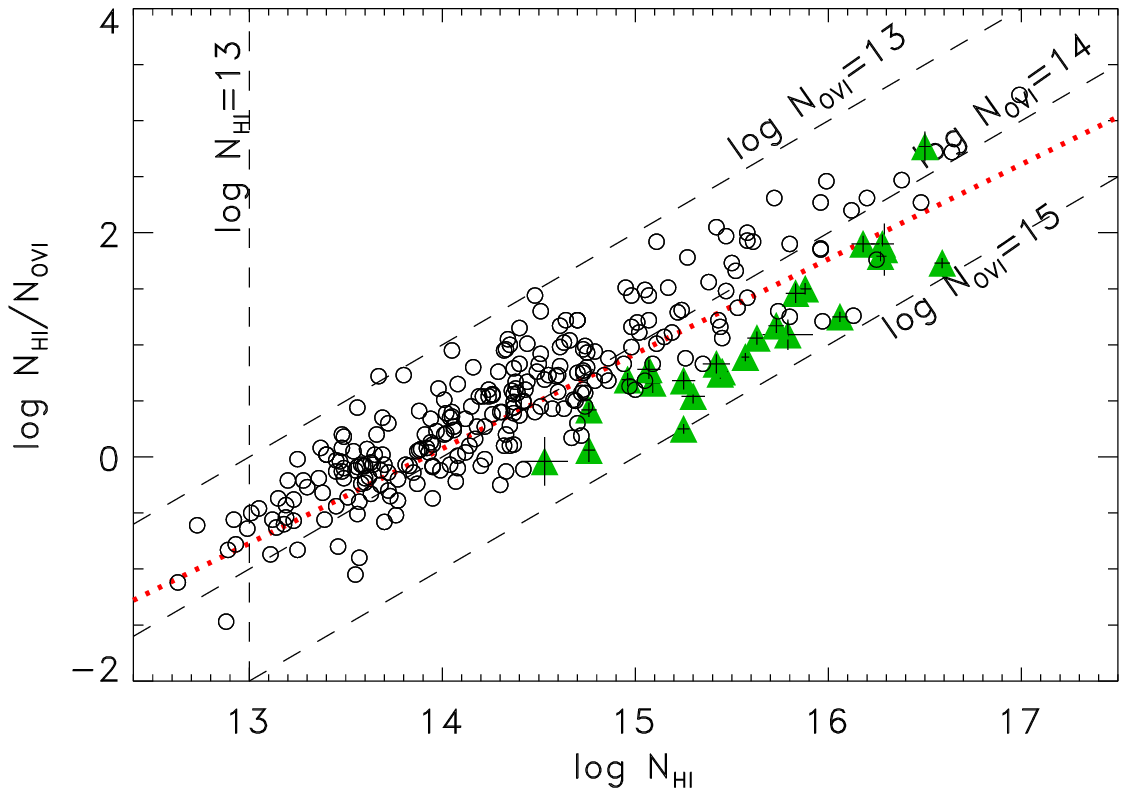}
  \caption{Column densities of H\,I and O\,VI are poorly correlated
  (left panel).  Green triangles show the absorbers from the COS-Halos
  survey \citep{Tumlinson11}.  The fact that COS-Halos points occupy
  only one region of the larger parameter space suggests that not all
  O\,VI absorption is associated with the halos of L$^*$ galaxies.
  The multiphase ratio $N_{\rm HI}/N_{\rm OVI}$ in the right panel as
  a function of $N_{\rm HI}$ shows the poor column density correlation
  between neutral and highly-ionized systems which are kinematically
  related.  Dashed lines show values of constant $N_{\rm OVI}$ and
  $N_{\rm HI}$.  A power-law fit to the data (red dotted line) has
  index $0.86\pm0.01$ showing that the two column densities are poorly
  correlated.  Components with $N_{\rm OVI}\la10^{13}$~\cd\ are not
  typically detectable at our survey S/N.}  \label{fig:multiphase}
\end{figure*}

\begin{figure*}
  \epsscale{1.2}\plotone{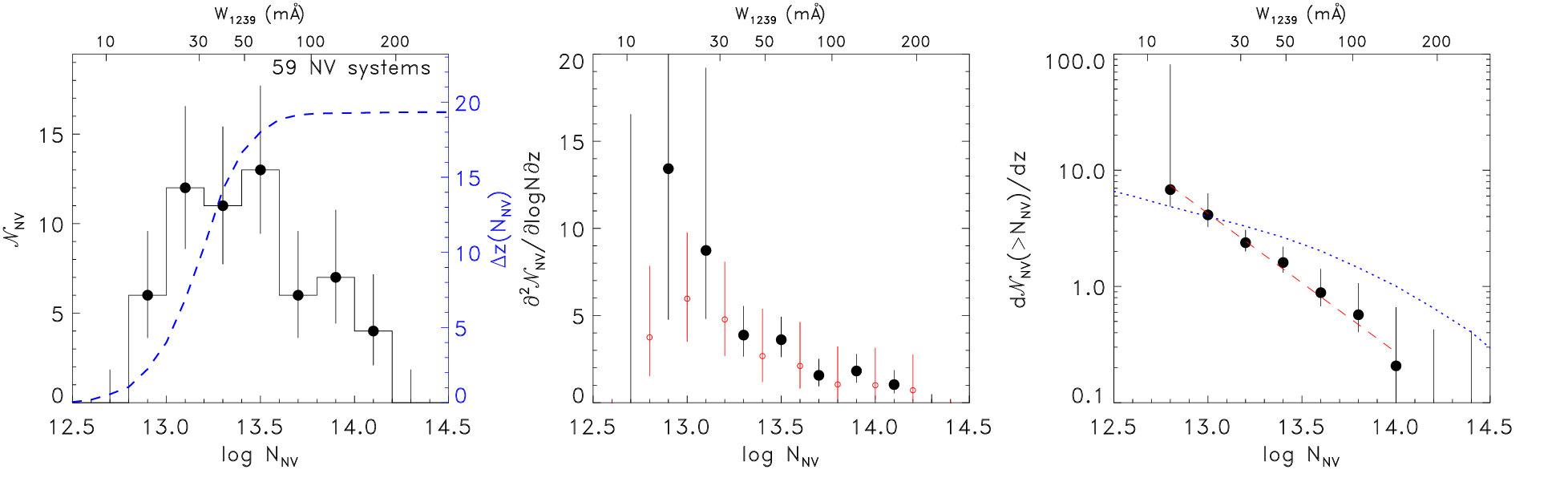}
  \epsscale{1.2}\plotone{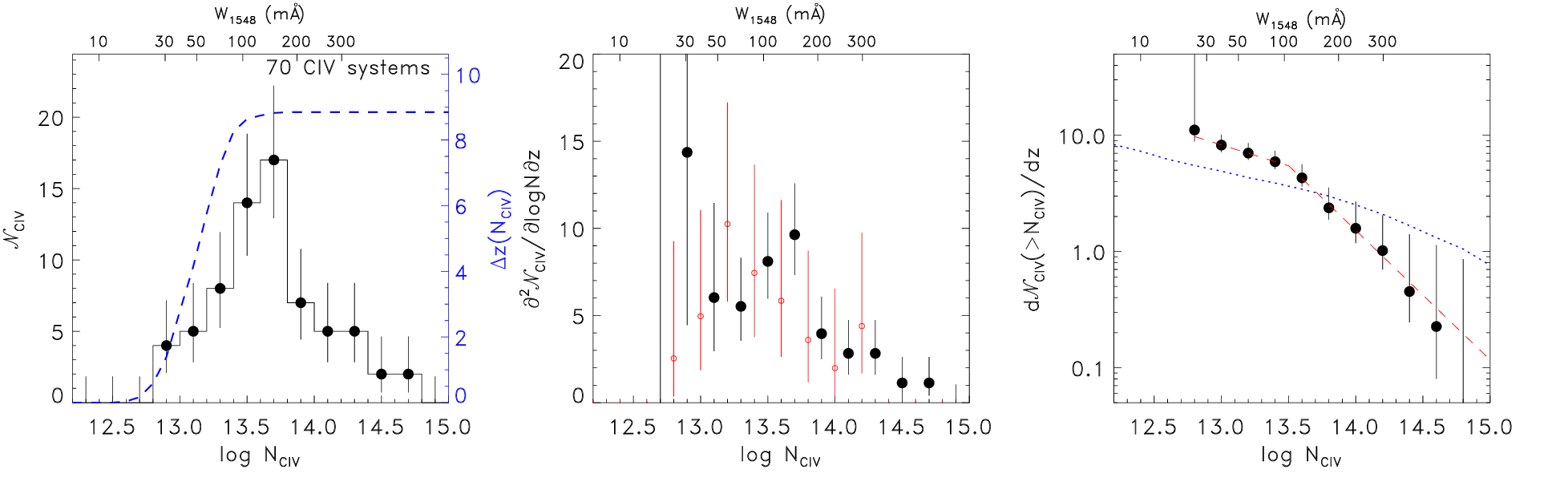}
  \caption{Same as Figure~\ref{fig:o6_dndz} but for 59 N\,V systems
  (top panels) and 70 C\,IV systems (bottom panels). }
  \label{fig:n5c4_dndz}
\end{figure*}

\begin{figure*}
  \epsscale{1.2}\plotone{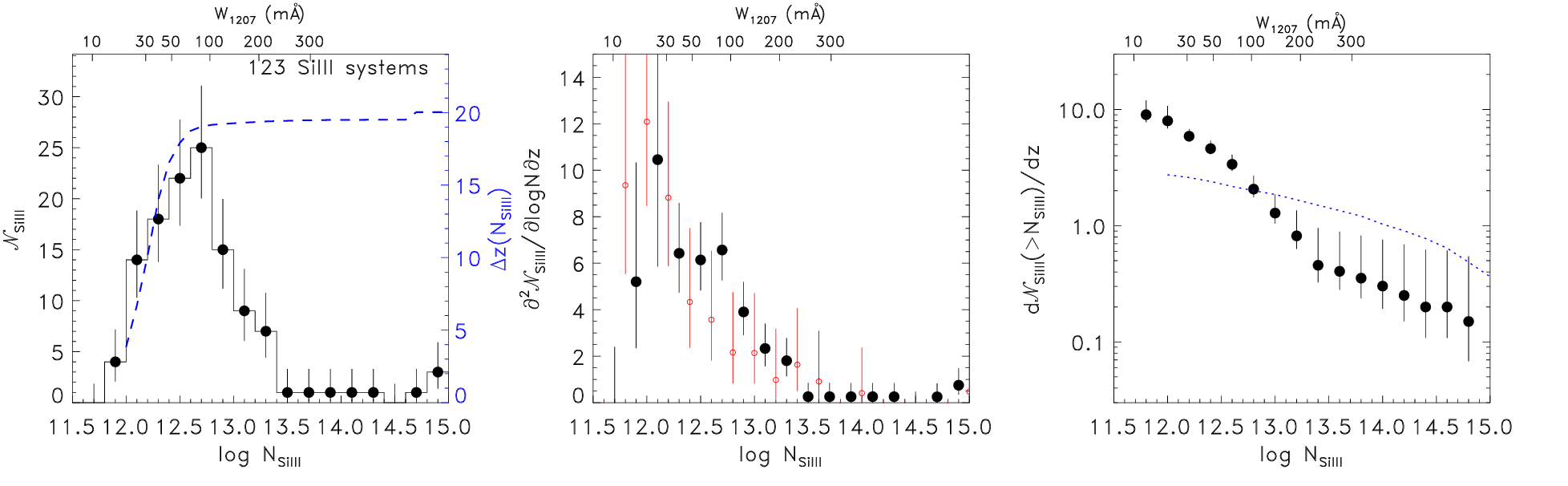}
  \epsscale{1.2}\plotone{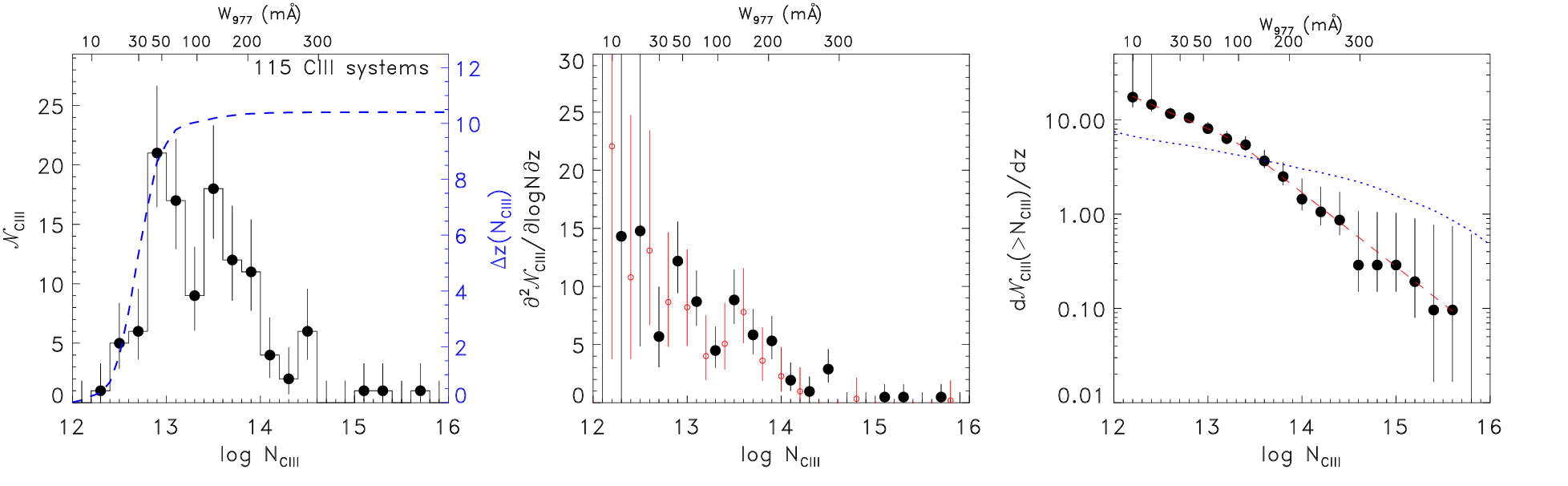}
  \epsscale{1.2}\plotone{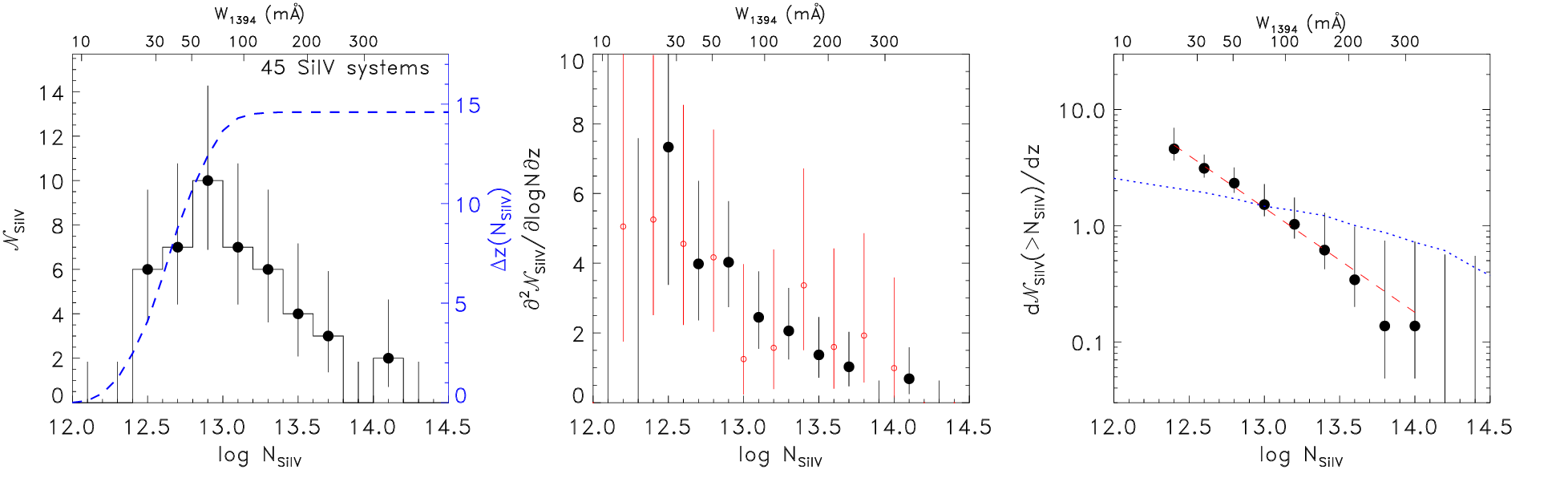}
  \caption{Same as Figure~\ref{fig:o6_dndz} for 123 Si\,III systems
  (top panels), 115 C\,III systems (middle panels), and 45 Si\,IV
  systems (bottom panels).}  \label{fig:dndz_lowions}
\end{figure*}

\subsubsection{\OVI}
The \OVI\ 1031.93, 1037.62~\AA\ doublet is a strong ($f=0.1325, 0.0658$), resonance ($2s-2p$) transition of lithium-like oxygen.  \OVI\ receives a great deal of attention not only because it is the most commonly seen metal ion in the low-redshift IGM and CGM, but because it is thought to be a tracer of the warm-hot intergalactic medium (WHIM) at $T>10^5$~K \citep{Tripp01,Danforth05,DS08,ThomChen08,Danforth09}.  High-quality COS observations of \OVI\ and \HI\ absorbers have been used to constrain the physical properties in a number of IGM systems already \citep{Savage10,Savage14,Stocke14,Muzahid15}.  Galaxy redshift surveys around IGM sight lines \citep{Stocke06,ChenMulchaey09,Prochaska11,Johnson15} have shown that \OVI\ systems are found $\la1$ Mpc from the nearest $L^*$ galaxies and thus closer to galaxies than typical \Lya\ forest lines.  Studies of strong \OVI\ systems ($\log N_{\rm OVI}\ga14.3$) in low-S/N COS data \citep[e.g. the COS-Halos project][]{Tumlinson11,Tumlinson13} have shown a strong correlation between the specific star formation rate and the presence of \OVI.

\OVI\ absorption is detected in COS/FUV data at $0.1\la z\la0.74$ because the G130M grating is only sensitive at $\lambda\ga1135$~\AA\ and G160M ends at 1796 \AA.  We find 280 \OVI\ systems out of a total of 1658 IGM systems in this redshift range ($\sim17$\%).  This is a factor of two more \OVI\ systems than the \citet{Tilton12} synthesis of STIS and \FUSE\ surveys, enabling us to address the low-$z$ \OVI\ properties with more statistical rigor.  Observational \OVI\ results are shown in Figure~\ref{fig:o6_dndz}.  For comparison, we include equivalent values derived from the \citet{Tilton12} absorber list.  In this and all subsequent plots that show absorbers from that study, we have combined all components within 30 \kms\ of each other to create absorption systems analogous to those we use in the COS survey.

The left panel of Figure~\ref{fig:multiphase} shows that \OVI\ and \HI\ column densities are essentially uncorrelated.  What's more, we see \OVI\ systems ranging over a factor of $\sim100$ in column density ($13\la\log N_{\rm OVI}<15$) while the \HI\ column density varies by a factor of $10^6$ in this survey ($12<\log N_{\rm HI}<18$) but has been observed at much higher column densities in other extragalactic contexts.  This lack of correlation has been noted before \citep{Danforth05} where we defined a multiphase ratio, $N_{\rm HI} / N_{\rm OVI}$, to track the relative amounts of warm, photoionized and highly-ionized gas which often are kinematically associated.  The right panel of Figure~\ref{fig:multiphase} shows the multiphase ratio as a function of \HI\ column density.  Because of S/N limitations, we are not sensitive to the weakest systems, which imposes approximate sensitivity limits $N\ga10^{13}$~\cd\ for both \OVI\ and \HI\ (upper dashed line in Figure~\ref{fig:multiphase}).  However, we see a sharp lower bound to the multiphase ratio, implying a maximum \OVI\ column density of a few times $10^{14}$~\cd\ which is only weakly correlated with \NHI.  A fit to the multiphase ratio (dotted line) shows the relationship $N_{\rm HI}/N_{\rm OVI}\propto(N_{\rm HI}/10^{14}~{\rm cm}^{-2})^{0.86\pm0.01}$.  This slope is consistent with that of \citet{Danforth05} who found $N_{\rm HI}/N_{\rm OVI}\propto N_{\rm HI}^{0.9\pm0.1}$.

Green triangles in Figure~\ref{fig:multiphase} show the equivalent measurements drawn from the COS-Halos survey \citep{Tumlinson11}.  These are absorbing systems associated with the halos of large ($>L*$) star-forming galaxies.  The COS-Halos surveys were not as sensitive to column densities of \HI\ or \OVI\ as is the current survey, but it is apparent that most star-forming galaxies can be associated with \OVI\ absorption.  However \citet{Stocke06} find strong \OVI\ absorbers at impact parameters out to $\sim0.8$~Mpc which is $2-3~R_{vir}$ for $L>L*$ galaxies.  So, while $L*$ galaxies have strong \OVI\ absorption, the converse is not necessarily true.

Calculating \dndz\ for \OVI\ in the same manner as we used in Section~\ref{sec:h1dndz} for \HI, we see a distinct ``knee'' in the distribution at $\log\,N\approx14$ (Figure~\ref{fig:o6_dndz}, right panel).  A broken power-law fit to the cumulative distribution of \OVI\ absorbers (dashed curve) shows a break at $\log\,N=13.9\pm0.1$ with indices $\beta_{\rm weak}=0.55\pm0.16$ and $\beta_{\rm strong}=2.5\pm0.2$ and a normalization at the break of $C_{\rm break}=9.7\pm1.3$.  There is curvature in the distribution not accounted for in the fit.  A power-law fit is probably not the most appropriate in this context, but we present it here for comparison with theory and previous observational work.  This two-slope distribution is in marked contrast with the results of DS08 who fitted the entire distribution (comprised of only 83 \OVI\ absorbers) with a single power-law with $\beta\sim2.0$ over the range $13.2<\log\,N_{\rm OVI}<14.8$.  

We note that the slope break at $\log N\approx14$ is the dividing line of the population of circumgalactic \OVI\ absorbers seen in the large COS-HALOS project.  \citet{Tumlinson11} find a strong correlation between star-forming galaxy halos and \OVI\ absorption at $\log N_{\rm OVI}\ga 14.2$, while passive red galaxies tend to have \OVI\ upper limits at this level or below.  The weaker systems seen in this and previous surveys may correspond to more diffuse gas, possibly intra-group \citep{Stocke14} or a true intergalactic medium.  ``Circumgalactic'' \OVI\ gas may lie outside the virial radius of galaxies and still be flowing out to enrich the IGM \citep{Stocke06,Stocke13,Shull14a}.

\subsubsection{\CIV\ and \NV}

\OVI, \NV\ and \CIV\ are all lithium-like tracers of highly-ionized gas, which is either collisionally ionized to $T\ga10^5$~K or photoionized by photons with $\sim50-100$ eV.  Due to the lower cosmic abundances of carbon and nitrogen, as well as the limited redshift range over which \CIV\ can be observed in COS/FUV spectra ($z_{\rm abs}\la0.16$), the number of IGM systems in which these ions are observed (\NV: 59 systems, \CIV: 70 systems) is considerably smaller than the 280 detected in \OVI.  Nevertheless, this is larger than previous surveys of these two ions in the low-redshift IGM.  Both species appear to follow similar distributions to \OVI, though without as much sensitivity to the weaker absorbers (Figure~\ref{fig:n5c4_dndz}).  The \NV\ distribution is consistent with a single power law with index $\beta=2.2\pm0.1$ and normalization $C_{14}=0.3$.  The cumulative distribution of \CIV\ shows a clear turnover and is fitted with a broken power law ($\beta_{\rm weak}=1.4\pm0.3$ and $\beta_{\rm strong}=2.1\pm0.3$) with a break at $\log\,N_{\rm CIV}=13.5\pm0.2$ and normalization $C_{\rm break}=5.5\pm1.5$. 

\subsubsection{\SiIII, \CIII, and \SiIV}

\SiIII\ and \CIII\ absorbers are relatively common in the IGM owing to the very high oscillator strengths of the 1206.5~\AA\ ($f=1.63$) and 977.0~\AA\ ($f=0.757$) transitions, respectively.  These systems in previous studies trace \HI\ systems to a much better extent than more highly-ionized species, probably because \CIII\ and \SiIII\ trace photoionized gas.  We detect \SiIII\ and/or \CIII\ absorption in 200 systems.  The \SiIV\ doublet (\lam1393.76, 1402.70) is also relatively strong, though the decreased abundance of Si compared with cosmic C/N/O means that these absorbers are somewhat less common; only 45 \SiIV\ systems are seen in this survey for a line frequency $d{\cal N}/dz\sim3$ for components with $W_\lambda\ga30$~m\AA. 

\SiIII, with an ionization range of 16.3-30.7 eV, traces low-ionization, metal-enriched gas, while \CIII\ (24.4-47.9 eV) and \SiIV\ (33.5-45.1 eV) trace an ionization middle-ground between \SiIII\ and the high-ionization species such as \CIV, \NV, and \OVI\ (50-150 eV).  The \dndz\ distribution of all three lower-ionization species is consistent with that seen in previous surveys (Figure~\ref{fig:dndz_lowions}).  The \SiIV\ distribution is fitted with $\beta=1.9\pm0.2$, $C_{\rm 14}=0.2$ for $\log\,N_{\rm SiIV}\le14.1$.  \CIII\ is reasonably fit with a two-slope power law ($\beta_{\rm weak}=1.4\pm0.1$ and $\beta_{\rm strong}=1.81\pm0.1$) with a break at $\log\,N_{\rm break}=13.3\pm0.3$ and normalization at the break of $C_{\rm break}=5.9\pm1.6$.  \SiIII\ does not follow an obvious power-law distribution.

\subsubsection{Are the \dndz\ turnovers real?}

The distributions of many of the metal ions shown in Figures~\ref{fig:o6_dndz}-\ref{fig:dndz_lowions}, particularly \OVI, \CIV, and \CIII, show a clear flattening in their numbers at lower column densities.  Because weaker systems are observed over smaller effective pathlengths, we investigated whether this apparent flattening in the \dndz\ values at lower column densities could be an effect of completeness.  The nominal $\Delta z(N)$ function (e.g., blue curve in Fig.~\ref{fig:o6_dndz}) is calculated for $3\sigma$ detections.  If this threshold is raised to $4\sigma$ or even $5\sigma$, there is relatively less effective pathlength at lower column densities.  The blue curve in the left-hand panels of Figures~\ref{fig:o6_dndz}, \ref{fig:n5c4_dndz}, \ref{fig:dndz_lowions} representing $\Delta z(N)$ moves to the right by $\sim0.1$ dex for each $1\sigma$ increase in significance level and the frequency of weak absorbers rises.  This results in a steepening of the weak-end slope of the cumulative distribution function (less pathlength for the same number of detections), but it does not eliminate the non-power-law nature of the distributions in any case.  Therefore, we believe that the turnovers seen in many of the metal absorber distributions are real, but the weak-end slopes should be treated with caution.

\subsubsection{\NeVIII}

The extreme-UV (2s-2p) doublet transition of \NeVIII\ ($\lambda=770.41, 780.32$~\AA, $f=0.1030, 0.0505$) is often presented as a less-ambiguous tracer of collisionally-ionized gas than \OVI\ or other highly-ionized, FUV metal transitions \citep{Savage05}.  Several compelling studies of warm-hot or multi-phase gas have been made using \NeVIII\ in conjunction with \HI, \OVI, and other ions to constrain gas temperature and density \citep{Savage05,Savage11,Narayanan09,Narayanan11,Tripp11,Meiring13,Hussain15}.  Unfortunately, the \NeVIII\ doublet only redshifts into the COS/G130M band at $z\ga0.47$, which limits us to the 13 highest-redshift AGN sight lines in this survey, usually without observations of the strong \HI\ counterparts.  We measure several dozen \NeVIII\ systems, but all are at redshifts similar to the background AGN and show the hallmarks of intrinsic, rather than intervening, absorption.  They are primarily strong, blended, multi-component absorption profiles in high ions, often lacking \HI\ absorption.  

There are three \NeVIII\ detections at $>3\sigma$ in our survey which can be reasonably identified as intergalactic rather than intrinsic systems.  The system detected toward PKS\,0405$-$123 at $z=0.49494$ in \NeVIII\ and \OVI\ by \citet{Narayanan11} shows $\log N_{\rm NeVIII}=13.5\pm0.2$ and $\log N_{\rm OVI}=14.32\pm0.05$.  In an earlier reduction of the data allowing for the contaminating effects of fixed pattern noise, \citet{Narayanan11} measured column densities of $\log N_{\rm NeVIII}=13.96\pm0.06$ and $\log N_{\rm OVI}=14.39\pm0.01$.  The origin of the \NeVIII\ and \OVI\ is consistent with collisionally ionized gas with $T\approx5\times10^5$~K and a baryonic column density of $N_H\sim10^{19}-10^{20}$~\cd.

Two other \NeVIII\ detections have no corresponding absorption in any other metal or \HI\ transition.  The first is a pair of broad absorption features toward PKS\,0637$-$752 consistent with \NeVIII\ doublet absorption at $z=0.60552$ with $\log\,N_{\rm NeVIII}=14.0$.  \Lyb\ and \OVI\ \lam1032 non-detections place $3\sigma$ column density upper limits of $\log\,N_{\rm HI}\le13.6$ and $\log\,N_{\rm OVI}\le13.4$, respectively.  A similar pair of weak absorption features in the HE\,0238$-$1904 sight line is consistent with a $z=0.50511$ \NeVIII\ doublet: $\log\,N_{\rm NeVIII}=14.1$, $\log\,N_{\rm HI}\le 13.4$, and $\log\,N_{\rm OVI}\le13.2$.  These two \NeVIII-only systems provide interesting limits on the temperature of the absorbing gas.  If in collisional ionization equilibrium (CIE) and with the solar Ne/O abundance, the \OVI\ non-detections imply a gas temperature of $T\ga3\times10^6$~K.  This is consistent with the broad line profiles observed in the \NeVIII\ lines ($b_{\rm thermal}\sim50$~\kms) and the \HI\ non-detection.


The effective pathlength for \NeVIII\ systems with $N_{\rm NeVIII}\approx10^{14}$~\cd\ in our survey is only $\Delta z\sim2$, quite small compared with the other species presented here.  With only three significant detections over this pathlength, only one of which is confirmed with absorption in other species, it is clear that \NeVIII\ absorbers detectable in data of modest S/N are rare.  Based on the detection of three \NeVIII/\OVI\ systems in the high-S/N spectrum of PG\,1148$+$549 ($z_{\rm em}=0.9754$), \citet{Meiring13} estimate $d{\cal N}/dz=7^{+7}_{-3}$ for \NeVIII\ systems with $W_r>30$ m\AA.  Our modest-S/N finds $d{\cal N}/dz\sim3$ with large uncertainties for \NeVIII\ systems with $W_r>30$ m\AA.  Effective surveys for a statistical sample of \NeVIII\ systems will require high-S/N observations of AGN at redshifts of $z\sim1$ or a deeper survey at slightly lower redshifts ($z_{\rm NeVIII}\ga0.41$) with the COS G130M/1222 setting.  


%

\subsection{Baryon Census} 
The majority of baryons at {\em all} epochs are not in the form of virialized, luminous matter \citep{Shull12a}.  Thus observations of the diffuse IGM are the most effective way of tracking the majority of normal matter across the history of the universe.  As in previous papers \citep{Danforth05,DS08,Tilton12}, we calculate the baryon content of the IGM systems, here observed with COS.  We compute two quantities: $\Omega_{\rm ion}$, the contribution to closure density by a particular element and ionization stage; and $\Omega_{\rm IGM}^{(ion)}$ which estimates the fraction of closure density represented by all the gas traced by absorption in a particular species.  $\Omega_{\rm ion}$ is a purely-observational quantity with no corrections, while $\Omega_{\rm IGM}^{(ion)}$ must include assumptions about metallicity $(Z/Z_\sun)$, solar elemental abundance $(M/H)_\sun$, and ion fraction $(f_{\rm ion})$.  

A full discussion of our methodology is presented in Section 2.4.1 of \citet{Tilton12} from which we use Eq.~(5) to calculate $\Omega_{\rm ion}$
\begin{eqnarray}
\Omega_{\rm ion}&=&(1.365\times10^{-23}~{\rm cm^2})h_{70}^{-1}(m_{\rm ion}/{\rm amu})\times\nonumber \\
&& \sum\limits_{i=\log N_{\rm min}}^{\log N_{\rm max}}\left[\frac{\partial^2{\cal N}(\log N)}{\partial \log N\,\partial z}\right]_i \langle N_i\rangle\Delta\log N_i.
\end{eqnarray}
For metal ions, we calculate $\Omega_{\rm IGM}^{(ion)}$ via Eq.~(6) of \citet{Tilton12}
\begin{eqnarray}
\Omega_{\rm IGM}^{\rm(ion)}&=&\frac{1.83\times10^{-23}h_{70}^{-1}{\rm cm^2}}{f_{\rm ion}(Z/Z_\sun)\,(M/H)_\sun}\times\nonumber\\
&& \sum\limits_{i=N_{\rm min}}^{N_{\rm max}}\left[\frac{\partial^2{\cal N}(\log N)}{\partial \log N\,\partial z}\right]_i \langle N_i \rangle\,\Delta\log N_i.
\end{eqnarray}
Instead of assuming a constant metallicity $Z$ and ion fraction $f_{\rm ion}$ as we have done in previous calculations \citep[][]{Penton04,Danforth05,DS08}, we take advantage of the covariance of the product $f_{\rm ion}\,(Z/Z_\sun)$ seen in in cosmological simulations.  We used parameterized fits of the form $f_{\rm ion}\,(Z/Z_\sun)=A\,(N_{\rm ion}/10^{14}~\rm cm^{-2})^B$.  Parametric coefficients for the species are as follows: \OVI: A=0.015, B=0.700 \citep{Shull12a}; \NV: A=0.036, B=0.617; \CIV: A=0.009, B=0.690 (B. Smith, 2013, priv. comm.).

The baryon fraction traced by photoionized \HI\ absorbers in the \Lya\ forest $\Omega_{\rm IGM}^{(HI)}$ is calculated via Eq.~(10) of \citet{Tilton12}
\begin{eqnarray}
\Omega_{\rm IGM}^{\rm(HI)}(z)&=&(9.0\times10^{-5})h_{70}^{-1}p_{100}^{1/2}T_{4.3}^{0.363}\times \nonumber\\
&& \frac{(1+z)^{0.2}}{[\Omega_m(1+z)^3+\Omega_\Lambda]^{1/2}}\times\nonumber\\
&& \sum\limits_{i=N_{\rm min}}^{N_{\rm max}}\left[\frac{\partial^2{\cal N}(\log N)}{\partial \log N\,\partial z}\right]_i\,\Delta \log N_i\,\left(\frac{\langle N_i\rangle}{10^{14}~{\rm cm^{-2}}}\right)^{1/2}
\end{eqnarray}
where the temperature $T$ is normalized at 20,000~K, column density is in units of $10^{14}~\rm cm^{-2}$, and the AGN sight line impact parameter $p_{100}$ is normalized to 100~kpc.  See full discussion in \citet{Tilton12} and \citet{Shull12a}.

We calculate $\Omega_{\rm ion}$ and $\Omega_{\rm IGM}^{(ion)}$ values for \HI, \OVI, \NV\ and \CIV\ (Table~\ref{tab:omegas}).  Uncertainties are the per-bin errors added in quadrature.  The dominant source of random error for all $\Omega$ calculations is small number statistics.  For this reason we sum over the column density range of typical, weak and moderate lines ($N_{\rm HI}<10^{16}$~\cd, $N_{\rm metal}<10^{15}$~\cd) and do not include the rare, high-column density absorbers.  These may add an additional $6-8$\% to $\Omega_b^{\rm (HI)}$.  The lower limit to the column density range is observationally motivated and corresponds to $W_r=30$~m\AA.  The sample of \HI\ absorbers is large enough to divide the sample into redshift bins of $\Delta z\approx0.1$ and maintain reasonable statistics.  Metal ions lack sufficient detections to subdivide the same way, and thus $\Omega_{\rm ion}$ and $\Omega_{\rm IGM}^{(ion)}$ for each metal ion are calculated over a single redshift range.  

The $\Omega_b$ analysis finds that the photoionized \Lya\ forest at $z\la0.5$ can account for $\sim20-25$\% of the baryons while WHIM gas as traced by \OVI\ can account for $\sim11$\%.  These fractions are smaller than those found in our previous surveys \citep[][]{Penton04,Danforth05,DS08} since we use the more realistic ionization corrections and $f_{\rm ion}\,(Z/Z_\sun)$ method described above while previous surveys used a constant metallicity $Z/Z_\sun=10$\% and an ionization fraction near the peak of the CIE abundance for that ion.  However, these results are consistent with those of \citet{Tilton12} and \citet{Shull14b} both of which apply the same covariance technique applied here to the STIS and FUSE data of \citet{Tilton12}.  For a more detailed comparison of the evolution of $\Omega_{\rm metal}$ in the IGM with previous work including higher-redshift studies \citep[see][]{Shull14b}. 

\subsection{Metal-ion Abundances}

Over the past two decades, there have been numerous studies of the metallicity evolution of the IGM, probed by the strong metal-ion absorbers in the rest-frame ultraviolet (\CIV, \SiIV, \OVI, and a few others).  These metal abundances are parameterized by their densities relative to the cosmological closure density, for example $\Omega_{CIV}$.  \citet{Shull14b} discussed the metal-line measurements from 49 low-redshift ($z<0.4$) HST/COS absorbers to find $\Omega_{CIV}=10.1^{+5.6}_{-2.4}\times10^{-8}\,(h_{70}^{-1}$.  In their Table~2, they compare this value to updated measurements from HST/STIS (Tilton et al. 2012).  The \citet{Tilton12} number was revised by \citet{Shull14b} to $\Omega_{CIV}=8.1^{+4.6}_{-1.7}\times10^{-8}\,h_{70}^{-1}$ which is not statistically different from the COS measurement.  The adjustments were based on three corrections to the method: (1) computation of absorbers over a consistent range of column density ($12.87<\log N_{CIV}<14.87$); (2) re-evaluation of the absorber sensitivity and effective redshift pathlength; (3) fitting the distribution $f(N,z)$ to an analytic form and integrating over the assumed column-density range.  The first restriction eliminated a few strong, uncertain absorbers that biased the computation.  The second and third adjustments resulted in a more robust computation, as discussed extensively in Section 2.2 and Table~2 of \citet{Shull14b}.  There is now general agreement on the low-redshift value for the \CIV\ metal abundance, $\Omega_{CIV}\sim10^{-7}~(H_{70})^{-1}$. 

\begin{deluxetable*}{lccrccc}
\tabletypesize{\footnotesize}
\tablecolumns{7} 
\tablewidth{0pt} 
\tablecaption{$\Omega$ Values}
\tablehead{
  \colhead{Species}  &
  \colhead{$\log N$\tnma} &
  \colhead{$z_{\rm abs}$} &
  \colhead{${\cal N}_{\rm abs}$}&
  \colhead{$\Omega_{\rm ion}$} &
  \colhead{$\Omega_{\rm IGM}^{(ion)}$} &
  \colhead{$\Omega_{\rm IGM}^{(ion)}/\Omega_b$\tnmb} \\
  \colhead{} &
  \colhead{} &
  \colhead{} &  
  \colhead{} &  
  \colhead{($\times10^{-8}$)} &
  \colhead{($\times10^{-3}$)} & 
  \colhead{(\%) }    
          }     
\startdata
H\,I  & $12.8-16$ & $ 0-0.1$  & 801 &$ 18.01^{+ 2.49}_{-2.06}$&$  7.70^{+0.43}_{-0.38}$&$16.9^{+0.9}_{-0.8}$ \\
H\,I  & $12.8-16$ & $ 0.1-0.2$& 660 &$ 24.14^{+ 4.21}_{-3.36}$&$  8.82^{+0.60}_{-0.51}$&$19.4^{+1.3}_{-1.1}$ \\
H\,I  & $12.8-16$ & $ 0.2-0.3$& 369 &$ 20.89^{+ 4.69}_{-3.49}$&$  8.14^{+0.71}_{-0.58}$&$17.9^{+1.6}_{-1.3}$ \\
H\,I  & $12.8-16$ & $ 0.3-0.4$& 262 &$ 32.95^{+ 7.34}_{-5.59}$&$ 11.66^{+1.18}_{-0.95}$&$25.6^{+2.6}_{-2.1}$ \\
H\,I  & $12.8-16$ & $0.4-0.47$& 140 &$ 40.47^{+14.66}_{-9.80}$&$ 14.19^{+2.32}_{-1.66}$&$31.2^{+5.1}_{-3.7}$ \\
&&&&&&\\                    
H\,I  & $12.8-16$ & $ 0-0.4  $&2092 &$ 24.00^{+ 4.99}_{-3.84}$&$  9.08^{+0.78}_{-0.64}$&$20.0^{+1.7}_{-1.4}$ \\ 
O\,VI & $13.4-15$ & $ 0.1-0.75 $& 255 &$ 40.37^{+4.54}_{-3.07} $&$  4.96^{+0.37}_{-0.31}$&$10.9^{+0.8}_{-0.7}$ \\
N\,V  & $13.2-15$ & $ 0-0.45   $&  46 &$  2.10^{+0.53}_{-0.35} $&$  1.57^{+0.32}_{-0.23}$&$ 3.4^{+0.7}_{-0.5}$ \\
C\,IV & $12.9-15$ & $ 0-0.16   $&  68 &$  9.89^{+3.24}_{-1.73} $&$  3.63^{+0.71}_{-0.45}$&$ 8.0^{+1.6}_{-1.0}$ \\
\enddata
\tablenotetext{a}{Additional baryons may be present outside the range ($12.8\le\log N_{\rm HI}\le 16.0$) listed here.}
\tablenotetext{b}{$\Omega_b=0.0455\pm0.0028$ \citep{Larson11}.}
\label{tab:omegas}
\end{deluxetable*}

\clearpage
\section{Discussion}

\subsection{Evolution of the Low-$z$ IGM}

\begin{figure*}
  \epsscale{1}\plottwo{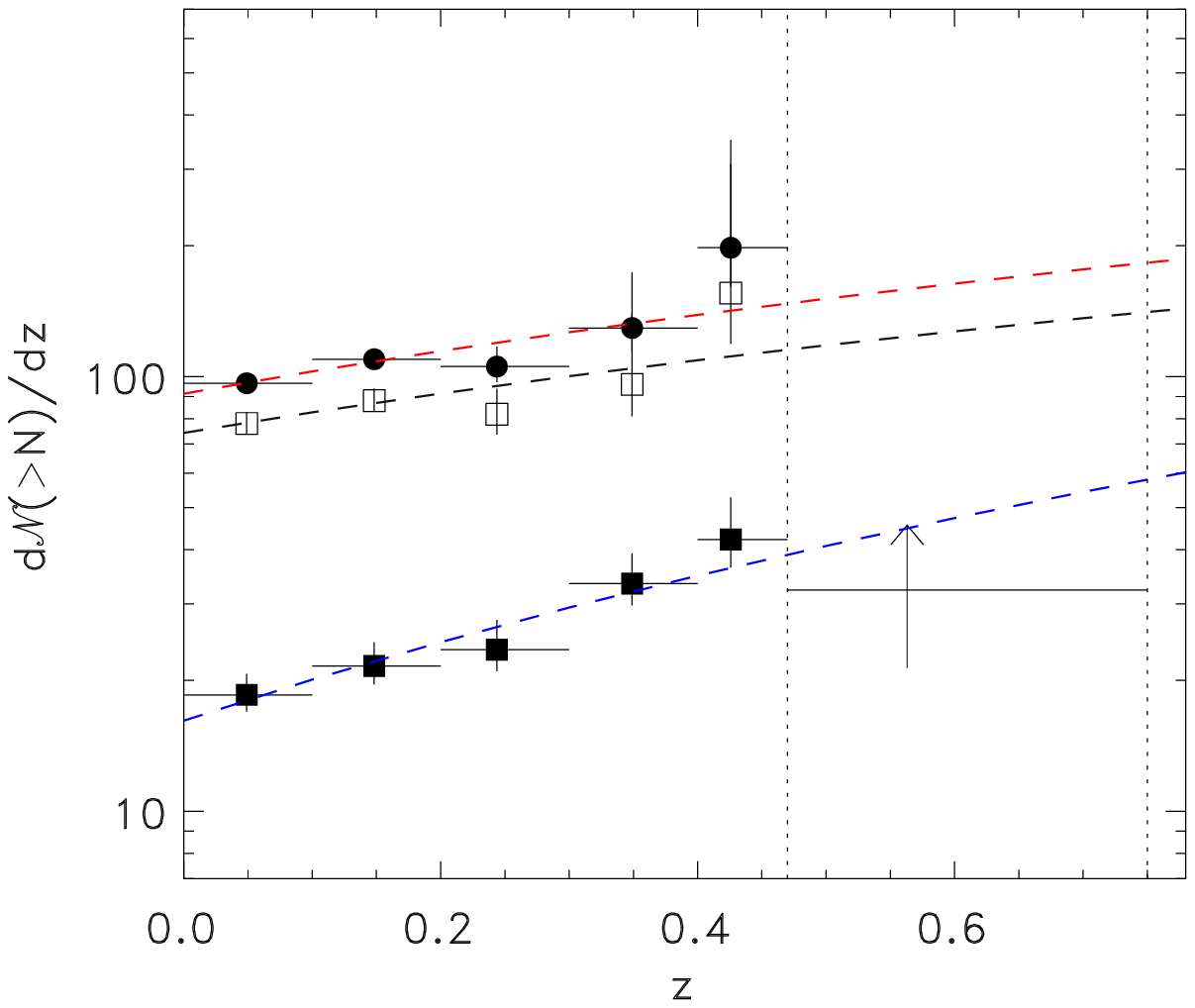}{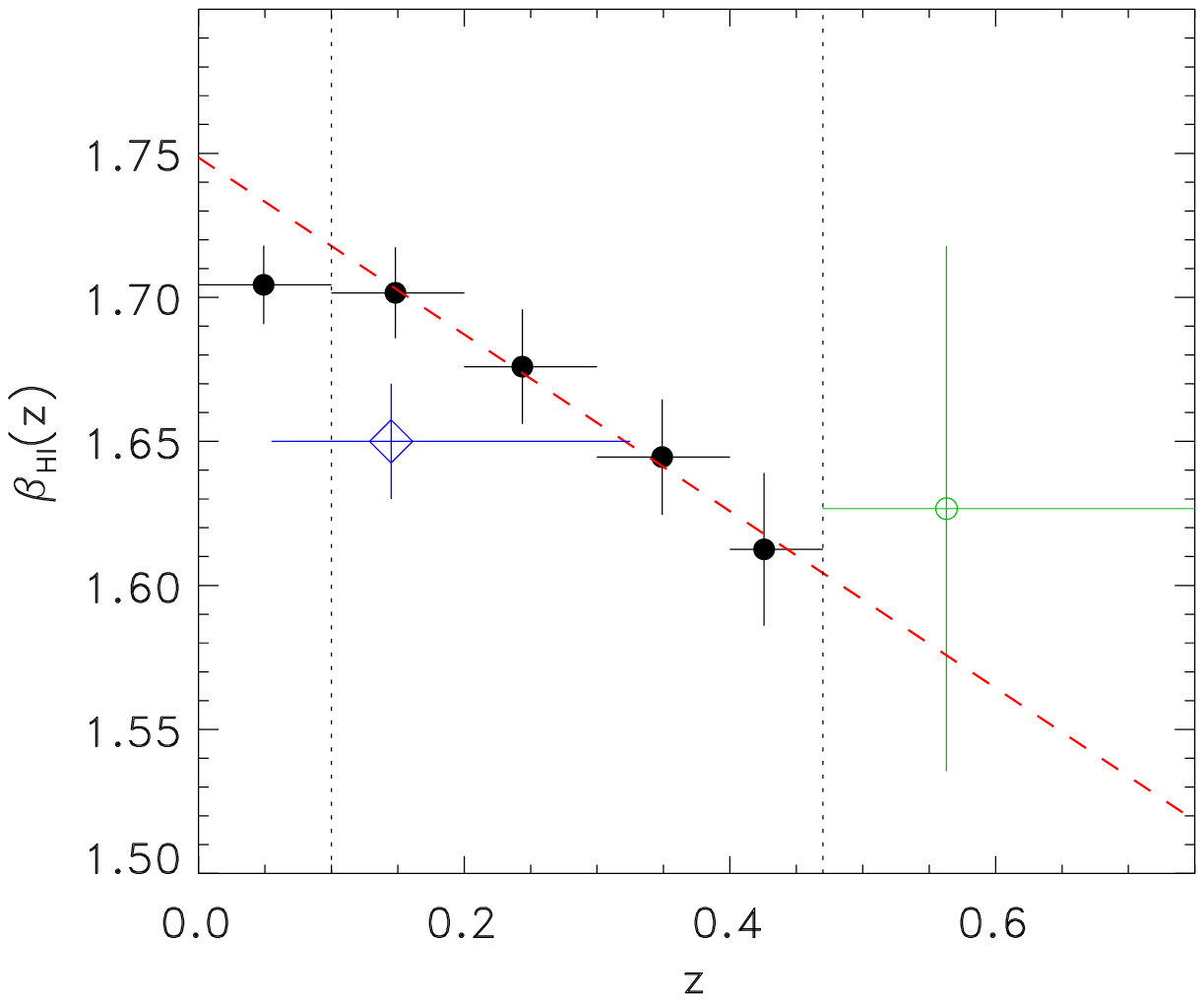}
  \caption{Evolution in the \Lya\ forest.  Left panel: integrated
  \dndz\ values for all ($N_{\rm HI}>10^{13}~\rm cm^{-2}$, filled
  circles), strong ($N_{\rm HI}>10^{14}~\rm cm^{-2}$, filled squares),
  and weak ($10^{13}~{\rm cm^{-2}}\le N_{\rm HI}<10^{14}~\rm cm^{-2}$,
  open squares) H\,I systems as a function of redshift.  The evolution
  of the stronger systems can be fitted by the relationship $d{\cal
  N}(>N,z)/dz= C_0\,(1+z)^\gamma$ with $\gamma_{\rm
  strong}=2.3\pm0.1$, $C_{0, \rm strong}=16\pm1$ (blue dashed line).
  The total and weaker distributions are fitted with $\gamma_{\rm
  all}=1.24\pm0.04$, $C_{0,\rm all}=91\pm1$ (red) and $\gamma_{\rm
  weak}=1.15\pm0.06$, $C_{0,\rm weak}=74\pm1$ (black).  Vertical
  dotted lines show the redshift limits of the \Lya\ and \Lyb\ forests
  which can be observed in COS/FUV data.  Right panel: power law index
  $\beta(z)$ as a function of redshift.  A linear fit to the
  $0.1<z_{\rm abs}<0.47$ bins (dashed red line) shows a steep
  evolution of $\beta(z) = (1.75\pm0.03)-(0.31\pm0.10)\,z$.  A
  high-redshift ($0.47<z<0.75$, green circle) data point based on
  \Lyb$+$\Lyg\ H\,I detections is included in the Figure, but not used
  in the fit.  The full-sample value of $\beta=1.65\pm0.02$ is shown
  as a blue diamond at the median absorber redshift along with the
  $\pm1\sigma$ range.}  \label{fig:lyaevol}
\end{figure*}

The COS IGM survey samples a large fraction of the history of the universe with a statistically-significant number of absorbers seen along many sight lines.  Our \Lya\ forest sensitivity with the COS G130M/G160M gratings ends at $z\approx0.47$, equivalent to a lookback time of 4.9~Gyr or 35\% of the age of the universe, so it is logical to look for changes in the overall sample properties (such as \dndz) over that time.  The sight lines are biased toward low-redshift AGN targets (median $z_{\rm AGN}=0.19$) and the absorbers are similarly biased toward lower redshifts (the median and $\pm1\sigma$ redshift sensitivity of our survey is $z_{\rm abs}=0.14^{+0.19}_{-0.10}$).  We calculate effective pathlength as a function of column density as above, but use only the portion of each sight line which probes a particular redshift range.  A limit and final point are included in Figure~\ref{fig:lyaevol} the bin at $0.47<z<0.75$ in which systems are found via \Lyb\ and \Lyg\ absorption.  However, the uncertainties in pathlength for this range, as well as the small sample size in this redshift bin, make any firm conclusions beyond $z\approx0.47$ difficult.

The left panel of Figure~\ref{fig:lyaevol} shows how the observed \dndz\ changes as a function of redshift in our sample.  We fit the cumulative frequency of lines above a certain column density as $d{\cal N}(>N,z)/dz=C_0\,(1+z)^\gamma$.  The strong sample ($N_{\rm HI}>10^{14}$~\cd) shows clear evolution with $\gamma=2.3\pm0.1$, $C_0=16\pm1$.  Evolution in the weaker sample ($10^{13}$~\cd $<N_{\rm HI}<10^{14}$~\cd) can be fitted with $\gamma=1.15\pm0.05$, $C_0=74\pm1$, but there is an increase in weak systems at $z\ga0.3$, which suggests that a simple power law may not be appropriate for weak system evolution.  The observed $\gamma>0$ means that the frequency of IGM absorbers is lower at $z=0$ than at higher redshift, while the difference $\Delta\gamma\approx1$ between strong and weak systems means that weaker systems become relatively more dominant at lower redshifts.  A fit to the $N_{\rm HI}\ge 10^{13}$~\cd\ sample gives $\gamma=1.24\pm0.06$ and $C_0=91\pm1$.

The difference in evolution indices $\gamma$ for strong and weak \HI\ systems implies that the slope $\beta$ of the \dndz\ distribution should become steeper with decreasing $z$.  This is observed in the data as well (right panel of Figure~\ref{fig:lyaevol}).  The evolution at $z\le0.47$ of systems in the range $13\le\log N_{\rm HI}<17$ is fitted by $\beta(z)=(1.70\pm0.01)-(0.15\pm0.06)\,z$.  However, given the heterogenous method of \HI\ column density determination between the $z<0.1$ and $z>0.1$ systems, we prefer the steeper fit $\beta(z)=(1.75\pm0.03)-(0.31\pm0.10)\,z$ (dashed red line).  

In a comparable set of IGM observations at with \HST/STIS in the near-UV, \citet{Janknecht06} measured $\beta=1.60\pm0.03$ for IGM absorbers at $0.5<z<1.9$.  \citet{Rudie13} find $\beta=1.65\pm0.02$ for $\log N>13.5$ \HI\ absorbers at $z\sim2-3$.  However, they note that the slope is shallower ($\beta=1.447\pm0.033$) for absorbers within 700 \kms\ of a galaxy, which they interpret as circumgalactic gas rather than IGM.  Extrapolating the low-$z$ fit to the redshifts of the comparable near-UV and optical studies produces slopes much shallower than are observed; $\beta(z=1)\approx1.4$, $\beta(z=2.5)\approx1.0$.  Thus, our linear relationship should not be extrapolated to redshifts beyond $z\approx0.5$.  

We can now modify Eq.~(\ref{eq:dndzfit}) to
\begin{equation}
\frac{d{\cal N}(>N,z)}{dz}=C_0\,(1+z)^\gamma\,\left(\frac{N}{10^{14}~\rm cm^{-2}}\right)^{-[\beta(z)-1]}.\label{eq:evol}
\end{equation}
Here we adopt $C_{14,0}=16$ and $\gamma=2.3$, and the second fit, $\beta(z)=1.75-0.31\,z$, which predicts stronger evolution.  Theoretically, we would expect the \Lya\ forest to evolve with redshift due to the rapid drop in photoionizing background at $z<2$ and the evolution of density and mass in the large-scale structure of the gaseous filaments.  Because the \HI\ neutral fraction depends on the ratio of ionizing flux to density (the photoionization parameter $U$), the \HI\ fraction will evolve in redshift, and the column-density distribution will shift \citep{Dave10,Smith11,Penton04}.

\subsubsection{The \Lya\ Decrement}

\begin{figure}
  \epsscale{1.2}\plotone{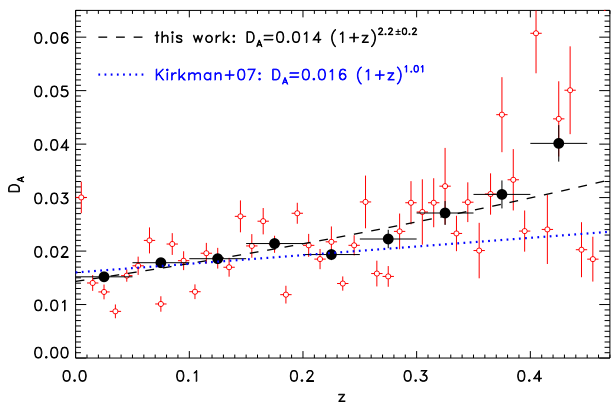}
  \caption{The observational \Lya\ decrement $D_A$ as a function of
  redshift.  The observed frame equivalent widths of all \Lya\
  components within a particular redshift range are summed and divided
  by the clear pathlength in that range.  Cosmic variance contributes
  significant scatter to $D_A(z)$, but the data, binned to $\Delta
  z=0.05$ (black data points), can be fitted with the form
  $D_A=0.014\, (1+z)^{2.2\pm0.2}$ (dashed line).  The dotted line
  shows the HST/FOS fit of \citet{Kirkman07};
  $D_A=0.016\,(1+z)^{1.01}$.}  \label{fig:lydec}
\end{figure}

Another way of assessing the evolution of the \Lya\ forest is to measure the \Lya\ decrement $D_A(z)$, \begin{equation}
D_A(z)=\frac{\sum\limits_i{W_{r,i}\,(1+z_i)}}{\lambda_0\, \Delta z_i(z)},
\end{equation}
the fraction of light removed from the continuum by the \Lya\ forest at any given redshift.  At high redshift, the decrement approaches 100\%, but the modern universe is much more transparent to light at 1215.67 \AA.  Measuring the \Lya\ decrement from the COS survey is a relatively simple matter.  Each absorption component has a measured rest equivalent width $W_{r,i}$.  Summing the observed equivalent widths $W_{\rm obs}=W_r\,(1+z)$, of all the \Lya\ absorption components in a particular redshift range and then dividing by total clear path length $\Delta z_i$ gives the decrement (Eq.~8).  

The \Lya\ decrement is dominated by \HI\ absorbers of column density $13.5<\log N_{\rm HI}<14.5$ which are both common and relatively strong.  Figure~\ref{fig:lydec} shows the decrement in $\Delta z=0.01$ bins (small open circles) and larger $\Delta z=0.05$ bins (filled circles).  There is considerable variation from one redshift bin to the next, but the typical decrement is a few percent.  The larger spread in data points at $z\ga0.35$ is probably due to the increasing effects of cosmic variance as a result of the smaller number of sight lines probing this redshift range.  A power law fit to the data gives $D_A=(0.014\pm0.001)\,(1+z)^{2.2\pm0.2}$ and $D_A=(0.013\pm0.001)\,(1+z)^{2.1\pm0.2}$ for the $\Delta z=0.05$ and $\Delta z=0.01$ binnings, respectively.  This is significantly steeper than the fit of \citet{Kirkman07} who used a pixel-optical-depth technique with data from HST/FOS to find $D_A=0.016\,(1+z)^{1.01}$. 

The distribution of $D_A(z)$ has been used as a constraint on the ionizing radiation field at redshifts $z<0.4$ \citep{Shull15}.  For absorbers that evolve in line frequency as $(1+z)^{\gamma}$, the flux decrement should increase with redshift as $D_A \propto (1+z)^{\gamma+1}$.  The extra factor of $(1+z)$ arises from the fact that the absorption-line equivalent width $W_{\lambda}$ increases with $(1+z)$.  We observe $D_A\propto(1+z)^{2.1-2.2}$, which is consistent with our index $\gamma=1.17\pm0.06$ for unsaturated lines ($13<\log N<14$). 

\subsubsection{Metal Evolution}

\begin{figure}
  \epsscale{1.2}\plotone{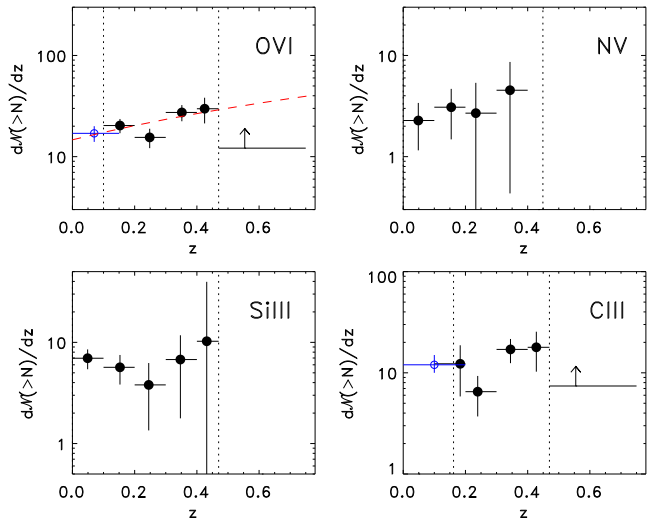}
  \caption{Evolution of metal systems in four species.  Cumulative
  \dndz\ as a function of redshift is shown for systems with
  rest-frame equivalent width $W_\lambda>30$~m\AA.  Lower limits are
  shown with arrows.  Vertical dotted lines indicate the redshift
  range over which the species appears in COS data.  O\,VI (upper
  left) and C\,III (lower right) lack COS coverage at $z_{\rm
  OVI}<0.10$ and $z_{\rm CIII}<0.16$, so literature values (blue, open
  symbols) from DS08 and references therein are plotted in place of
  the lowest-redshift bins.  A dashed line shows the best fit to the
  O\,VI data; all other ions are consistent with no evolution.}
  \label{fig:metal_evol}
\end{figure}

The sample of metal-line systems is much smaller than the IGM \HI\ sample, but we can still use it to constrain the evolution of different metal-ion systems across a significant redshift range.  We group metal systems into the same six redshift bins used for \HI\ and measure the cumulative $d{\cal N}(>N)/dz$ at the observational threshold $W\ge30$~m\AA\ for metal species with both significant redshift coverage and detection statistics.  The results are shown in Figure~\ref{fig:metal_evol} for \OVI, \NV, \CIII, and \SiIII.  \CIV\ is sampled only over a small redshift range, and there are not enough \SiIV\ detections to provide any reliable statistics when spread over multiple redshift bins.  We follow the same procedure as used above for \HI, calculating integrated \dndz\ profiles in up to five redshift bins.  \OVI\ systems at $z<0.1$ and \CIII\ systems at $z<0.16$ are not observed in COS data, so comparable literature values from \citet{Danforth05} and \citet{Danforth06} are used for the lowest-redshift data point.

Statistics are poor even for \OVI, the most numerous of the metal-ion detections.  \OVI\ appears to evolve with redshift in the same sense at \HI.  A fit of the form $d{\cal N}(>N)/dz\propto(1+z)^\gamma$ gives $\gamma_{\rm OVI}=1.8\pm0.8$ at $z<0.47$ for $W_\lambda>30$~m\AA\ including the $z<0.16$ value from \citet{Danforth05} at the lowest redshift bin.  Evolution in the other three ions is poorly constrained.  \NV\ is fitted with $\gamma_{\rm NV}=2.4\pm2.9$.  \SiIII\ and \CIII, which typically show properties correlated closely with \HI\ (DS08), are also poorly constrained ($\gamma_{\rm SiIII}=-1\pm2$, $\gamma_{\rm CIII}=1.7\pm1.3$).  Much larger samples of metal-ion absorbers are required before metal evolution can be measured with any precision.


\subsection{Clustering of IGM Absorbers}

The two-point correlation function (TPCF) and its amplitude $\xi$ is often used as a measure of clustering in the universe \citep{Peebles80}.  A value of $\xi>0$ means there is greater-than-random clustering, while $\xi<0$ may indicate anti-clustering (voids).  Since our sight lines sample physically unrelated regions in almost all cases, we can use absorbers along them to measure the velocity-space clustering of absorbing materials in the IGM.  

\begin{figure}
  \epsscale{1.1}\plotone{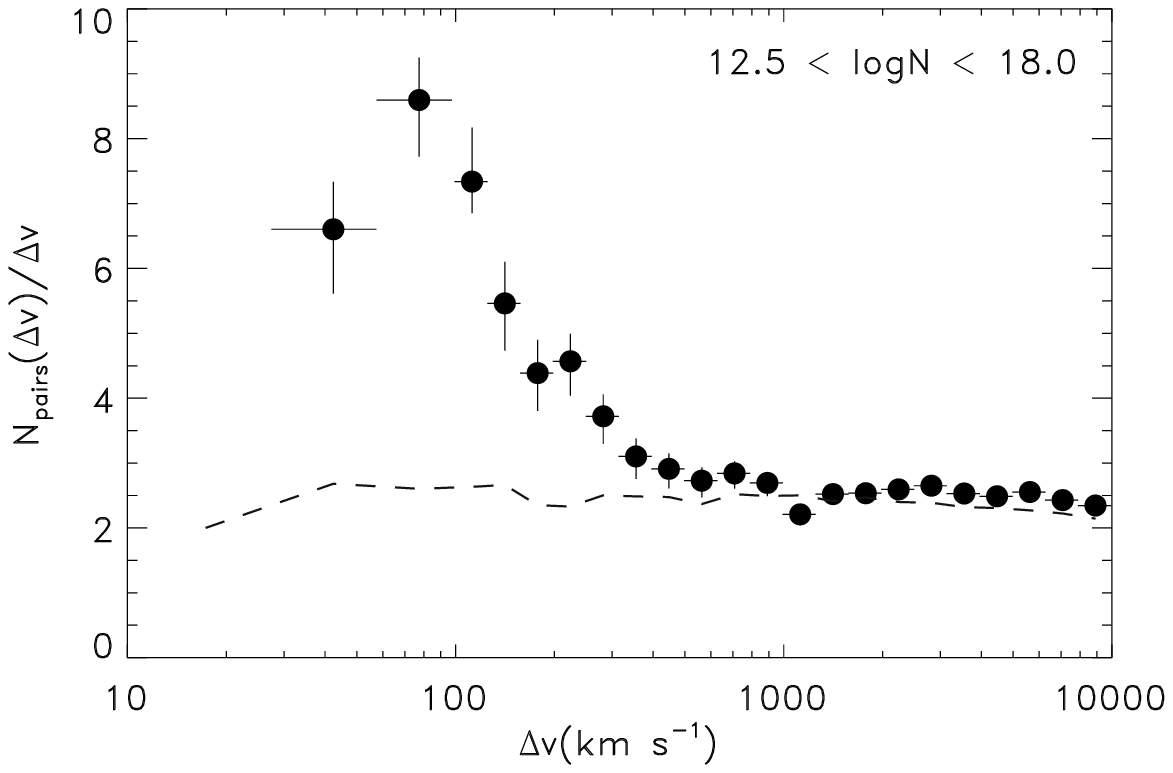}
  \epsscale{1.1}\plotone{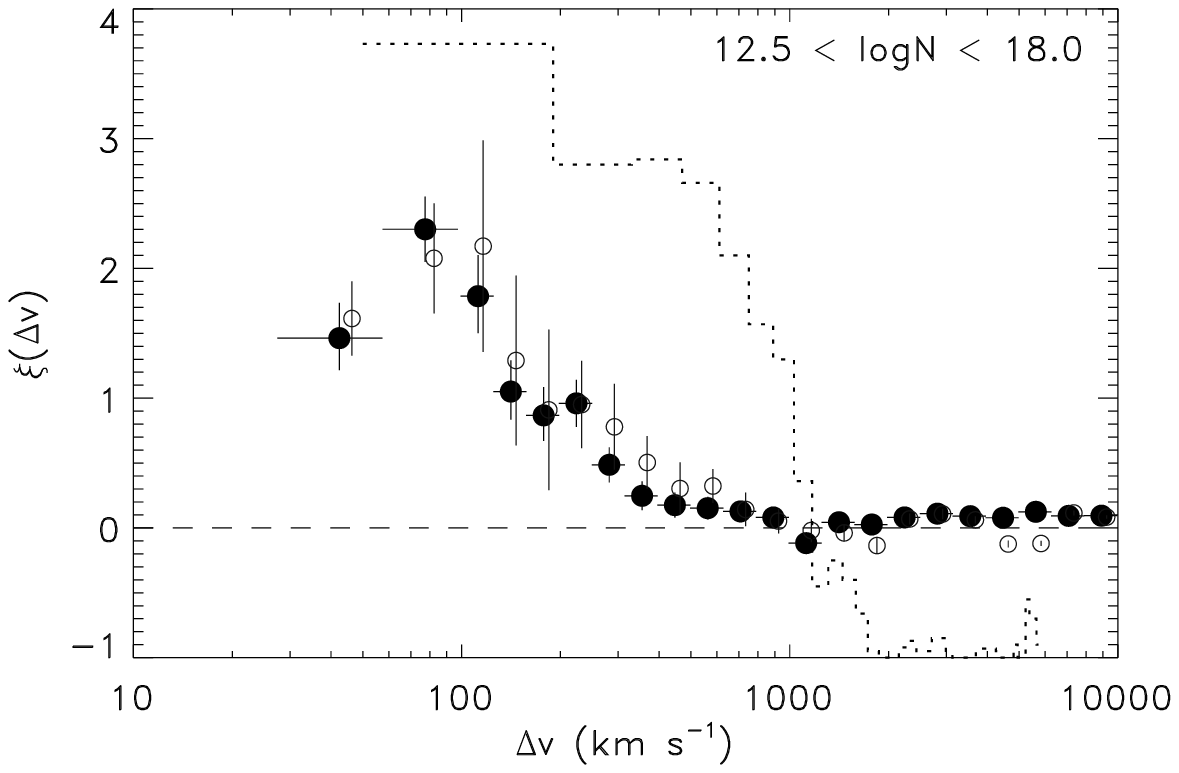}
  \caption{Two-point correlation function in the low-$z$ \Lya\ forest.
  Observed component pairs (top) show significant signal at $\Delta
  v\sim100$ \kms\ compared with a sample of randomly-placed components
  with the same overall \dndz\ behavior (dashed line).  The two-point
  correlation function, $\xi(\Delta v)$ (bottom) shows the same
  clustering at $\Delta v\sim100$ \kms, and no clustering at higher
  velocity separations.  This is in contrast to the TPCF of galaxies
  to galaxies \citep[dotted line;][]{Penton04} which showed
  significant correlation at $\Delta v<1000$ \kms.  \HST/STIS data
  from \citet{Tilton12} show a similar behavior (open circles)
  including the smaller TPCF at low velocity
  separations.}\label{fig:tpcf}
\end{figure}

\begin{figure}
  \epsscale{1.1}\plotone{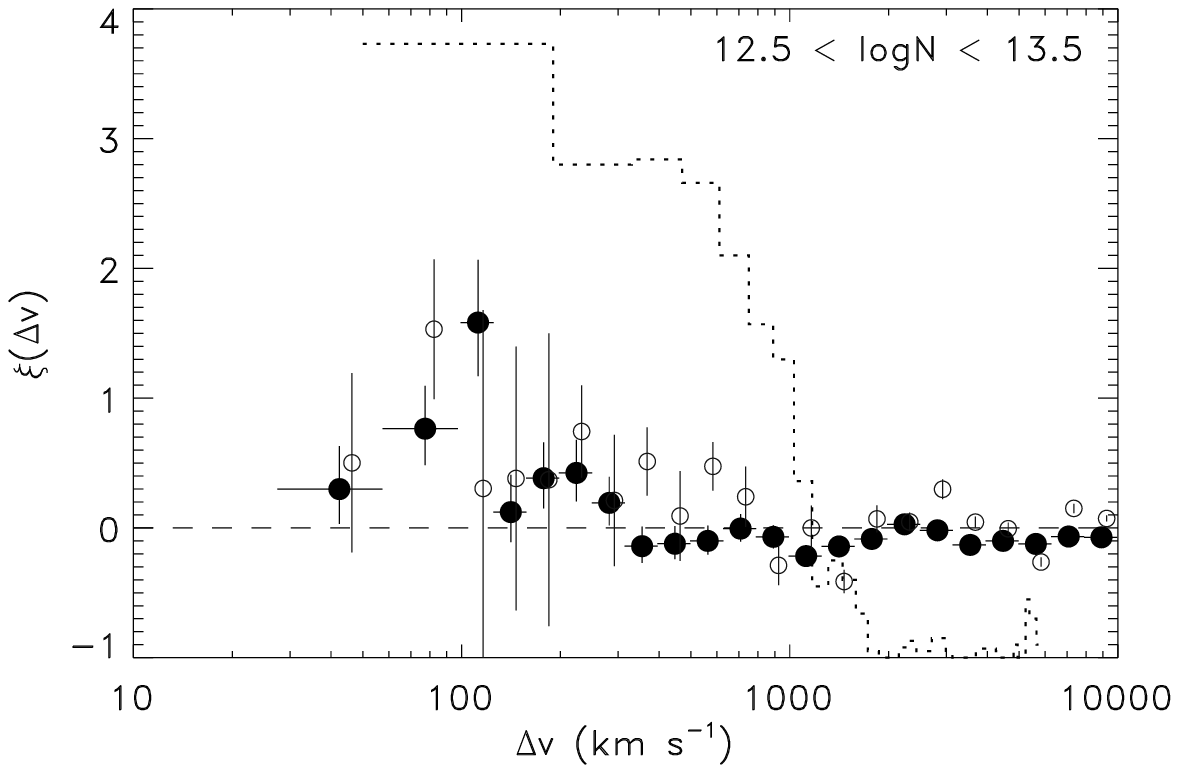}
  \epsscale{1.1}\plotone{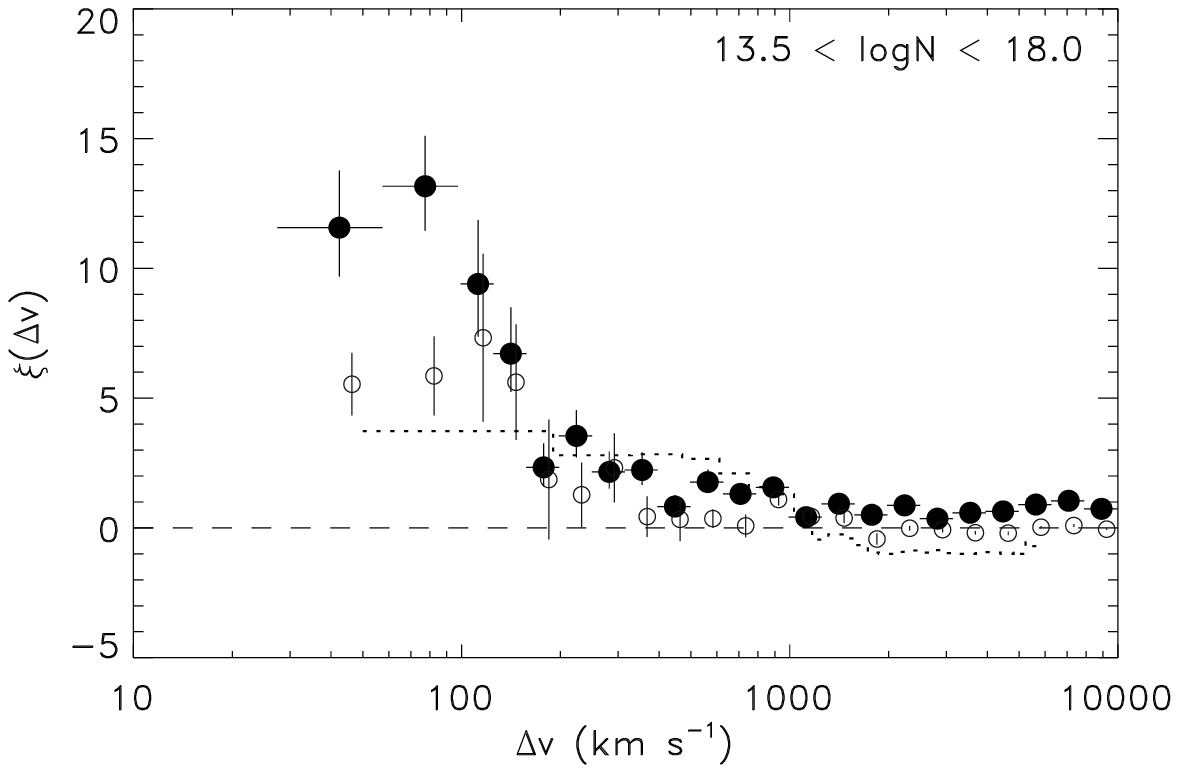}
  \caption{Variation in TPCF with component strength.  Weak \HI\
  absorbers (top, $12.5<\log N<13.5$) show only modest clustering,
  while stronger \Lya\ components (bottom, $\log N>13.5$) show
  significant clustering at $\Delta v\sim50-300$ \kms.  The same trend
  is apparent in the \HST/STIS absorbers from \citet{Tilton12} (open
  circles).  Dotted lines show the galaxy-galaxy TPCF of
  \citet{Penton04}.}  \label{fig:tpcf_bins}
\end{figure}

We calculate $\xi(\Delta v)$ for \Lya\ components in each sight line as 
  \begin{equation}
    \xi(\Delta v)=\frac{N_{\rm obs}(\Delta v)}{N_{\rm ran}(\Delta v)}-1,\label{eq:tpcf}
  \end{equation}
where $N_{\rm obs}$ and $N_{\rm ran}$ are the normalized number of absorption component pairs with a given velocity separation per $\Delta v$ \citep{Kerscher00} and where the velocity separation ($z_1>z_2$) is defined relativistically 
  \begin{equation}
    \frac{\Delta v}{c}\equiv\frac{(1+z_1)^2-(1+z_2)^2}{(1+z_1)^2+(1+z_2)^2}.
  \end{equation}
The observed redshifts $(z_1,z_2)$ are taken as any pair of \Lya\ components in the same sight line detected at $\ge4\sigma$.  For simplicity, and to avoid spurious clustering signal from velocity mis-matches between lines in different transitions, we use only \Lya\ components, not systems in which different closely-spaced components may be grouped together into the same system.  In principle, any absorption component could be used.  See \citet{Labatie10} and references therein for a discussion of biases inherent to this and other estimators of the TPCF.

The random absorber distribution is calculated with a Monte-Carlo simulation using the detailed fit to the observed IGM detection statistics and the actual data in each of the survey sight lines.  Random component locations are simulated 100 times in each sight line.  A similar technique was used by \citet{Penton2} and \citet{Penton04} to simulate absorbers in \HST/FOS and STIS data.  At each resolution element $\Delta v$ in each sight line, we calculate the 4$\sigma$ minimum equivalent width detection supported by the $S/N$ of the data.  The $W_{\rm min}(\lambda)$ vector is converted to $N_{\rm Ly\alpha,min}(z)$ and the probability of finding a component in velocity resolution element $dv$ is given from Eq.~(\ref{eq:evol}) as
\begin{eqnarray}
P_{\rm ran}(z)&=&C_0\,\frac{dv}{c}\,(1+z)^\gamma\, \\ \nonumber
&&\times\left[\left(\frac{N_{\rm min}(z)}{10^{14}~{\rm cm^{-2}}}\right)^{-\beta}-
      \left(\frac{N_{\rm max}(z)}{10^{14}~{\rm cm^{-2}}}\right)^{-\beta}\right].
\end{eqnarray}
Integrated over the \Lya\ pathlength in each component, $P_{\rm ran}(z)$ should equal the number of observed \Lya\ systems in that sight line, modulo cosmic variance.  If the probability at a given resolution element is greater than a randomly generated number, an absorption component is placed at this redshift position.  When the entire redshift pathlength of the sight line has been processed, the number of pairs in the randomly-distributed components is found, and the result $N_{\rm ran}(\Delta v)$ is added to a list of random pairs.  We produce strong and weak pairs by setting limits on $N_{\rm min}$ and $N_{\rm max}$ in both the observed and artificial component lists.

\begin{figure}
  \epsscale{1.2}\plotone{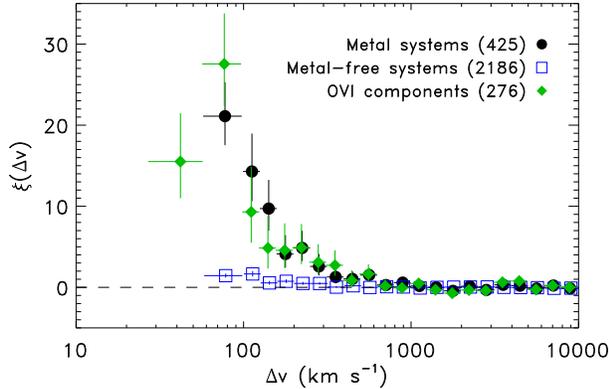}
  \caption{Two-point correlation function of metal/non-metal
  absorbers.  IGM systems with absorption in at least one metal ion
  (black filled circles) show a strong TPCF signal at $\Delta
  v\sim100$~\kms, while \HI-only systems (open squares) show little or
  no signal.  O\,VI \lam1032 components (green diamonds) show a
  similar TPCF behavior to metal-bearing H\,I
  systems.}\label{fig:tpcf_metals}
\end{figure}

Figure~\ref{fig:tpcf} shows the behavior of the observed and random component pairs, normalized to the velocity width in each bin in our sample (top panel).  As expected, the distribution of random pairs is flat as a function of $\Delta v$.  The observed pairs show a significant peak at $50\la\Delta v\la 300$ \kms.  The two-point correlation function (lower panel) shows a significant correlation at $\Delta v\la300$ \kms\ and no signal at much higher velocities.  This is in contrast to the galaxy-galaxy TPCF (dotted) from \citet{Penton04} which shows significant correlation at $\Delta v<1000$ \kms\ and significant anticorrelation at $\Delta v>1000$ \kms.  Whether this anticorrelation is due to the presence of voids in the galaxy distribution, or is an artifact of the TPCF methodology \citep{Kaiser87} is unknown.  We note that the random absorber population (responsible for the denominator in $\xi$) is extremely sensitive to the fit parameters assumed for the \Lya\ forest and its evolution.  This introduces a small uncertainty in the scaling of the $\xi$, but it does not change the overall flat nature of $N_{\rm ran}(\Delta v)$ seen in the top panel of Figure~\ref{fig:tpcf}.

Splitting the TPCF into strong and weak subsamples (Figure~\ref{fig:tpcf_bins}), we see that the strong components ($\log N>13.5$) show signal equal to or stronger than the galaxy-galaxy TPCF at $\Delta v\sim100$ \kms, while the weaker components ($12.5\le\log N<13.5$) show a much smaller clustering signal.  This is in keeping with the picture of strong \HI\ systems in and around galaxy halos, which are clustered \citep{Rudie13}, or at least associated with large-scale structures such as filaments.  Neither strong nor weak samples show significant TPCF at $300\la\Delta v\la1000$ \kms\ where the galaxies are still highly clustered ($\xi\ga3$).

The decrease in $\xi$ at $\Delta v<60$ \kms\ shown in Figures~\ref{fig:tpcf} and \ref{fig:tpcf_bins} is intriguing.  The velocity resolution of COS ($\sim17$ \kms) and the typical width of \Lya\ lines ($b\sim33$ \kms, $FWHM\sim55$ \kms) suggest that the low $\xi$ values in the lowest-velocity bins may be due to finite instrumental resolution, line blending, and related systematic effects.  Post-processing the \citet{Tilton12} catalog of \HST/STIS absorbers shows a similar turn-down at low velocities (open circles in Figures~\ref{fig:tpcf} and \ref{fig:tpcf_bins}).  Since the resolution of the STIS/E140M grating is $\sim7$ \kms\ (compared with $\sim17$ \kms\ for the medium-resolution COS gratings), this hints that the downturn at $\Delta v<100$ \kms\ may be a real effect.  If real, this lack of correlation at the smallest velocities may be indicative of the kinematics within galaxy halos.  However, differentiating blended components, especially in strong absorbers, is subject to quite a bit of systematic uncertainty and we view this apparent downturn in $\xi$ at the smallest $\Delta v$ as suggestive only.  Unfortunately, resolution limitations restrict the ability of modern cosmological simulations to track such small velocity separations \citep[e.g.,][]{CenChisari11}.

At smaller velocity separations, line blending introduces uncertainties and biases into our ability to separate and identify velocity components.  The median doppler parameters of \HI\ and \OVI\ absorbers are comparable, at $\langle b \rangle \approx 30$~\kms, and the full width at half maximum, $\Delta v_{\rm FWHM} \approx 1.67 b$ is approximately 50~\kms.  We therefore distrust any TPCF signal at those separations.

Next, we investigate the clustering properties of IGM systems with and without metal-ion absorption.  We approach this in two ways.  First, we analyze IGM systems rather than the individual \Lya\ components studied above, so that metal absorption can more easily be associated with \HI\ columns despite small velocity uncertainties.  Since systems have a minimum velocity half-width of 30 \kms\ and many systems (particularly metal systems) are broader than this, we restrict this analysis to velocity separations $\Delta v>60$ \kms.  To generate a TPCF, we assume that the distribution of randomly-placed systems in both metal and non-metal systems is flat with $\Delta v$, as in the top panel of Figure~\ref{fig:tpcf}, and that there is no clustering at $\Delta v>1000$~\kms.  Secondly, in order to eliminate bias due to system definitions, we investigate the clustering of metal-ion components themselves using the most commonly-seen transition (\OVI\ \lam1032).  This method is analogous to the \Lya\ TPCF shown in Figure~\ref{fig:tpcf}.  Again, we assume a flat distribution of randomly placed absorbers in each case.  

Figure~\ref{fig:tpcf_metals} shows the clustering properties of the non-metal systems (open circles) along with the metal systems.  Individual metal component clustering is shown in the various colored symbols.  The metal component sample sizes are quite a bit smaller than the \HI\ sample or even the metal/non-metal system samples, and thus the uncertainties are much larger.  However, \OVI\ \lam1032 components show a clustering signal at $\Delta v\approx100$~\kms\ which is considerably stronger than that seen in the metal systems.  This difference may be a result of our process of system definition: in many cases, closely-spaced metal components are grouped together into a single system which will systematically reduce the number of absorber pairs at close velocity separation.  

Qualitatively, metal-ion components and metal-bearing systems show strong clustering ($\xi\sim10-50$) peaked at $\Delta v\sim50-200$ \kms, albeit with substantial uncertainty due to the small size of the samples.  \citet{Pieri13} see a similar trend in a large sample of \Lya\ absorbers at $2.4<z<3.1$ in the BOSS survey, with a high degree of clustering and correlation of metal absorbers at scales down to $\Delta v\approx130$~\kms\ (the resolution limit of their data).  The non-metal systems show a peak $\xi\la1$ at the same velocity range.  This suggests that most of the radial-velocity clustering in the IGM can be attributed to strong, metal-bearing systems in the CGM, again consistent with the picture of strong, metal-enriched absorption being associated with galaxy halos.  

\section{Summary of Primary Results}

We present a high-quality, medium-resolution \HST/COS survey of the IGM along 82 UV-bright AGN sight lines.  Because the sight lines were chosen for sensitivity to weak IGM absorbers at low-redshift over the maximum pathlength, we favor targets observed with both the COS/G130M and G160M gratings with a typical $S/N\ga15$ per resolution element.  We limit the redshifts of the AGN to $0.05<z_{\rm AGN}<0.85$ to maximize IGM pathlength for species of interest (\HI, \OVI, etc.) while minimizing line confusion.  

The spectra were processed with semi-automated continuum fitting and line-finding/measurement routines to minimize the subjective bias associated with many previous IGM surveys.  The identity of absorption features is established through a manual process and lines are remeasured as necessary.  Galactic, instrumental, and probable AGN-intrinsic features are flagged.  In total, 5138 individual lines are identified as absorption from intervening material in the IGM.  This includes 4234 \Lya\ lines, 606 \Lyb\ lines, and 1633 metal-ion lines representing 25 metal ion species.  The median and $\pm1\sigma$ redshift of absorption systems is $z=0.14^{+0.18}_{-0.10}$.  

To better facilitate comparisons between species at the same redshift, the IGM lines are grouped by into 2611 distinct redshift systems of which 418 are detected in at least one metal line.  The most common metal species is \OVI\ (present in 280 systems) followed by \CIII\ (115), \SiIII\ (123), \CIV\ (70), and \NV\ (59).  \NeVIII\ is only detected at a significant level in three systems.  

We present below a summary of our primary science results:
\begin{itemize}
\item The fraction of IGM \HI\ systems detected in one or more metal ions is a strong function of \NHI.  Metals are rarely (less than 10\%) detected in $N_{\rm HI}\la 10^{13.5}$~\cd\ absorbers, but become nearly ubiquitous for strong systems ($N_{\rm HI}\ga10^{15}$~\cd).
\item The cumulative distribution of \HI\ absorbers at $z\le0.47$ in the sample (Figure~\ref{fig:dndz_h1}) follows a power law in \HI\ column density of the form $d{\cal N}(>N)/dz=C_{14}\,(N/10^{14}\rm~cm^{-2})^{-(\beta-1)}$ over the column density range $12\le \log N_{\rm HI}\le 17$ with normalization $C_{14}=25\pm1$ and differential index $\beta=1.65\pm0.02$.  
\item Dividing the sample into redshift bins of $\Delta z\approx0.1$ and analyzing the subsamples, we see clear evolution in both the slope $\beta$ and normalization $C_{14}$ of the distribution.  We parameterize the evolution of the diffuse IGM \HI\ absorbers as $d{\cal N}(>N)/dz=C_{0}\,(1+z)^\gamma\,(N/10^{14}\rm~cm^{-2})^{-[\beta(z)-1]}$ with $C_{0}=16\pm1$, $\gamma=2.3$, and $\beta(z)=(1.75\pm0.03)-(0.31\pm0.10)\,z$ for $z\le0.47$. 
\item Metal systems analyzed in the same manner as \HI\ suggest that \OVI\ evolves in the same sense as \HI\ with $\gamma\sim1.8\pm0.8$.  Smaller samples of \NV, \CIII, and \SiIII\ absorbers do not present clear evidence for evolution. 
\item We calculate the contribution to the closure density by a particular species, $\Omega_{\rm ion}$, and the contribution represented by gas which is traced by a particular species, $\Omega_{\rm IGM}^{\rm (ion)}$.  The values given in Table~\ref{tab:omegas} are consistent with previous surveys.
\item A two-point correlation function (TPCF) of \Lya\ components shows that there is significant clustering of IGM absorbers in radial velocity.  Figure~\ref{fig:tpcf} shows a significant clustering signal at $\Delta v=60-300$ \kms\ and little or no signal at higher velocities.  This is in contrast to the galaxy-galaxy radial velocity clustering found by \citet{Penton04} which shows significant clustering at $\Delta v<1000$~\kms.  Dividing the sample into strong ($\log N_{\rm HI}>13.5$) and weak ($\log N_{\rm HI}<13.5$) absorbers, we see that nearly all of the clustering signal is accounted for by the stronger systems.  Examining metal and non-metal systems reveals an extremely strong TPCF for metal systems as well as in components of common metal-ion transitions (\OVI\ 1032~\AA, etc.)
\end{itemize}

This paper represents many years of work by the COS Science Team to characterize the baryon content, structure, and metallicity of the low-redshift IGM.  At this point, it is worth summarizing the status of low-redshift IGM surveys and models:  What properties of the IGM are now well established?  What are the remaining uncertainties?  Where are avenues for future observations and theoretical work?  The COS (G130M/G160M) spectra had resolving power $R\approx18,000$ and identified $\sim2600$ IGM absorbing system and 418 metal-line systems (280 \OVI, 70 \CIV).  The total redshift path length, $\Delta z=21.7$, is four times larger than our previous low-redshift surveys with STIS \citep{DS08,Tilton12} which identified 650-750 \Lya\ absorbers.  For comparison, the Quasar Absorption Line Survey HST/FOS Key Project \citep{Jannuzi98} observed 83 AGN sight lines.  Although the FOS survey covered a larger path length, $\Delta z\approx49$, its line sample was smaller, with 1129 \Lya\ lines, 107 \CIV\ systems, and 41 \OVI\ systems, and the spectra were obtained at an order-of-magnitude lower resolution ($R\approx1300$).  

With the larger COS medium-resolution IGM survey, the bivariate distribution of \HI\ absorbers, $f(N_{\rm HI},z)$, is now well established at redshifts $z\le0.4$.  Its parameterization in column density and redshift are fitted to the form $N_{\rm HI}^{-\beta} (1+z)^{\gamma}$, with a differential low-redshift slope of $\beta=1.65\pm0.02$ over the range $12<\log N_{\rm HI}<17$, identical to the slope at $\langle z\rangle=2.4$ found by the Keck Baryonic Structure Survey for \Lya\ absorbers with $\log N_{\rm HI}>13.5$.  (An earlier survey with VLT/UVES by \citet{Kim02} found $\beta\approx1.5$ over the range $1.5<z<4.0$.)  The COS survey has also established that strong \HI\ absorbers evolve faster than weak absorbers, with redshift-evolution indices $\gamma_{\rm weak}=1.15\pm0.05$ for $13<\log N_{\rm HI}<14$ and $\gamma_{\rm strong}=2.3\pm0.1$ for $\log N_{\rm HI}>14$.  However, for the low column density absorbers, the $dN/dz$ values determined in this work are somewhat too high to join seamlessly with the higher redshift line densities found from ground-based spectroscopy using the slopes obtained by both the COS and ground-based studies.  For the higher column density absorbers, a slope change is {\em required} somewhere between $0.4\la z\la 2$.  The evolution of the IGM cannot be well-understood at all cosmic epochs simply by interpolating between the the low ($z\la0.4$) and high redshift ($z\ga2$) regimes.

For the cosmological baryon census, the COS survey confirms previous studies \citep{Shull12a,Tilton12} which found that $\sim20-25$\% of the baryons reside in the intergalactic \Lya\ forest, ranging in column density from $12.8<\log N_{\rm HI}<16.0$.  Additional matter exists in higher column density systems, extending up to $\log N_{\rm HI}\approx19$ \citep{Penton04}.  With nearly 2600 \HI\ absorbers, the COS survey provides better statistics.  Some uncertainties remain for weak \Lya\ lines with equivalent widths $W_{\rm \lambda}<30$~m\AA\ ($\log N_{\rm HI}<12.74$) and for rare high-column density systems.  For strong absorbers ($\log N_{\rm HI}>15$) our survey typically has fewer than 10 \Lya\ lines per column density bin ($\Delta \log N=0.2$).  In addition, the ionization corrections required to convert the absorber distribution, $f(N_{\rm HI},z)$, into a baryon density parameter, $\Omega_b$, involve uncertain physical parameters such as the ionizing UV background and hydrogen photoionization rate ($\Gamma_H$), characteristic absorbers sizes, and cloud temperature.  In fact, the column density distribution of \Lya\ absorbers can be used, together with cosmological simulations, to constrain the amplitude of the low-redshft ionizing background \citep{Kollmeier14,Shull15}.  

Further progress in characterizing the evolution of structure in the low-redshift IGM will likely occur along several fronts.  First, we need to characterize the \HI\ distribution at the low-column end, searching for the expected turnover in the distribution below $\log N_{\rm HI}<12.4$.  This will require high-S/N data to measure \Lya\ equivalent widths down to 5 m\AA\ ($\log N_{\rm HI}\approx 12.0$.)  Second, we have little knowledge of the evolution of the IGM between $z\approx1.5$ and $z\approx0.4$, the epoch when physical conditions change rapidly as star-formation rates decline.  Probing the IGM beyond $z>0.4$ will require extensive surveys in the ``near-ultraviolet desert" ($\lambda>1700$~\AA), both for \Lya\ absorbers as well as key lines of carbon, oxygen, nitrogen, and silicon that measure the extent of metal transport from galaxies into the IGM.


\medskip
\medskip

It is our pleasure to acknowledge fruitful discussions with many colleagues during the course of this work including Ben Oppenheimer, Devin Silvia, and Joshua Moloney.  Julie Davis (Colorado), Alex Filippenko, Brad Cenko, Weidong Li, Weikang Zhang (U. C. Berkeley), Meg Urry, Erin Bonning Wells, and Jedidah Isler (Yale) were instrumental in the acquisition of several of the datasets used in this study.  We also acknowledge the valuable contributions made by our anonymous referee who provided the thorough, critical, expert review and lead to a much-improved scientific paper.
This work was supported by NASA grants NNX08AC14G (COS Science Team), HST-AR-1243.06, HST-GO-12612.01-A, and HST-GO-13008.01-A to the University of Colorado at Boulder.  
BAK and JTS acknowledge support from NSF grant AST1109117.  
JMS acknowledges NSF grant AST07-07474 and thanks the Institute for Astronomy at Cambridge University for support as a Sackler Visiting Lecturer.  
Scott Fleming at MAST was instrumental in the production of the High-Level Science Products associated with this survey.  The authors made extensive use of the MAST, NED, and ADS Archives during this work. 

\medskip
{\it Facility: HST (COS), FUSE, HST (STIS)}


\end{document}